\documentclass[11pt]{article}
\usepackage[hidelinks]{hyperref} %hyperlink but without urgly boxes
\usepackage{setspace}
\usepackage[natbibapa]{apacite}
\usepackage[affil-it]{authblk}

\usepackage{scrextend}
\usepackage{multirow} %This then provides the command needed for spanning rows
\usepackage{booktabs,caption}
\usepackage[flushleft]{threeparttable}
\usepackage{natbib} % BibTex - Remember in configure TexMaker, set BibTex as bibtex  %.aux
\usepackage[capposition=bottom]{floatrow}
\usepackage{chngcntr} % to change equation numbering in appendix
\usepackage[english]{babel} % English language/hyphenation
\usepackage{amsmath,amsfonts,amsthm} % Math packages
\usepackage{microtype}
\usepackage[]{geometry}
\usepackage{mathtools} % for multiple lines in math equations
\usepackage{graphicx}

\usepackage[T1]{fontenc}

\usepackage{bookmark} % to show the bookmarks in output pdf file
\usepackage{sectsty} % Allows customizing section commands
\sectionfont{\fontsize{16}{16}\scshape}
\usepackage{tikz}
\usetikzlibrary{calc}
\usetikzlibrary{matrix}
\usetikzlibrary{positioning}
\usepackage{fancyhdr} % Custom headers and footers
\newtheorem{lemma}{Lemma}
\newtheorem{Proposition}{Proposition}
\newtheorem{Corollary}{Corollary}

\usepackage{multirow}
\usepackage{setspace}

\usepackage{pdfpages}
\begin{document}
\onehalfspacing
%\pagenumbering{gobble}

\selectlanguage{english}

\title{\textbf{Doing Less for More: Consumer Search and Undertreatment in Credence Service Markets}}
%\date{\nodate}
\date{}
\author{Xiaoyan Xu, Weishi Lim, Xing Zhang, Jeff Cai} %\thanks{}% 
%Xiaoyan Xu is an associate professor in marketing at Research Institute of Economics and Management, Southwestern University of Finance and Economics. Email: xuxy@swufe.edu.cn. Weishi Lim is an associate professor in marketing at NUS Business School, National University of Singapore. Email: 23weishi.lim@gmail.com. Xing Zhang is an assistant professor in marketing at Graduate School of Business, Sungkyunkwan University. Email: zhangxingis@gmail.com. Jeff Cai is an assistant professor in marketing at NUS Business School, National University of Singapore. Email: jeffccx@gmail.com}}

\maketitle

%\ARTICLEAUTHORS{%
%\AUTHOR{John Doe}
%\AFF{Department of Operations Management, Operations University, \EMAIL{jdoe@operations.edu}} %, \URL{}}
%\AUTHOR{Jane Doe}
%\AFF{Institute for Supply Chain Management, University of Management \EMAIL{jdoe@iscm.edu}}
%} % end of the block

\begin{abstract}
% 255 / 350
In many service markets, expert providers possess an information advantage over consumers regarding the necessary services, creating opportunities for fraudulent practices. These may involve overtreatment through unnecessary services or undertreatment with ineffective solutions that fail to address consumers' problems. When issues are resolved, consumers exit the market; when unresolved, they must decide whether to revisit the initial provider or seek a new one. Little is known about how repeated interactions and the consumer search process influence expert fraud and consumer welfare in such markets.
We develop a dynamic game-theoretic model to examine the role of consumer search behavior and repeated interactions between consumers and service providers. We find that overtreatment and undertreatment can arise simultaneously in equilibrium. That is, an opportunistic expert would prescribe a serious treatment to a minor problem and a minor treatment to a serious problem. The consumer tends to accept the prescribed treatment, resulting in overtreatment and undertreatment, respectively, in equilibrium. Interestingly, undertreatment---being less costly for the consumer---can initially act as a ``hook'' to induce acceptance of a minor treatment recommendation. When this minor treatment fails to resolve the issue, it can generate additional demand for a more expensive and serious treatment. This would arise when the cost of revisiting the intial provider is lower than that of searching for a new one. The extent of undertreatment exhibits a non-monotonic relationship with consumers' \textit{ex ante} belief about the nature of their problems and the market's ethical level. Our results can shed light on how market ethical levels, provider capabilities and capacities, and consumer privacy protection policies interact with undertreatment and affect consumer welfare. Specifically, consumer welfare can decrease as the market becomes more ethical. Enhancing providers' diagnostic capabilities and capacities can exacerbate undertreatment. Providing access to consumers' diagnosis histories can help mitigate the undertreatment issue and improve consumer welfare.
\end{abstract}

\noindent {\bf Keywords:} Credence services, Undertreatment, Overtreatment, Consumer search, Second opinion, Asymmetric information
%Use this for final submission
%\HISTORY{This paper was first submitted on January 1, 2021 and has been with the authors for 3 months for 2 revisions.}

%%%%%%%%%%%%%%%%%%%%%%%%%%%%%%%%%%%%%%%%%%%%%%%%%%%%%%%%%%%%%%%%%%%%%%

\section{Introduction}

In many service markets, consumers lack information about the type of service they need and must rely on the expertise of service providers. Expert providers may exploit this informational advantage by recommending services that benefit themselves more than the consumers. Consider the following scenarios illustrating the challenges consumers face in decision-making:

\begin{itemize}           
    \item \textbf{Auto Repair}: A consumer brings her car to a mechanic because it occasionally fails to start. Possible solutions range from a simple fix, such as replacing the faulty wiring, to a costly engine overhaul. After inspecting the car, the mechanic suggests replacing the wiring, and the consumer accepts the recommendation. However, a few days later, the problem persists. Now, the consumer must decide whether to return to the same mechanic or search for a different one.
    
    \item \textbf{Veterinary Care}: A dog owner consults a veterinarian because his dog persistently limps. As a pet owner, he is unable to assess whether the issue is a minor sprain requiring a simple anti-inflammatory treatment, or a more serious condition like a torn ligament needing surgery. The veterinarian recommends antibiotics, but the owner is concerned about the sufficiency of the treatment. 
    
    \item \textbf{Technology Consulting}: A company's internal database experiences frequent crashes. The company's manager is unsure whether the problem is due to minor inefficiencies or a fundamental flaw in the system. The manager visits a technology consultant who insists that a costly extensive upgrade of the software and the hardware is required, but the manager wonders whether a cheaper, quick fix, such as optimizing the existing system, would be equally effective in resolving the issue.

\end{itemize}

Some common themes emerge from these examples, which are the key features that we will capture in the paper:

\begin{itemize}
    \item \textbf{Information Asymmetry}: Consumers are uncertain about the severity of their issues and the type of service they require, giving expert service providers an informational advantage that they can exploit. Expert fraud occurs when service providers recommend an overly serious treatment for a minor issue (overtreatment) or prescribe a minor solution for a serious problem (undertreatment).
    \item \textbf{Costly Search}: Upon the service providers' recommendation, consumers can either accept it or reject it and seek a second opinion, but searching for a new service provider involves additional costs, such as time, money, and effort.
    \item \textbf{Repeated Interactions}: If the problem is resolved, consumers exit the market. If it remains unresolved, they face the decision of either revisiting the initial provider or searching for a new one.
\end{itemize}    

Markets with these characteristics are known as credence service markets. Substantial anecdotal evidence and studies demonstrate that expert fraud is both common and costly in such markets \citep{schneider2012agency,balafoutas2013drives,beck2014car, gottschalk2020health}. For instance, \cite{schneider2012agency} found widespread instances of under- and over-treatment in the auto repair market through a field experiment. When the potential for future interactions exists, an opportunistic expert faces a tradeoff: capitalize on immediate gains by recommending a more expensive treatment, or offer a low-cost initial treatment as a ``hook'' and then upsell a more expensive treatment when the initial fix proves insufficient. It is unclear how the consumer’s search process and the nature of repeated interactions influence the expert’s incentive to misrepresent treatment needs. This paper addresses the following research questions: First, how does consumer search, in the context of repeated interactions, affect expert fraud? Specifically, which type of fraud is more likely to emerge: overtreatment, undertreatment, or both simultaneously. Second, how do the severity of the issue and the ethical level of the market affect the degree of expert fraud and consumer welfare? Third, would expert fraud be mitigated or exacerbated if service providers face limited capacity and capability constraints?

We construct a dynamic game-theoretic model focusing on the role of consumer search and repeated interactions between the consumers and the expert service providers. The market consists of an infinite sequence of short-lived consumers and a large number of long-lived expert service providers, who are either honest or opportunistic. Honest experts consistently recommend the appropriate treatment for the problem, whereas opportunistic experts seek to maximize their expected profits, which may involve recommending inappropriate treatments. Therefore, the proportion of honest experts indicates the ethical level of the market. Consumers face uncertainty regarding the type of expert they encounter and the nature of their problem, which can be either minor or serious. A consumer begins by visiting an expert and upon receiving the recommendation, can accept it or incur a cost to search for another expert. If the consumer accepts the recommendation, the treatment may either resolve the problem or not. If the problem is resolved,  consumers leave the market. If not, the consumer may search for a new expert or return to the initial one for further treatment. Experts have perfect information about the consumer’s problem and can resolve both minor and serious issues. Opportunistic experts’ strategies involve the probability of recommending a truthful treatment conditional on the nature of the problem. 

We find that overtreatment and undertreatment can occur simultaneously in equilibrium. Specifically, the opportunistic expert prescribes a serious treatment to a minor problem (i.e., an excessive treatment), and prescribes a minor treatment to a serious problem (i.e., an inadequate treatment). The consumer tends to accept the excessive or inadequate treatment recommendations, resulting in overtreatment and undertreatment, respectively, in equilibrium. The rationale behind the overtreatment is simply that the expert can exploit the information advantage and immediately gain a higher profit margin. However, undertreatment by the expert is influenced by the nature of repeated interactions and the costly search process. First, by prescribing a minor treatment, say, replacing the wiring instead of replacing the engine, the expert is inducing acceptance of this seemingly less costly recommendation. Second, this can increase the subsequent demand for the expert if the problem is not resolved. This is because the consumer tends to return to the initial expert as it is less costly than searching for a new one. As a result, an opportunistic expert can capitalize on the additional treatment in the subsequent visit, upon which he would then prescribe the (accurate) serious treatment. 

Interestingly, our comparative statics show that consumer welfare can decrease as the proportion of honest experts increases. This is because as the market becomes more ethical, consumers are more willing to accept the recommendation, giving opportunistic experts more incentives to defraud consumers, resulting in lower consumer welfare. We also find that as the probability of a minor issue increases, the consumer is not necessarily better off. When the probability of a minor issue is low and the market is unethical, the consumer is more concerned with undertreatment. As the probability of a minor issue increases, the consumer is more likely to accept a minor treatment (undertreatment), which reduces her welfare.

We extend the analysis in the following ways: First, we investigate how imperfect diagnostic capability and capacity shock impact undertreatment. Interestingly, we find that increasing the expert's diagnostic capability and capacity can paradoxically lead to a higher tendency of intentional undertreatment. Second, we examine a scenario where consumers' diagnosis histories are confidential and inaccessible to providers. Our findings suggest that restricting information sharing about diagnosis history with opportunistic providers can actually harm consumer welfare. Third, we explore several model variants, including an alternative contract, undetected undertreatment in the short run, consumer resentment, heterogeneity in treatment capability, and endogenized price. These variants illustrate the boundary conditions of the main insights, and help further understand the mechanisms of undertreatment. 

The paper is organized as follows. In \S\ref{sect:lit}, we discuss our contribution to the literature. In \S\ref{sect:msetup}, we set up the model. We provide the analysis and the main results in \S\ref{sect:analysis}. In \S\ref{sect:exten}, we examine several extensions. We conclude in \S\ref{sect:discuss}.

\section{Related Research and Contributions}
\label{sect:lit}

We contribute to the theoretic literature on expert fraud in the credence goods market (e.g., \cite{darby1973free,emons1997credence,marty1999expert,fong2005experts,dulleck2006doctors, alger2006theory}). In their work, the expert service providers either overtreat or undertreat the consumers, but never concurrently. However, in our equilibrium, undertreatment arises simultaneously with overtreatment. This is because earlier papers are largely silent about what would happen when the problem is undertreated and remains unresolved. In contrast, our work incorporates more realism in which a consumer can reject a recommendation and seek a second opinion; or, when the problem persists after an inadequate treatment, a consumer may seek a second treatment from either the same or a different service provider depending on the search cost. The expert has to make a tradeoff between reaping the short-run benefit now by recommending a more expensive treatment, and inducing acceptance first and generating a subsequent demand. The thrust of our work lies in providing the underlying mechanism between consumers and service providers that can help explain when an expert can be concurrently untruthful on two fronts, namely, undertreatment {\em and} overtreatment. In addition, we examine the implications of limited capability and capacity, the two key challenges service providers are facing, and how they moderate the incentive to undertreat consumers.

Our work is related to the research on strategic under-provision of products and services. In many cases, under-provision is largely due to non-strategic reasons, such as capacity and capability constraints \citep{g2018matching, keskinocak2020review}. For example, according to a systematic analysis of amenable deaths around the globe, poor quality of health care was a major driver of excess mortality \citep{kruk2018mortality}. Several prior papers showed even without capacity and capability constraints, the service provider would strategically withhold the provision of service or provide suboptimal service to the consumers. For example, strategic under-provision of service can arise due to service providers' future liability \citep{dai2023artificial} and reputational concern \citep{ely2003bad,durbin2009corruptible,dai2020conspicuous}, consumer's observational learning \citep{debo2008queuing}, and managing strategic consumers \citep{allon2011buying, xu2020withholding}. Our paper is closely related to \cite{wu2018matchmaker}, who showed that matchmaking platforms, such as \textit{eharmony.com}, have incentives to provide suboptimal matchmaking services to prolong user participation and increase fees. In contrast, we examine on a different mechanism: opportunistic experts strategically use undertreatment initially to build trust and later ``lock-in'' consumers for more expensive treatments.

Our work adds to the research on consumer search (e.g., \cite{wolinsky1993competition}, \cite{Cachon2005assort}, \cite{wu2022bundling}, \cite{fong2022trust}, and \cite{guan2024tele}). The papers most closely related to ours are \cite{wolinsky1993competition} and \cite{fong2022trust}. Similar to our setup, \cite{wolinsky1993competition} constructs a model where consumers can search for second opinions. The key difference from our model is \cite{wolinsky1993competition} assumed that the expert would never undertreat the consumers. \cite{fong2022trust} studied trust building in credence goods markets in a repeated game framework where consumers can monitor an expert's honesty by rejecting his services and experiencing the actual loss. Our work differs from their investigation in two important aspects: First, we explicitly model the search process and belief updating regarding what happens if the issue is not resolved after the initial treatment. Specifically, the consumer has the incentive to return to the initial expert due to lowered search costs. Second, we incorporate heterogeneity in the expert's type. That is, the expert can be an opportunistic or an honest type. Hence, the motivation underlying the consumer's rejection is different. Here, the consumer rejects the recommendation because she intends to search for an honest expert. Consequently, while \cite{fong2022trust} finds that an expert either overtreats or undertreats the consumer, our model shows the coexistence of both overtreatment and undertreatment by the same expert in equilibrium. As the proportion of honest experts indicates the ethical level of the market, we examine the implications for consumer welfare and find a more ethical market can result in lower consumer welfare. 

\section{Model Setup}
\label{sect:msetup}

\subsection{Consumers}

Consider a market where an infinite sequence of short-lived consumers interact with a large number of long-lived expert service providers. The consumers are short-lived in the service market, which is to capture the reality that the consumers would leave the market if the problem is resolved (e.g., her car is repaired) or suffers the related loss (e.g., the car breaks down on the highway).  In Period $t\in \{1, 2, ..., \infty\}$, a consumer, denoted by $C_t$, arrives in the market and becomes aware of the existence of a problem but is uncertain about its nature. She holds the {\em ex ante} belief that the problem $i \in \{m, s \}$ is {\em minor} ($i=m$) with probability $\mu \in (0,1) $, or {\em serious} ($i=s$) with probability $(1-\mu)$. The parameter $\mu$ also reflects the degree of uncertainty the consumer faces, which may be a result of information-gathering activities before entering the market, such as self-diagnosis through online resources. If the problem is left untreated, $C_t$ suffers a loss $l_i$ ($l_s> l_m >0$) at the end of Period ($t+1$), i.e., $C_t$ suffers a greater loss from a serious problem than from a minor problem, and has limited time, specifically two periods, to resolve the problem. In Period $(t+1)$ a new consumer, denoted as $C_{t+1}$, arrives in the market. As $C_{t+1}$ at Period $(t+1)$ faces the same decision problem as $C_t$ in Period $t$, we focus on Consumer $C_t$. For ease of exposition, we refer to a consumer as ``she'' and an expert as ``he''. 

At the beginning of Period $t$, Consumer $C_t$ can decide whether to incur a search cost to sample an expert. The search cost, denoted as $k>0$, encompasses the mental and physical effort required to find the expert, as well as the traveling cost to visit the expert. The cost resembles the `barrier' that a consumer faces in consulting an expert \citep{wolinsky1993competition}. %\footnote{According to \cite{wolinsky1993competition}'s interpretation, the search cost $k$ also includes the diagnosis cost incurred by the experts but borne by the consumer. Here we assume the diagnosis cost is included in the treatment price.}
Once an expert is drawn, the service provider will make a treatment recommendation to the consumer based on her condition. If the consumer accepts the recommendation and the problem is resolved, she leaves the market. However, if the consumer accepts the recommendation but the problem persists, the consumer decides whether to search for a new expert or return to the first expert for further treatment; otherwise, if Consumer $C_t$  rejects the recommendation in Period $t$, the consumer searches a new expert in Period $(t+1)$. When searching for a new expert, the consumer has to incur another search cost $k$. We assume the search cost of returning back to the first expert is $k^{'}\geq 0$, and returning to the first expert usually requires a lower effort than finding a new expert (i.e., $k^{'} < k$). Note that $k^{'} > k$ would resemble resentment towards the initial expert which is examined in \S \ref{subsubsec:resentment}. During the second visit in Period $(t+1)$, the consumer gets another recommendation, either from the previous expert or from a new one depending on her search decision, and decides whether to accept or reject the recommendation.

\subsection{Expert Service Provider}

In our main model, we assume that the expert service providers have perfect information about $C_t$'s problem and are capable of resolving both minor and serious problems (see \S \ref{sec:limited_ability} and \S \ref{sec:capacity} where these assumptions are relaxed). The market consists of two types of experts, an {\em honest} type, with probability $h \in (0,1)$, or an {\em opportunistic} type, with probability $(1-h)$. An honest expert recommends Treatment $i$ if and only if a consumer has Problem $i$. %and receives utility $(p_i-c_i)$ when the recommendation is accepted. 
Honesty can be driven by the experts' altruism or professionalism \citep{durbin2009corruptible,morris2001political,inderst2012competition}. An opportunistic expert, however, aims to maximize his expected profits, and may behave dishonestly to exploit consumers. The degree of $h$ captures the ethical level of the market. In our paper, we assume that all the experts are identical in their skills, and that the experts are only differentiated in their honesty (see \S \ref{subsubsec:heter_ability} where the assumption is relaxed). An expert knows his own type, but consumers know only the probability distribution of the expert's type, which is also common knowledge. 

We refer to the recommendation of Treatment $i$ to Problem $i$ as a {\em truthful} recommendation. If a serious treatment is recommended for a minor problem, we refer to it as an {\em excessive} recommendation, since a minor treatment would suffice to resolve a minor problem. {\em Overtreatment} occurs if the consumer accepts the excessive recommendation, and the problem is always resolved. If a minor treatment is recommended for a serious problem, we refer to it as an {\em inadequate} recommendation, and {\em undertreatment} occurs when the consumer accepts the inadequate recommendation and the problem remains. We assume that a minor problem can be resolved by either a minor or a serious treatment, while a serious problem requires a serious treatment. For example, the replacement of the engine (serious treatment) can resolve the issue of ignition failure, even if cleaning the engine (minor treatment) suffices to resolve it. %\footnote{We acknowledge the reality that there might be instances when an expert `plays it safe' and recommends a serious treatment for a minor problem, or he might be negligent and recommends inadequately. While these may be commonplace, our focus in this paper is on showing that an expert's financial motivation itself is sufficient to result in both (untruthful) inadequate and excessive recommendation in equilibrium.} 

Upon consultation by $C_t$, the expert identifies the nature of the problem, and recommends a treatment that is either minor or serious at prices $p_m$ and $p_s$, respectively. As our focus is on the role of consumer search when the consumers face uncertainty about both the experts' type and the nature of the problem, we assume that the prices of treatments are exogenous (also see \cite{allon2011buying}).\footnote{In Extension \S \ref{sec:endo_price}, we examine the cases where the price is endogenized.} %This resembles the practices in certain industries such as healthcare services where the recommended prices for various treatments are publicly available and regulated by government authorities. 
The minor and serious treatments cost the expert $c_m$ and $c_s$ respectively, with a serious treatment being more profitable than a minor treatment $(p_s - c_s>p_m-c_m)$ for the expert. The expert, regardless of the type, receives a payoff of zero if the consumer rejects the recommendation.

\subsection{Information and Belief Updating}

As we discussed earlier, in our main model, the experts are fully informed about their own types, honest or opportunistic, and they have the perfect ability to diagnose and resolve the problem. The proportion of honest and opportunistic experts, and the price and the cost of the treatment are given and are common knowledge. The consumer $C_t$ is uncertain about the nature of her problem and the type of expert she is interacting with. Upon receiving the recommendation, the consumer updates her belief in a Bayesian fashion. 

Since Consumer $C_t$ can visit the experts at most two times, at Period $(t+1)$, her belief about her condition and the type of expert is updated after her first visit. In Period $(t+1)$, a new consumer --- $C_{t+1}$ arrives and visits the expert as her first visit. Note that our model resembles an overlapping generation model. In the main model, we assume that when a consumer is present, the experts are aware of the consultation history of the consumer. That is, the expert knows whether the consumer who visits is $C_t$ on her second visit or $C_{t+1}$ on her first visit Period $(t+1)$.\footnote{In the extension \S \ref{sec:inaccess_history}, we investigate the case where the expert cannot access the consumer's diagnosis history.} 

\subsection{Sequence of Events}

\begin{figure}[ht]
    \centering
\caption{Timeline of the Game}
    \includegraphics[scale=0.55]{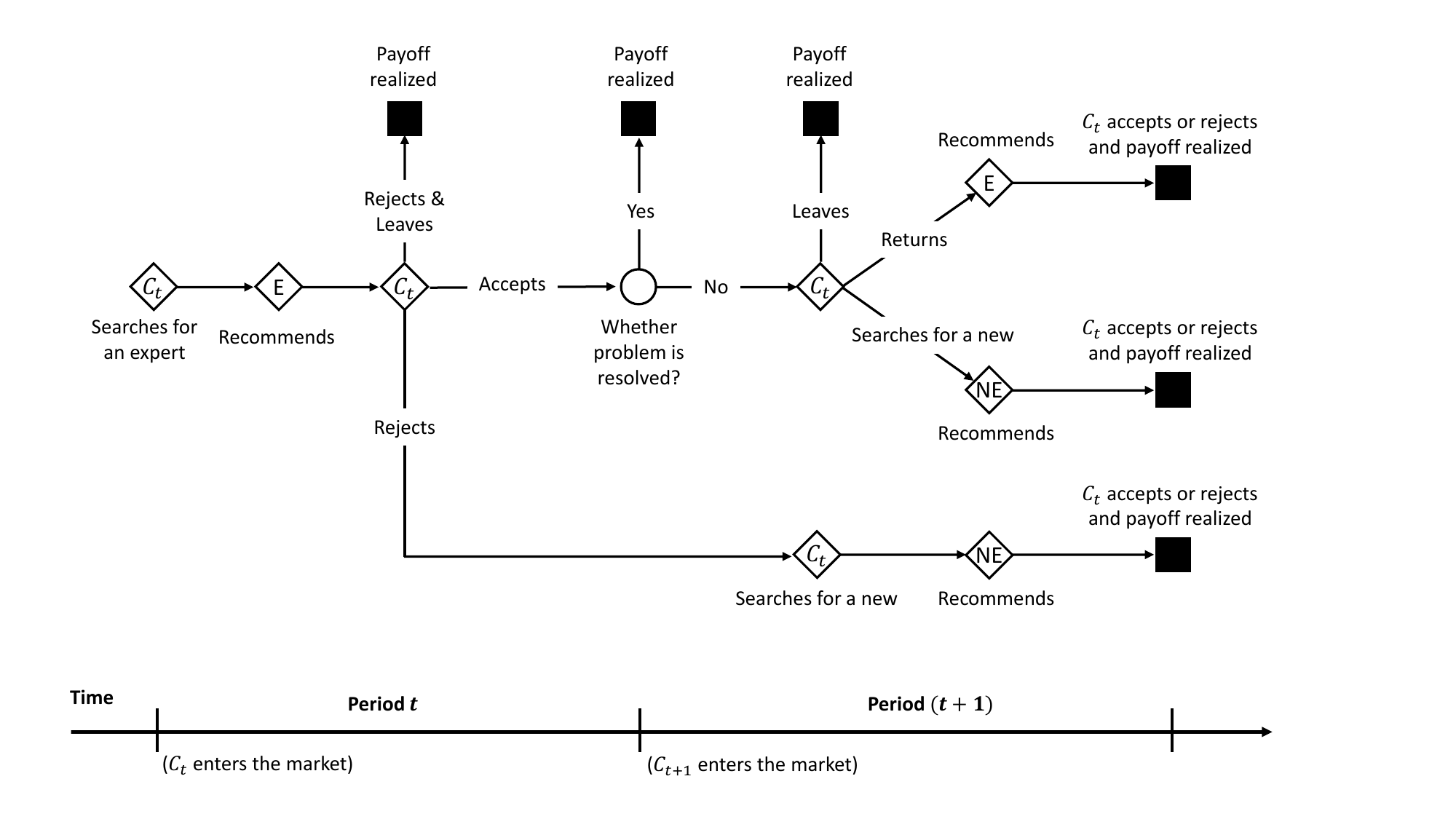}
     \label{fig:timeline}
      \parbox{\linewidth}{   
  {\textit{\underline{Note}: The decision nodes for the consumer, the expert, and the new expert are labeled as ``$C_t$'', ``E'' and ``NE'', respectively. }}
  }
\end{figure}

Here we summarize our model by demonstrating the timeline of events (illustrated in Figure \ref{fig:timeline}) and relevant payoff in each period. Since Consumer $C_t$ can visit at most two experts, we focus on Periods $t$ and $t+1$. Table \ref{tbl:summary} provides a summary of the notations.

\textbf{Period $t$}: At the beginning of Period $t$, Consumer $C_t$ is drawn and nature determines her condition, either serious or minor, according to probability $\mu$. She decides whether to incur a search cost $k$ to consult an expert. If she chooses not to, she leaves the market and suffers a loss $l_i$ at the end of Period $(t+1)$; otherwise she draws an expert and proceeds to consult an expert.

The expert diagnoses the problem and recommends Treatment $i$. $C_t$ decides whether to accept the recommendation. If she chooses to accept the recommendation, she pays $p_i$. If her problem is resolved, she leaves the market without incurring any further loss. If she chooses to reject the recommendation, she can either choose to leave the market and suffer a loss $l_i$ later, or she can search for a second provider at the cost of $k$. If she does not leave the market at the end of Period $t$, she proceeds to the next stage. If she accepted a recommendation in Period $t$ but the problem persists, she enters the next period.

\textbf{Period $(t+1)$}: Given the problem is not resolved, $C_t$ has three options: (i) leaving the market resulting in loss $l_i$; (ii) returning back to the expert she visited before at a cost of $k^{'}\geq 0$; (iii) searching for a new expert at a cost of $k$. In the case of (ii) or (iii), she receives a second treatment recommendation and she can decide whether to accept the recommendation. 

At the end of Period $(t+1)$, the payoffs for the consumer $C_t$ and the expert are realized based on the consumer's decision and treatment outcome.

\subsection{Strategy and Equilibrium}

The strategy of $C_t$ is represented by $A_{ij}$ where $A_{ij} \in [0,1]$ denote the probability of accepting the recommendation $i\in \{s, m\}$ on her $j$th visit ($j\in \{1, 2\}$). We define an opportunistic expert's strategy as $T_{ij}$ where $T_{ij} \in [0,1]$ is the probability that an opportunistic expert recommends a {\em truthful} treatment to a consumer with problem $i$ on Consumer $C_t$'s $j$th visit. Note that the honest expert's strategy is $T_{ij}=1$, for $\forall i$ and  $\forall j$. Put differently, ($1-T_{mj}$) is the probability of overtreatment and ($1-T_{sj}$) the probability of undertreatment on Consumer $C_t$'s $j$th visit. %, and $y$ is the probability that an opportunistic expert recommends a {\em truthful} minor treatment when the problem is minor. 
For example:
\begin{itemize}
\item If $T_{s1} = 0$, when a consumer with a serious problem comes to the expert on her first visit, the opportunistic expert behaves completely untruthfully and would recommend inadequately a minor treatment with probability one, and undertreatment occurs if the consumer accepts the inadequate minor recommendation. 
\item If $T_{m2} = 0$, when a consumer with a minor problem comes to the expert on her second visit, the opportunistic expert would always recommend %, which is, 
excessively a serious treatment with probability one, and overtreatment occurs if the consumer accepts the excessive recommendation. 
\end{itemize}

Therefore, we define $T_{sj}=0$ as a \textit{full undertreatment} and $T_{mj}=0$ as a \textit{full overtreatment}. Moreover, we define $0<T_{sj}<1$ as a \textit{partial undertreatment} and $0<T_{mj}<1$ as a \textit{partial overtreatment} since the expert would under- or over-treat probabilistically.

We focus on the decisive equilibria, in which consumers will accept the recommendations with a positive probability. As in \cite{fong2005experts} and \cite{fong2014role}, the attention is restricted to undominated equilibria, i.e., equilibria in which players do not play weakly dominated strategies.

\begin{table}[]
\caption{Summary of Notations}
%\footnotesize
\centering
{\small
\renewcommand{\arraystretch}{1.5} % <-- Adjust this value as needed
\begin{tabular*}{0.9\textwidth}{@{\extracolsep{\fill}}cl}
\hline
\hline
     {\bf Notation} & {\bf Description} \\
\hline
$h$& Proportion of honest experts in the market, $h \in (0,1)$ 
\\
\hline
$\mu$ & Consumer's {\em ex ante} belief that her problem is minor, $\mu \in (0,1)$
\\
\hline
$l_i$& Consumer loss if Problem $i$ is unresolved, $i =\{m,s\}, l_s > l_m > 0$
\\
\hline
$p_i$& Price for Treatment $i$, $p_s > p_m > 0$
\\
\hline
$c_i$ & Treatment cost incurred by the expert for Treatment $i$, $c_s > c_m \geq 0$
\\
\hline
$k$ & Consumer search cost in finding a new expert, $k>0$
\\
\hline
$k^{'}$ & Consumer search cost when returning to the previous expert, $k>k^{'} \geq 0$
\\
\hline
$T_{ij}$& Probability of recommending a truthful treatment on the consumer's 
\\
& $j$th visit ($j\in\{1, 2\}$), given that the problem is $i$, $T_{ij} \in [0,1]$
\\
\hline
$A_{ij}$& Probability of accepting the treatment $i$ on the consumer's $j$th visit, $A_{ij} \in [0,1]$ 
%\\
%\hline
%$\gamma_i$ & Probability that a consumer is visiting the first expert given Problem $i$
\\
\hline
$\tau_i$ & Consumer's posterior belief that recommendation of treatment $i$ is truthful
\\
\hline
$\epsilon$ & Error rate in diagnosis, $\epsilon \in [0, 0.5)$
\\
\hline
$\chi$ & The probability of capacity shock,  $\chi \in [0, 1]$
\\
\hline
\hline
\end{tabular*}
}
\label{tbl:summary}
\vspace{1em}
\end{table}

\section{Analysis}
\label{sect:analysis}
%We begin by examining a simplified case in which the game consists of only one period. In this case, the consumer conducts a single search, and no future interactions between the consumer and experts are possible after the decision is made. Consequently, the consumer must either accept or reject the recommendation. If the recommendation is rejected, a loss  $l_i$ will be incurred.

%\begin{lemma}
%\label{lem:One_search}
%In a one-period game, when $p_s + k < l_m$, opportunistic experts invariably recommend a serious treatment (i.e., $T_{s}=1$ and $T_{m}=0$) and consumer will accept any recommendation from an expert (i.e., $A_{s}(t)=1$ and $A_{m}(t)=1$). 
%\end{lemma}

%As Lemma \ref{lem:One_search} shows, in a one-period game, when the loss from an unresolved problem is sufficiently high, only the serious treatment exists in equilibrium. In our subsequent analysis, we will focus on the parameter range where $p_s + k < l_m$.

%Now we consider the case where the game has two periods. After the first visit, the consumer can decide whether to accept the recommendation, or to have a second visit. 

%We provide the proofs of Lemmas $1$, $2$ and Propositions $1$-$4$ in the Online Supplement. All other proofs are given in the Online Appendix.

We adopt backward induction and start by examining Consumer $C_t$ and the opportunistic expert's decision in Period $(t+1)$. 

\subsection{Consumer and Expert's Decision in Period $(t+1)$}

We first look at the consumer and expert's decision conditional on the consumer's visit to the second expert in Period $(t+1)$. Note that Period $(t+1)$ is the last period for the consumer to conduct a search, and no future interaction between the consumer and experts is possible after the decision is made. 
When the loss of the problem is sufficiently high (i.e., $p_s<l_m$), facing a serious treatment recommendation, it is a dominant strategy for a consumer to accept it rather than leave the problem unresolved ($A_{s2}=1$). Then, it is a dominant strategy for an opportunistic expert to recommend a serious treatment invariantly since $(p_s-c_s)A_{s2}>(p_m-c_m)A_{m2}$ holds with strict inequality. On the other hand, if it is a minor recommendation, a consumer can immediately infer it is a truthful recommendation from an honest expert, and the consumer is also willing to accept it ($A_{m2}=1$). In such a case, there is no undertreatment in the equilibrium. Therefore, we have the following lemma:

% We first look at the consumer and expert's decision conditional on the consumer's visit to the second expert in Period $(t+1)$. Note that Period $(t+1)$ is the last period for the consumer to conduct a search, and no future interaction between the consumer and experts is possible after the decision is made. If it is a serious recommendation, by accepting it, the consumer pays $p_s$, and her problem is resolved; by rejecting it, the consumer gets $-l_m$ if the problem is minor, and gets $-l_s$ if the problem is serious. Therefore, if $p_s<l_m<l_s$, it is a dominant strategy for a consumer to accept a serious recommendation. On the other hand, if it is a minor recommendation, a consumer can immediately infer it is a truthful recommendation from an honest expert, and the consumer is also willing to accept it. Thus, $A_{m2}=1$ and $A_{s2}=1$. From the opportunistic expert's perspective, invariably recommending a serious treatment to a consumer is a dominant strategy because $p_s-c_s>p_m-c_m$ and $A_{m2}=A_{s2}=1$. The results resemble the case where the consumer and the expert only interact for one period. In such a case, there is no undertreatment in the equilibrium. Therefore we have the following lemma:

\begin{lemma}
\label{lem:period2_strat}
In Period $(t+1)$, when $p_s < l_m$ and $p_s-c_s>p_m-c_m$, the opportunistic expert invariably recommends a serious treatment and the consumer will accept any recommendation from an expert, i.e., $T_{m2}^*=0$, $T_{s2}^*=1$, $A_{m2}^*=1$, and $A_{s2}^*=1$.
\end{lemma}

Now we move one step back and consider $C_t$'s decision when she enters Period $(t+1)$. If $C_t$ has received and accepted a serious treatment recommendation in Period $t$, the problem would have been resolved regardless of the true nature of the problem and she will not be in Period $(t+1)$. $C_t$ appears in Period $(t+1)$ because she either (a) received and accepted a minor treatment recommendation in Period $t$, but the problem remains; or (b) rejected the first recommendation she received. In case (a), as $C_t$'s problem remains after receiving a minor treatment in Period $t$, she then deduces that her problem must be a serious one. In such a case, $C_t$ can either choose to return to the first expert or search for a new expert.  

\begin{lemma}
\label{lem:consumer_return}
If $C_t$'s problem remains after a minor treatment, it is a dominant strategy for consumer $C_t$ to return to the previous expert in Period $(t+1)$ when $k>k'$.
\end{lemma}

Lemma \ref{lem:consumer_return} holds because since the problem is a serious one and Period $(t+1)$ is the last period the consumer will stay in the market, both honest and opportunistic experts will recommend a serious treatment on her next visit, and she will accept the recommendation anyway (Lemma \ref{lem:period2_strat}). However, returning to the first expert entails a lower search cost $k^{'}$, even though she realizes that the first expert is an opportunistic type who undertreated her. %\footnote{In Extension $1$, we examine the case where once a consumer infers from the unresolved problem that the first expert is opportunistic and has undertreated, sentiments of injustice and anger drive her to visit a second (different) expert upon persistence of the problem, despite incurring an additional search cost of $k$ when searching for a second expert. We shall show that different search cost in dealing with undertreatment affects the extent of inadequate treatment recommendation in equilibrium and the underlying mechanism.} 

In case (b), $C_t$ shows up in Period $(t+1)$ because she rejects the first recommendation and seeks a second opinion in Period $(t+1)$. In such a case, she would like to accept the recommendation from the second expert (according to Lemma \ref{lem:period2_strat}). %The rationale is as follows. If $C_t$ rejects the first recommendation and is willing to incur an additional search cost, she must have an expected benefit from a {\em different} recommendation by the second expert. When the loss from the unresolved problem is sufficiently high, as assumed earlier, she does not want to leave the problem unresolved. Thus, $C_t$ would like to accept the recommendation from the second expert.

In summary, $C_t$ in Period $(t+1)$ either returns to the first expert for a serious treatment (when the problem remains after a minor treatment), or accepts the recommendation of the second expert she visits in Period $(t+1)$. The opportunistic expert always recommends a serious treatment in Period $(t+1)$.

In the following, we examine the strategy of a typical consumer who is visiting an expert for the first time in Period $t$.

\subsection{Consumer's Decision in Period $t$}

In Period $t$, $C_t$ becomes aware of her problem and searches for an expert to resolve her problem. On her first visit, with a probability of $h$ she will sample an honest expert who would always give a truthful recommendation, and with a probability of $(1-h)$ she will sample an opportunistic expert who would give a truthful recommendation with a probability of $T_{i1}$.  We define $\bar{T}_s$ and $\bar{T}_m$ as the {\em expected} probability of a truthful serious and minor recommendation, respectively.

\begin{eqnarray}
\label{eqn:bar}
\bar{T}_s =h + (1-h) T_{s1}, && \bar{T}_m= h + (1-h) T_{m1}.
\end{eqnarray}

After the expert gives the recommendation, she updates her belief about the nature of her problem according to Bayes' rule. %\footnote{In the Online Supplement Proof \uppercase\expandafter{\romannumeral2}, we consider the case when the consumer updates her belief about the type of the expert (honest or opportunistic) based on the recommendation. We show that the results are robust, %as opportunistic experts do not always recommend untruthfully. As a result, 
%i.e., when a consumer updates her belief on the type of the expert, she makes her decision based on the likelihood of a truthful recommendation as an opportunistic expert may also recommend truthfully.} 
Let $\tau_i$ denote the updated belief that $C_t$ has Problem $i$, conditional on a recommendation of Treatment $i$. That is, $\tau_i$ is consumer's posterior belief that the recommendation of Treatment $i$ is truthful after she gets the recommendation from an expert. 

Hence, the updated beliefs are:

\begin{eqnarray}
\label{eqn:post}
\tau_m= \frac{\mu \bar{T}_m}{\mu \bar{T}_m+(1-\mu)(1-\bar{T}_s)}, && \tau_s =\frac{(1-\mu) \bar{T}_s}{(1-\mu) \bar{T}_s+\mu (1-\bar{T}_m)}.
\end{eqnarray}

where $\mu \bar{T}_m$ and $(1-\mu) (1-\bar{T}_s)$ are the probabilities that the minor treatment recommendation is truthful and inadequate respectively, and $(1-\mu) \bar{T}_s$ and $\mu (1-\bar{T}_m)$ are the probabilities that the serious treatment recommendation is truthful and excessive respectively. $C_{t}$ decides whether to accept the recommendation provided by the first expert in Period $t$ based on the expected payoff according to $\tau_m$ and $\tau_s$.

Facing a serious recommendation, by accepting it, Consumer $C_t$ pays a price $p_s$ and ensures her problem is resolved by the first expert she visits. By rejecting a serious treatment recommendation, consumer $C_t$ searches another expert with an additional cost $k$. Note Lemma \ref{lem:period2_strat} shows that a consumer accepts any recommendation on the second visit to an expert. If Consumer $C_t$'s problem is serious (with probability $\tau_s$) or Consumer $C_t$'s problem is minor but meets an opportunistic expert (with probability $(1-\tau_s)(1-h)$), the consumer receives a serious treatment recommendation and accepts it. Only when the Consumer $C_t$'s problem is minor and the consumer meets an honest expert (with probability $(1-\tau_s)h$) will the consumer receive and accept a minor treatment. As a result, we present the incentive compatibility constraint for Consumer $C_t$ to accept a serious treatment recommendation in Period $t$ as follows:

%\begin{eqnarray}
%\mbox{IC(CS)}: \hspace{0.5cm}
%-p_s \geq \tau_s (-p_s-k) + (1-\tau_s)h (-p_m-k) + (1-\tau_s)(1-h)(-p_s-k)  
%\label{eqn:csar}
%\end{eqnarray}
%\normalsize%

\begin{eqnarray}
%\mbox{IC(CS)}: \hspace{0.5cm}
-p_s \geq \underbrace{\tau_s (-p_s-k)}_{\text{condition is serious}} + 
\underbrace{(1-\tau_s)h (-p_m-k)}_{\scriptsize \shortstack{\text{condition is minor and }\\ \text{meets an honest expert}}} + 
\underbrace{(1-\tau_s)(1-h)(-p_s-k)}_{\scriptsize \shortstack{\text{condition is minor and }\\ \text{meets an opportunistic expert}}} 
\label{eqn:csar}
\end{eqnarray}

The left-hand side of Inequality (\ref{eqn:csar}) is the price the consumer has to pay when accepting a serious recommendation. The right-hand side is the expected payoff when rejecting the recommendation. If Inequality (\ref{eqn:csar}) holds with strict inequality, then $A_{s1}=1$. If the inequality is binding, she adopts a mixed-strategy, i.e., $0<A_{s1}<1$, and rejects ($A_{s1} = 0$) otherwise.

By the same token, the incentive compatibility constraint for $C_t$ to accept a minor treatment recommendation in Period $t$ is:

%\begin{equation}
%    \label{equ:ut_m_rej}
%    \tau_m (-p_m) + (1-\tau_m) (- p_m - p_s-k^{'}) \geq \tau_m h (-p_m-k) + \tau_m (1-h) (-p_s-k) + (1-\tau_m) (-p_s - k). 
%\end{equation}
\begin{align}
%\begin{equation}
    \label{equ:ut_m_rej}
   \overbrace{\tau_m (-p_m)}^{\text{condition is minor}} + \overbrace{(1-\tau_m) (- p_m - p_s-k^{'})}^{\scriptsize \shortstack{\text{condition is serious and }\\ \text{returns to the initial expert}}} &\geq \nonumber \\
& \underbrace{\tau_m h (-p_m-k)}_{\scriptsize \shortstack{\text{condition is minor and }\\ \text{ meets an honest expert}}}  + \underbrace{\tau_m (1-h) (-p_s-k)}_{\scriptsize \shortstack{\text{condition is minor and }\\ \text{ meets an opportunistic expert}}} + \underbrace{(1-\tau_m) (-p_s - k)}_{\scriptsize \shortstack{\text{condition is serious }}}
%\end{equation}
\end{align}

The left-hand side of Inequality (\ref{equ:ut_m_rej}) is the expected payoff when accepting a minor recommendation. Note that the consumer will return to the initial expert and pay an extra $- p_s-k^{'}$ if the problem is not resolved because it is a serious one. The right-hand side is the expected payoff when rejecting the recommendation. Consumer $C_t$ accepts the recommendation, i.e., $A_{m1}=1$ if Inequality (\ref{equ:ut_m_rej}) holds with strict inequality. If the inequality is binding, she adopts a mixed strategy (i.e., $0<A_{m1}<1$); otherwise, she rejects the recommendation.

We next examine the opportunistic expert's strategy.

\subsection{Opportunistic Expert's Decision in Period $t$}

In Period $t$, if the consumer's condition is minor, the incentive compatibility constraint for an opportunistic expert to recommend an excessive treatment, i.e., $T_{m1}=0$, is:

\begin{align}
& IC(Excessive):  (p_s-c_s)A_{s1} \geq (p_m-c_m)A_{m1}. \label{eqn:sm}
\end{align}

The left-hand side is the expected profits of recommending a serious treatment --- overtreatment. The right-hand side of Inequality (\ref{eqn:sm}) is the expected profits of recommending a minor treatment if the consumer's condition is minor. If Inequality (\ref{eqn:sm}) holds with strict inequality, the opportunistic expert would give an excessive recommendation. If the inequality is binding, he adopts a mixed-strategy (i.e., $0<T_{m1}<1$). Otherwise, the expert will provide a truthful recommendation to the consumer (i.e., $T_{m1}=1$).

If the consumer's condition is serious, the incentive compatibility constraint for an opportunistic expert to recommend an inadequate treatment, i.e., $T_{s1}=0$, is:

\begin{align}
& IC(Inadequate): (p_s-c_s+p_m-c_m)A_{m1} \geq (p_s-c_s)A_{s1}  \label{eqn:ss}
\end{align}

The left-hand side is the expected profits of recommending a minor treatment if the consumer's condition is serious. Note that $p_s - c_s$ is the extra profits the expert gets in Period $(t+1)$ by undertreating the consumer in Period $t$. The right-hand side of Inequality (\ref{eqn:ss}) is the expected profits of recommending a serious treatment. If Inequality (\ref{eqn:ss}) holds with strict inequality, then the expert would make a recommendation that would undertreat the consumer. If the inequality is binding, he adopts a mixed-strategy (i.e., $0<T_{s1}<1$). Otherwise, the expert will provide a truthful recommendation to the consumer (i.e., $T_{s1}=1$).  

\subsection{Equilibrium Strategy}

As the strategy of the consumer and the expert in Period $(t+1)$ characterized in Lemma \ref{lem:period2_strat} and \ref{lem:consumer_return} are dominant strategies, which are always played in equilibrium, for the ease of exposition, we focus on the equilibrium strategy in Period $t$ throughout the paper. Next, we show that the equilibrium strategy depends on the ethical level of the market $h$ and the probability of the minor problem $\mu$. To summarize, we have the following proposition:

\begin{Proposition}
\label{prop:MainEqbm}

In Period $t$, for a consumer who is on her first visit,

i) \textbf{Full overtreatment and full undertreatment (FOFU) equilibrium:} If the probability of a minor issue is moderate (i.e., $\mu \in (\mu_1^*,\mu_2^*)$), there exists a pure strategy equilibrium that the opportunistic expert will fully overtreat and fully undertreat the consumer (i.e, $T_{m1}^*=0$, $T_{s1}^*=0$, $A_{m1}^*=1$, and $A_{s1}^*=1$).

%For $\mu \in (\mu_1^*,\mu_2^*)$, there exists a pure strategy equilibrium $E_1 = (T_{m1}=0, T_{s1}=0, A_{m1}=1, A_{s1}=1)$, i.e., the opportunistic expert provides an excessive recommendation for a minor problem and an inadequate recommendation for a serious problem, and a consumer accepts any recommendations.

ii) \textbf{Partial overtreatment and full undertreatment (POFU) equilibrium:} If the probability of a minor issue is high (i.e., $\mu \in (\mu_2^*, 1)$), there exists a mixed-strategy equilibrium that the opportunistic expert will partially overtreat and fully undertreat the consumer (i.e., $T_{m1}^*<1$, $T_{s1}^*=0$, $A_{m1}^*=1$, and $A_{s1}^*<1$).

%For $\mu \in (\mu_2^*, 1]$, there exists a mixed-strategy equilibrium $E_2= (T_{m1}=T_{m1}^*, T_{s1}=0, A_{m1}=1, A_{s1}=A_{s1}^*)$, i.e., the opportunistic expert provides an inadequate recommendation for a serious problem with a probability 1, and an excessive recommendation with a probability $1-T_{m1}^*$; a consumer will always accept a minor recommendation while accepting a serious recommendation with a probability $A_{s1}^*$.

iii) \textbf{Full overtreatment and partial undertreatment (FOPU) equilibrium:} If the probability of a minor issue is low (i.e., $\mu \in (0, \mu_1^*)$), there exists a mixed-strategy equilibrium that the opportunistic expert will fully overtreat and partially undertreat the consumer (i.e., $T_{m1}^*=0$, $T_{s1}^*<1$, $A_{m1}^*<1$, and $A_{s1}^*=1$).\footnote{In the Appendix, we show that if $\mu_2^*<\mu_1^*$, the equilibrium will be POFU when $\mu>\frac{p_m-c_m}{p_s-c_s}$; otherwise, the equilibrium will be FOPU when $\mu<\frac{p_m-c_m}{p_s-c_s}$.}

%iii) For $\mu \in [0, \mu_1^*)$, there exists a mixed-strategy equilibrium $E_3 = (T_{m1}=0, T_{s1}=T_{s1}^*, A_{m1}=A_{m1}^*, A_{s1}=1)$, i.e., the opportunistic expert provides an excessive recommendation with a probability 1, and an inadequate recommendation with a probability $1-T_{s1}^*$; a consumer will always accept a serious recommendation while accepting a minor recommendation with a probability $A_{m1}^*$. 
\end{Proposition}

Here in the POFU equilibrium, $T_{m1}^*=1-\frac{(1-\mu)hk}{\mu(1-h)(h(p_s-p_m)-k)}$, $A_{s1}^*=\frac{p_m-c_m}{p_s-c_s}$. In the FOPU equilibrium, $T_{s1}^*=1-\frac{\mu h((1-h)(p_s-p_m)+k)}{(1-\mu)(1-h)(p_m+k^{'}-k)}$, and $A_{m1}^*=\frac{p_s-c_s}{p_s-c_s+p_m-c_m}$. In addition, $\mu^*_1=\frac{(1-h)(p_m+k^{'}-k)}{hk+(1-h)h(p_s-p_m)+(1-h)(p_m+k^{'}-k)}$, while $\mu^*_2=1$ when $h\leq\frac{k}{p_s-p_m}$ and $\mu^*_2=\frac{hk}{hk + (1-h)h(p_s - p_m) - (1-h)k}$ when $h>\frac{k}{p_s-p_m}$.

 \begin{figure}[ht]
  \centering
\caption{Equilibria and Expert's Strategy}
    \includegraphics[scale=0.6]{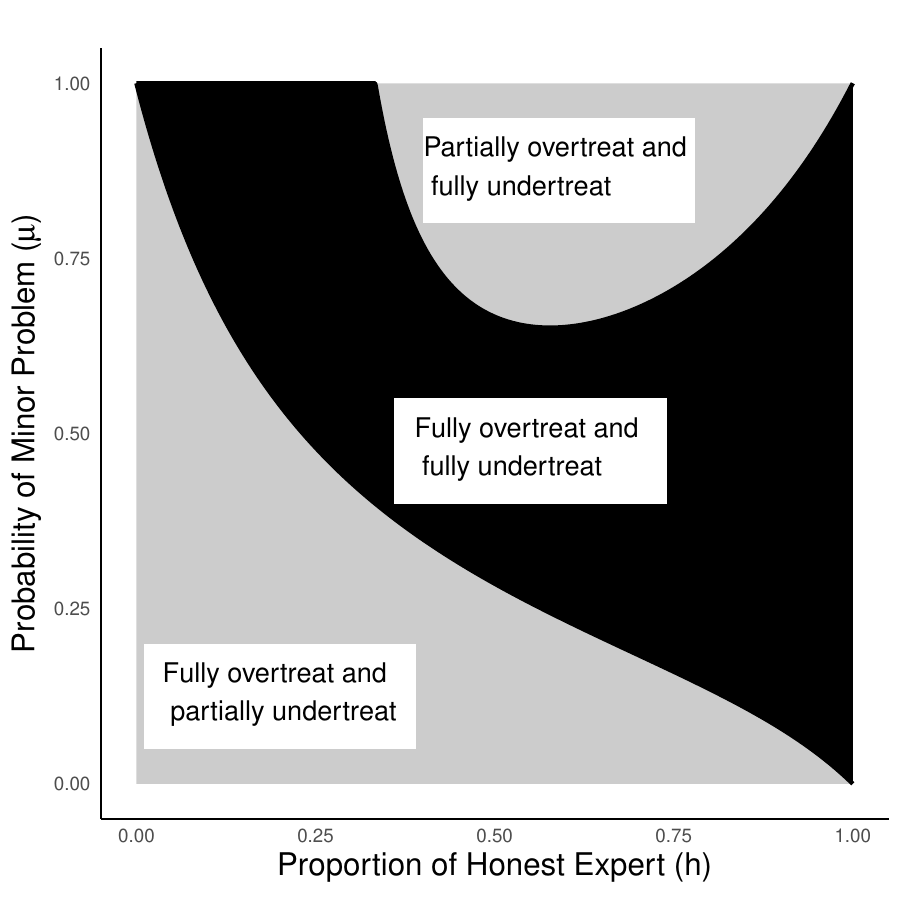}
  \label{fig:Eqm_knownhistory}
        \parbox{\linewidth}{  
  {\textit{\underline{Note}: The three equilibria are presented on the graph. The parameters used in drawing the graph are $k=1$, $k'=0$, $p_s=5$, and $p_m = 2$. The upper region in grey corresponds to $\mu \in (\mu_2^*, 1)$ where the expert undertreats with certainty, and overtreats probabilistically (i.e., $T_{m1}^*<1$, $T_{s1}^*=0$). The middle region in black corresponds to $\mu \in (\mu_1^*,\mu_2^*)$ where the expert undertreats and overtreats with certainty (i.e., $T_{m1}^*=0$, $T_{s1}^*=0$). The lower region in grey corresponds to $\mu \in (0, \mu_1^*]$ where the expert overtreats with certainty, and undertreats probabilistically (i.e.,$T_{m1}^*=0$, $T_{s1}^*<1$).}}
  }
  
  %In $E_1$, the expert undertreats and overtreats with certainty. In $E_2$, the expert undertreats with certainty, and overtreats probabilistically. In $E_3$, the expert overtreats with certainty, and undertreats probabilistically.
\end{figure}

The proof is in Appendix. We plot the equilibria in Figure \ref{fig:Eqm_knownhistory}. As Proposition \ref{prop:MainEqbm} shows, the opportunistic expert is never fully truth-telling, i.e., $T_{m1}<1$ and $T_{s1}<1$. When $\mu$ is in the moderate range $(\mu_1^*,\mu_2^*)$, any recommendation (minor or serious) by the opportunistic expert is untruthful, i.e., he recommends excessively and inadequately with probability one. A consumer always accepts the recommendation from the first expert in the equilibrium. This equilibrium can arise because when $\mu$ is in the moderate range, the consumer is facing high uncertainty about the nature of her problem. The opportunistic expert can exploit the most rent from the information advantage.

Interestingly, when the probability of getting a minor problem is high, i.e., $\mu >\mu_2^*$, the opportunistic expert plays a mixed strategy in terms of giving a truthful recommendation conditional on a minor problem. This is because when $\mu$ is high, the consumer is suspicious upon receiving a serious treatment recommendation and hence only accepts it with a probability of $A_{s1}^*<1$. As a result, the opportunistic expert has to randomize between giving a serious treatment recommendation and a minor one. However, conditional on a serious problem, the expert plays a pure strategy in giving an inadequate minor recommendation. This occurs because, with a high $\mu$, the consumer is more inclined to accept a minor treatment recommendation. Undertreatment not only generates a higher profit margin but also serves as a tactic to conceal untruthful recommendations, enabling the opportunistic expert to pressure the consumer into accepting the recommended treatment.

When the probability of getting a serious problem is high, i.e., $\mu < \mu_1^*$, the consumer is less suspicious upon receiving a recommendation for a serious treatment. As a result, conditional on a minor problem, the opportunistic expert always overtreats by recommending a serious treatment. However, when $\mu$ is low, the consumer is suspicious upon receiving a minor treatment recommendation. As a result, the opportunistic expert plays a mixed strategy, randomizing between recommending a truthful minor and an excessive serious treatment.

\begin{Corollary}
\label{cor:Equ_kost}
i) Both excessive and inadequate recommendations are more likely to occur as the search cost $k$ increases.

ii) As the serious treatment becomes more profitable for the expert, there will be more undertreatment and less overtreatment. 

\end{Corollary}

Corollary \ref{cor:Equ_kost} i) highlights the impact of the search cost $k$ on the strategic behavior of the opportunistic expert. In general, a higher search cost gives the opportunistic expert more incentives to recommend untruthfully. This is because with the increase of the search cost $k$, the consumer becomes more inclined to accept the recommendation from the first expert and searches with a low  er likelihood. As a result, the opportunistic expert takes advantage of the higher search cost and recommends untruthfully with a higher likelihood to exploit a higher profit. As can be seen in Proposition \ref{prop:MainEqbm}, as $k$ increases, the range for the existence of pure strategy (i.e., FOFU) equilibrium increases, (i.e., $\mu^*_2$ weakly increases and $\mu^*_1$ decreases). In the mixed strategy equilibria, the probability that the opportunistic expert recommends untruthfully increases with $k$, i.e., $T_{m1}^*$ in POFU equilibrium decreases with $k$ and $T_{s1}^*$ in FOPU equilibrium decreases with $k$.

As the serious treatment becomes more profitable for the expert, e.g., as $p_s$ increases, the consumer becomes more concerned with getting overtreated and becomes more likely to reject the serious treatment. Therefore, the minor treatment can become more attractive to the consumer. As a result, the increase in price for a serious treatment increases the tendency of undertreatment (i.e.,  $\mu^*_1$ decreases and $T^*_{s1}$ in FOPU decreases) and decreases the tendency of overtreatment (i.e., $\mu^*_2$ weakly decreases and $T^*_{m1}$ in POFU increases).

\subsection{Profits}

Next we examine the implications for the profits of the opportunistic expert. First, we look at the profit implications of $h$, which indicates the ethical level of the market.

\begin{Proposition}
\label{prop:profit_ethics}
i) When the probability of the serious problem is high (i.e., $\mu \leq \mu_3^*$), the opportunistic expert's profits weakly increase with $h$. 

ii) When the probability of the minor problem is high (i.e., $\mu > \mu_3^*$), the opportunistic expert's profits first weakly decrease and then increase with $h$.
\end{Proposition}

In Proposion \ref{prop:profit_ethics}, $\mu^*_3=\frac{k}{p_s-p_m+2k-2\sqrt{k(p_s-p_m)}}$, and the detailed proof is provided in the E-Companion.

When the probability of a serious problem is high (i.e., $\mu$ is low), the consumer's primary concern is undertreatment. Facing a minor treatment recommendation, a consumer is likely to reject it and seek another expert to avoid undertreatment, because even if she draws an opportunitisc expert, the opportunistic expert will never undertreat on the second visit. An increase in the proportion of honest experts boosts consumer confidence in accepting the first minor treatment, thereby increasing the opportunistic experts' expected profits.

When the probability of a minor problem is high  (i.e., $\mu$ is high), the consumer's main concern shifts to overtreatment. In such a case, the proportion of honest experts has a non-linear effect on the consumer's beliefs: if $h$ is low, the consumer is aware that a serious treatment is likely to be an overtreatment but has little chance of receiving a truthful second opinion given a low $h$, thus leading the consumer to accept any recommendation on the first visit. Conversely, when $h$ is high, the consumer trusts the expert's recommendation and is more willing to accept it. Only when the proportion of honest experts is moderate will the consumer reject a serious treatment and seek a new expert, resulting in lower expected profits for the opportunistic expert.

Conditional the ethical level of the market $h$, the opportunistic expert's profits also depend on the probability of the minor problem $\mu$.

\begin{Proposition}
\label{prop:profit_probminor}
The opportunistic expert's profits first weakly increase and then weakly decrease in $\mu$. 
\end{Proposition}

Intuitively, one may expect that the increase of $\mu$ will lead to lower expected profits, since a serious treatment is more profitable than a minor treatment. However, we find that the opportunistic expert's profit can increase with $\mu$ when $\mu$ is low. This is because when the probability of a minor problem ($\mu$) is low, the key concern of the consumer is undertreatment. When facing a minor treatment recommendation, a consumer is suspicious and likely to reject it. As a result, it is difficult for an opportunistic expert to exploit the consumer by recommending inadequately. However, when $\mu$ increases from low to moderate, the consumer is more uncertain about the true nature of her problem and becomes more willing to accept a minor treatment (potential undertreatment) on the first visit to an expert. As a result, the increase of $\mu$ enlarges the information asymmetry between consumers and opportunistic experts, leading to an increase in the opportunistic expert's expected profits.

%When the proportion of honest experts is low, visiting a second expert reduces the chance of undertreatment but not overtreatment, as a selfish expert never undertreats a second-visit consumer. When $\mu$ (the probability of a minor problem) is low, the consumer's primary concern is undertreatment, leading them to likely reject the initial recommendation. As $\mu$ increases, the consumer's concern shifts to overtreatment. However, visiting a new expert will not resolve the overtreatment problem, resulting in the consumer accepting any recommendation from the first expert they consult.

%When the proportion of honest experts is high, a second expert is likely to be honest, prompting the consumer to seek a new expert if the first recommendation is for a serious treatment, especially when 
%$\mu$ is high. Only when $\mu$ is moderate will the consumer accept any recommendation due to the high uncertainty about the nature of the problem, in which case the selfish expert achieves the highest expected payoff.

%At last, the increase in search cost $k$ increases the expert’s expected profits. This is because a higher search cost prevents the consumer from searching for a second opinion, and the opportunitistc expert can exploit the consumer more by recommending untruthful treatment. 

\subsection{Consumer Welfare}

We next examine consumer welfare. By investigating the comparative statics on consumer welfare in each equilibrium, we demonstrate how consumer welfare changes with the market ethics level ($h$) and the probability of getting a minor problem ($\mu$).

\begin{Proposition}
\label{prop:cw_ethic}
Consumer welfare can decrease with the market ethics level $h$ when both $\mu$ and $h$ are sufficiently low (i.e., $\mu\leq \mu^*_3$ and $h \leq min\{h^*_1, h^*_2\}$); 

% ii) $\mu$ is sufficiently high and $h$ is moderate (i.e., $\mu>\mu^*_3$ and $ h^*_4\leq h\leq min\{h^*_3,h^*_5\}$). 
\end{Proposition}

In Proposion \ref{prop:cw_ethic}, $h^*_1=\frac{\mu (p_s-p_m+k)+(1-\mu)(p_m+k^{'}-k)}{2\mu(p_s-p_m)}-\frac{\sqrt{[\mu(p_s-p_m+k)+(1-\mu)(p_m+k^{'}-k)]^2-4\mu(1-\mu)(p_s-p_m)(p_m+k^{'}-k)}}{2\mu(p_s-p_m)}$, $h^*_2=\frac{(p_s+k^{'})k}{2(p_s-p_m)(p_m+k^{'})}$, and the detailed proof is provided in the E-Companion. 

Interestingly, we find that consumer welfare can decrease in the proportion of honest experts ($h$). This occurs in the mixed strategy equilibrium FOPU. In FOPU equilibrium whereby both $\mu$ and $h$ are low, the key concern for a consumer is undertreatment, and the consumer would reject minor treatments with a probability $1-A_{m1}^*$ to decrease the motivation for opportunistic experts to recommend inadequately. However, with the increase of $h$, the consumer becomes more confident in accepting a minor recommendation, rendering more incentives for the opportunistic expert to undertreat. As a result, the probability that the opportunistic expert recommends inadequately increases, thus leading to lower consumer welfare. 

%Similarly, in equilibrium POFU equilibrium whereby $\mu$ is high while $h$ is moderate, the key concern for the consumer is overtreatment. When the proportion of honest experts shifts from low to moderate, a consumer becomes more likely to accept a serious treatment recommendation on the first visit to an expert, leading to a higher probability of overtreating and lower consumer welfare. 

%Interestingly, consumer welfare weakly decreases with $h$, the proportion of honest experts. This decline occurs because, when the probability of a serious problem is high (i.e., $\mu<\mu_1^*$), as in equilibrium $E_2$, the primary concern of consumers is undertreatment. With an increase in the proportion of honest experts, the consumer is more likely to accept minor recommendations. This acceptance encourages opportunistic experts to undertreat more frequently, resulting in a lower consumer surplus. Conversely, when $\mu$ is high, as in equilibrium $E_3$, consumers are primarily concerned with overtreatment. When the proportion of honest experts is moderate, an increase in $h$ reduces consumers' incentives to seek second opinions since the first serious treatment is likely to be truthful. Consequently, opportunistic experts have a higher probability of providing overtreatment as $h$ increases, leading to a decrease in consumer surplus.

\begin{Proposition}
\label{prop:mu_cw}
Consumer welfare decreases with the probability of a minor problem $\mu$ when both $\mu$ and $h$ are sufficiently low ($\mu\leq \mu^*_1$ and $h \leq h^*_3$).
\end{Proposition}

In Proposition \ref{prop:mu_cw}, $h^*_3=\frac{k(p_s+k^{'})}{(p_s-p_m)(p_m+k^{'})}$ and the detailed proof is provided in the Appendix. 

When $\mu$, the probability of the minor problem, increases, one might think the consumer would be better off. However, Proposition \ref{prop:mu_cw} shows that consumer welfare can decrease in $\mu$ when both $\mu$ and $h$ are low. This is because in FOPU equilibrium, the primary concern of a consumer is undertreatment. Given the low proportion of honest experts, a consumer has high incentives to reject a minor recommendation, since it is very likely to be an undertreatment from an opportunistic expert. As $\mu$ increases, the uncertainty in the nature of the problem increases, and an opportunistic expert will take advantage of the information asymmetry, and undertreat the consumer with a higher probability. As a result, consumer surplus decreases with $\mu$.

In other cases, when $\mu$ or $h$ is sufficiently high, not surprisingly, the consumer welfare increases in $\mu$. Compared with a serious problem, a minor problem incurs a lower overall cost, and the increase of $\mu$ can reduce the expected payment of a consumer, leading to higher consumer welfare.

%This is because in such a case, the opportunistic expert will also recommend truthful treatment with some probability, and the problem is more likely to be resolved at a low price on the first visit.

\section{Extentions}
\label{sect:exten}

%In this section, we present several extensions of the model. First, we consider the case where the expert has limited ability to diagnose the consumer. Second, we investigate the situation where the experts face a negative exogenous shock to the capacity. Third, we examine an information-sharing problem, i.e., how the accessibility of consumer diagnosis history affects the equilibrium strategy and consumer welfare. In the fourth extension, the prices are endogenously determined by the experts. Finally, we considered other variants of the model to check the boundary conditions of the main insights. These extensions demonstrate how undertreatment can be moderated, and they further reveal the underlying mechanisms of undertreatment. All details of the proofs are in the Online Supplement.

In this section we analyze several extensions to the basic model. The first three extensions examine how limited capability, capacity constraints, and accessibility of a consumer's diagnosis history interact with undertreatment. Finally, we considered other variants of the model to check the boundary conditions of the main insights. These extensions further reveal the underlying mechanisms of undertreatment. All details of the proofs are in the E-Companion.

\subsection{Limited Capability in Diagnosis}
\label{sec:limited_ability}
 % (I am not sure whether we should discuss the treatment here.)

In the base model, we assume that  experts' diagnosis is always accurate and the expert has perfect information about the nature of the consumer's problem. In reality, the experts have limited ability in diagnosis, which can influence the consumer's tendency to accept the treatment and the expert's incentive to provide untruthful recommendations. Does limitation in capability exacerbate or mitigate undertreatment? Here we assume with a probability of $\epsilon \in [0,\frac{1}{2})$, the experts will make an error in diagnosis. Specifically, when the consumer is diagnosed with a minor (serious) problem, the problem is indeed minor (serious) with probability $1-\epsilon$, and with probability $\epsilon$ the problem is serious (minor). The experts and the consumer are fully aware of the error. The experts' strategy is consistent with his own diagnosis, even though the diagnosis might be wrong.

%Due to the uncertainty in diagnosis, we define a {\em{truthful}} recommendation as one which is consistent with the expert's own diagnosis, and an {\em{accurate}} recommendation as one where the nature of the problem and the recommended treatment is the same. Proposition $\ref{prop:ability}$ summarizes the equilibrium outcome when the expert has limited capability in diagnosis. 

In this setting, undertreatment and overtreatment can occur because of a wrong diagnosis. For example, when a consumer has a serious issue, the expert might misperceive it as a minor issue and recommend a minor treatment. To clarify the mechanisms of undertreatment, we define undertreatment due to a wrong diagnosis as {\em non-strategic undertreatment}, whereas undertreatment due to the expert's intentional fraud as {\em strategic undertreatment}. We focus on the strategic undertreatment. That is, the expert diagnoses the problem as a serious one, and recommends a minor treatment. Note that given the setup, strategic undertreatment does not necessarily result in an undertreated outcome. With a probability of $\epsilon$, the problem is actually a minor, and the minor treatment is actually appropriate. The following proposition summarizes the main findings if the expert has limited diagnostic capability.

\begin{Proposition}
\label{prop:ability}
For $\epsilon<\epsilon^*$, (i) if the probability of a minor issue is moderate (i.e., $\mu \in (max\{\mu^{\epsilon}_1,\mu^{\epsilon}_3\},\mu^{\epsilon}_2)$), there exists a pure strategy equilibrium that the opportunistic expert will fully overtreat and fully undertreat the consumer (i.e, $T_{m1}^{*}=0$, $T_{s1}^{*}=0$, $A_{m1}^{*}=1$, and $A_{s1}^{*}=1$).
% $\mu^{\epsilon}_1 = (max\{\mu^{UD*}_1,\mu^{UD*}_3\}

(ii) The error rate $\epsilon$ increases the tendency of overtreatment and undertreatment when the market's ethical level is low (i.e., $h<\frac{1}{2}$), and reduces overtreatment and undertreatment when the market's ethical level is sufficiently high (i.e., h approaches $1$).  

Here $\epsilon^* = min\{\frac{(1-\mu)k}{(1-\mu)k+\mu h(p_s-p_m+k)},\frac{\mu(1-h)(p_s-p_m+k)}{\mu(1-h)(p_s-p_m+k)+(1-\mu)(p_m+k^{'}-k)}\}$.

\end{Proposition}

% (Do we need to insert a figure to show the equilibrium distribution here?)

Our result shows that the equilibrium outcome in our base model remains robust with the presence of limited diagnostic capability. The full overtreatment and full undertreatment equilibrium exists when the probability of the minor issue is moderate.  In addition, we find that the impact of error rate on expert recommendation strategies depends on the market's ethical level; when the market's ethical level is low, the increase in error rate will lead to more overtreatment and undertreatment in the marketplace. This is because even if opportunistic experts intend to recommend untruthfully to a consumer, with the increase in error rate, more untruthful recommendations will turn out to be appropriate treatment, which would increase the consumer's trust towards the recommendation more. As a result, the expert becomes more likely to overtreat and undertreat the consumers. On the other hand, when the market's ethical level is high, the increase in error rate will lead to less overtreatment and undertreatment since the diagnosis error reduces the consumer's trust in the expert's recommendation, which in return pushes opportunistic experts to recommend more truthfully. 

Importantly, the analysis of limited diagnostic capability can shed light on the liability issues if the consumers are undertreated. When the diagnosis is perfectly accurate, if the expert fails to resolve the issue, the consumer can pursue legal action against the expert. However, the limited diagnosis ability can, to some extent, excuse the expert from liability. The failed treatment outcome can be either due to honest diagnosis mistakes instead of fraud.

\subsection{Limited Capacity}
\label{sec:capacity}
In service markets, limited capacity imposes challenges for service providers to offer adequate treatment to the consumers. Our main model focuses on the case where the expert has no capacity constraint when treating the consumer, and thus, undertreatment occurs due to strategic reasons. Here we relax the assumption, and we examine how capacity influences equilibrium outcomes. We assume, in each period, there is a probability $\chi \in [0, 1]$ that the experts would face a negative shock to the capacity such as a shortage of manpower, break-down of the equipment, etc. When the capacity shock occurs, the experts only have the resources to deal with a minor treatment but are unable to deal with a serious treatment. Note that here, the capacity shock occurs independently across periods. The consumer does not observe the capacity facing the expert. A higher value of $\chi$ indicates a higher likelihood that the experts would face the shock. When the capacity limitation occurs, an honest expert will reject a consumer with a serious problem, while an opportunistic expert chooses to recommend a minor treatment to a consumer with a serious problem. To better pinpoint the mechanism of undertreatment, we define the undertreatment due to capacity constraints as {\em non-strategic undertreatment}, whereas undertreatment without limited capacity as {\em strategic undertreatment}. Note that the honest experts will reject the consumer if the capacity is limited, and thus non-strategic undertreatment can only come from opportunistic experts. Since capacity constraint occurs exogenously, we focus on strategic undertreatment in the equilibria. The following proposition states how the presence of limited capacity would moderate the experts' tendency to undertreat consumers.

\begin{Proposition}
\label{prop:capacity}
(i) When the probability of limited capacity is low (i.e., $\chi < \chi^* $), strategic undertreatment always occurs in all equilibria (i.e., $T_{s1}^*<1$).

(ii) When the probability of limited capacity is high (i.e., $\chi\geq  \chi^*$), strategic undertreatment occurs when the probability of a minor problem $\mu$ is high (i.e., $\mu \geq \mu^{\chi*}$), but it does not occur in the equilibria when $\mu$ is low.

%i). When the capacity constraints are not severe ($\chi\leq \frac{p_m-c_m}{p_s-c_S}$), both strategic undertreatment and non-strategic undertreatment in all equilibria.

%ii). When the capacity constraints are severe ($\chi\geq \frac{p_m-c_m}{p_s-c_S}$),  both strategic undertreatment and non-strategic undertreatment occur when the probability of a minor problem ($\mu$) is high. When the probability of a minor problem is low, strategic undertreatment no longer constitutes an equilibrium. 
\end{Proposition}
%Here, $\chi^*=\frac{p_m-c_m}{p_s-c_S}$ and $\mu^{\chi*}=\frac{k+h\chi(l_s-p_s)+(1-h)\chi(l_s-p_s+p_m)}{hk+h\chi(l_s-p_s)+(1-h)\chi(l_s-p_s+p_m)+(1-h)(\chi+h(1-\chi))(p_s-p_m)}$.

Here, $\chi^*=\frac{p_m-c_m}{p_s-c_S}$ and $\mu^{\chi*}=\frac{k+h\chi(l_s-p_s)+(1-h)\chi(l_s-p_s+p_m)}{hk+h\chi(l_s-p_s)+(1-h)\chi(l_s-p_s+p_m)+(1-h)(\chi+h(1-\chi))(p_s-p_m)}$. When $\chi$ is low, the equilibrium is similar to that in the base model. Facing a consumer with a serious problem, opportunistic experts recommend inadequately even if the capacity is not limited since the opportunistic expert has a high chance (with probability $1-\chi$) to conduct a serious treatment to the under-treated consumer on the second visit and benefit from two treatments. 

However, when the probability of limited capacity is high, the opportunistic expert has more incentive to recommend truthfully to a consumer with a serious problem. As a result, the strategic undertreatment no longer exists in equilibrium when $\mu$ is low. This is because when the capacity constraints are severe, non-strategic undertreatment occurs with a high probability. Facing a minor treatment recommendation, a consumer is ready to reject it since the consumer would perceive a minor treatment recommendation to be inadequate, especially when $\mu$ is low. In addition, from the opportunistic expert's perspective, limited capacity also constrains the opportunistic expert from making profits from a serious treatment in the next period. As a result, it is a dominant strategy for opportunistic experts to recommend truthfully to a consumer with a serious problem when $\mu$ is low while the capacity constraints are severe. Our results here show that both overtreatment and undertreatment occur, no matter whether capacity constraints are present. Besides, we identify different mechanisms of undertreatment in the credence good market: experts may undertreat due to strategic reasons (benefit from two treatments) or non-strategic reasons (limitation of resources). Interestingly, we also find that the limited capacity and non-strategic undertreatment have a crowd-out effect on strategic undertreatment: when the capacity constraints are severe, a consumer perceives a minor treatment to be inadequate and is likely to reject it, which in turn encourages the opportunistic expert to recommend truthfully on a serious problem when the capacity is not limited. Our finding is related to \cite{pacc2015false}, who demonstrated that congestion can mitigate overtreatment problems. We have different settings, and we show that capacity shock can mitigate the strategic undertreatment problem.

\subsection{Restricted Access to Consumers Diagnosis History}
\label{sec:inaccess_history}
In the benchmark model, we examined a credence service market where consumers' diagnosis history is fully accessible to the experts. For example, in the healthcare industry, information technology efforts facilitate the exchange of patients' past histories across hospitals, specialists, and laboratories. However, in many cases, consumers' diagnosis history is confidential and inaccessible to the experts. Here, we relax the assumption about the accessibility of consumers' diagnosis history. We consider a case where the expert cannot distinguish whether the consumer is on her first or second visit due to the inaccessibility of the consumer's diagnosis history. Given the incentive of fraud by the opportunistic expert, does prohibiting information sharing about the diagnosis history mitigate the problem of overtreatment and undertreatment? What are the implications for consumer welfare if consumers' diagnosis history is inaccessible to the experts?

In Period $t$, the visiting consumer can be either $C_t$ or $C_{t-1}$. If the consumer is $C_t$, she is on her first visit. Then, the consumer's decision process is similar to the main model, and she might reject the recommendation and seek a second opinion. If the consumer is $C_{t-1}$, she either rejected the previous recommendation or her issue is not resolved in Period $t$, and thus is on her second visit. The expert cannot distinguish whether the consumer is $C_t$ or $C_{t-1}$ and forms a belief about the type of the consumer, and the belief is consistent with the consumer's equilibrium strategy. Note that $C_{t-1}$ has a higher chance of rejecting the minor treatment than $C_t$ because Period $t$ is her last period in the game.

\begin{Proposition}
\label{prop:diag_history}
i) \textbf{Full overtreatment and full undertreatment (FOFU) equilibrium:} If the probability of a minor issue is moderate (i.e., $\mu \in (max\{\mu^{\gamma}_1,\mu^{\gamma}_3\},\mu^{\gamma}_2)$), there exists a pure strategy equilibrium that the opportunistic expert will fully overtreat and fully undertreat the consumer (i.e, $T_{m1}^*=0$, $T_{s1}^*=0$, $A_{m1}^*=1$, and $A_{s1}^*=1$).

ii) When the probability of minor issue is moderately high or moderately low (i.e., $\mu^*_2\leq\mu\leq \mu^{\gamma}_2$ or $max\{\mu^{\gamma}_1, \mu^{\gamma}_3\}\leq\mu\leq \mu^{*}_1$), the consumer is worse off if the access to the consumers' diagnosis history is restricted.

\end{Proposition}

%The values of thresholds are provided in E-Companion. 
Here, $\mu^{\gamma}_1=0$ when $h\leq\frac{l_s-p_s+k-k^{'}}{l_s-p_s+p_m}$; otherwise $\mu^{\gamma}_1=\frac{(1-h)(h(p_m+k^{'}-k)-(1-h)(l_s-p_s+k-k^{'}))}{(1-h)(h(p_m+k^{'}-k)-(1-h)(l_s-p_s+k-k^{'}))+h((1-h)(p_s-p_m)+k)}$. $\mu^{\gamma}_2=1$ when $h\leq\frac{h}{p_s-p_m}$; otherwise $\mu^{\gamma}_2=\frac{h^2k+h(1-h)(l_s-p_s+p_m+k)}{h^2k+h(1-h)(l_s-p_s+p_m+k)+(1-h)(h(p_s-p_m)-d)}$. $\mu^{\gamma}_3=0$ when $h\leq\frac{l_s-p_s-p_m+k-k^{'}}{l_s-p_s}$; otherwise $\mu^{\gamma}_3=\frac{(1-h)((p_m+k^{'}-k)-(1-h)(l_s-p_s))}{(1-h)((p_m+k^{'}-k)-(1-h)(l_s-p_s))+h^2(l_m-p_m+k)+(1-h)h(p_s-p_m+k)}$.

Similar to the rationale of the benchmark model, the higher profit margin of serious treatment ($p_s-c_s>p_m-c_m$) induces opportunistic experts to overtreat, while the lower search cost of returning the previous expert ($k^{'}<k$) induces opportunistic experts to undertreatment. 

One may expect that full knowledge of consumers' diagnosis history may facilitate the opportunistic experts' exploitation of the consumers. Hence, limiting the access to information would improve consumer welfare. We show that when the uncertainty in the nature of the consumer's problem is moderate ($\mu^*_2\leq\mu\leq \mu^{\gamma}_2$ or $max\{\mu^{\gamma}_1, \mu^{\gamma}_3\}\leq\mu\leq \mu^{*}_1$), consumer welfare can be better off if the experts have information about the consumer's search history. This is because when the experts have more information about consumers, opportunistic experts will recommend a serious problem invariantly to a consumer on the second visit to an expert, and it reduces the chance of undertreatment. When the consumer is visiting the first expert, the consumer becomes more likely to reject a recommended treatment since the consumer knows that there will be no undertreatment on the second visit and her problem can be resolved for sure. Conversely, if the consumer's search history is not known to the experts, the consumer faces the risk of being undertreated if she visits the second expert, and the consumer becomes more conservative and likely to accept any recommendation on the first visit to an expert, which in return induces more undertreatment and overtreatment in the marketplace. Our analysis provides insights into the credence good marketplace; the practice that facilitates information sharing across different institutes, such as electronic healthcare records in the healthcare system, can be beneficial for consumers because it reduces consumers' distrust, which eventually increases consumer welare.

%When the consumer diagnosis history is not known to the experts, there is a different equilibrium in which opportunistic experts recommend excessively with probability one, albeit the consumers on the first visit reject any serious treatment recommendation. This is because the market consists of both first-visit and second-visit consumers, the opportunistic experts recommend excessively to target those consumers on the second visit to an expert. 

\subsection{Other Variants of the Model}
\label{subsect:other_variants}
We consider various model variants as robustness checks.

\subsubsection{Alternative contract}
\label{subsubsec:alternative_contract}

In some cases, if the problem is unresolved, the service provider might have to refund the fee for the initial treatment before charging for the second treatment. Otherwise, the provider may face legal repercussions. We examine an alternative contract under which a consumer pays for $p_s-p_m$ for the second serious treatment if she received a minor treatment but the problem persists. Put differently, the alternative contract removes the direct financial motivation for the experts, and an expert at most gets $p_s$ from two treatments by recommending a minor treatment to a consumer with a serious problem. %We investigate the equilibrium outcome under this alternative contract and clarify how financial incentives impact opportunistic experts' behavior and whether the undertreatment issue can be resolved by adopting an alternative contract. 
Under the alternative contract, we found that full undertreatment remains in equilibrium when the probability of a minor problem and the proportion of honest experts are high, and the pricing gap between the minor and serious treatment is sufficiently large (i.e., $\mu > \frac{k}{h k+(1-h)h(p_s-p_m)}$, $h>\frac{k}{p_s-p_m}$ and $p_s-p_m>c_s$).

The rationale is as follows, given a large pool of honest experts and a high likelihood of minor problems, a consumer would be likely to perceive a serious treatment to be excessive; combined with a high price for a serious treatment, the consumer has incentives to search for another opinion when she is recommended to a serious problem. As a response, an opportunistic expert strategically recommends inadequately to a consumer with a serious problem, and when the consumer returns to the previous expert after undertreatment, the opportunistic expert ensures his payoff from a serious treatment $p_s$. The analysis of the alternative contract helps to disentangle the two motivations behind undertreatment: first, it is used as a “hook” to induce the acceptance of the cheaper, minor treatment; second, it generates a subsequent demand for a more profitable treatment because the undertreatment does not resolve the issue. The analysis shows that when we eliminate the second motivation, the first motivation is sufficient to generate undertreatment.

\subsubsection{Undertreatment is not Fully Discovered in the Short Run}
\label{subsubsec:short_discovery}

Undertreatment may not be discovered immediately by the consumer until the very end. For example, an auditing agency might provide inadequate service to the firm. The firm might only discover the undertreatment by the agency until a bankruptcy. We examine situations where undertreatment is not fully discovered in the short run. Extending our main model, here we assume that with probability $\delta\in(0,1)$, the consumer does not discover whether her problem is resolved or not after receiving a minor treatment in Period $t$. As a result, the consumer's problem can remain unresolved at the end of the game. That is, the consumer is uncertain whether the minor treatment she received in Period $t$ is undertreatment in Period $(t+1)$. 

Our analysis shows that when the probability that an undertreatment is uncovered is high ($\delta\geq\frac{p_m-c_m}{p_s-c_s}$), opportunistic experts engage in undertreating only when the probability of a minor problem ($\mu$) is high; when the probability of a minor problem is low, an opportunistic expert recommends truthfully on a serious problem. The rationale is that the presence of undiscovered undertreatment daunts the opportunistic expert to undertreat, since with a probability $\delta$, the consumer cannot discover her problem and return to the experts, realizing no additional profits for the experts. For the consumers,  there is a risk with a probability $\delta$ that her problem will remain undiscovered and unresolved at the end of the game with a loss of $l_s$.  When the probability of a minor problem is low, a consumer facing a minor treatment recommendation is highly vigilant that the recommendation is inadequate, while the experts realize the chance of getting additional profits from the next period is low, and undertreatment does not occur in equilibrium. 

\subsubsection{Consumer Resentment After an Undertreatment}
\label{subsubsec:resentment}

In our main model, if the problem persists, the decision facing the consumer is whether to sample a new expert (incurring a search cost $k$) or to return to the initial expert (incurring a lower search cost $k'< k$). Since in the next period, both the honest and opportunistic expert will recommend a serious treatment conditional on a serious issue, the consumer would prefer to return to the initial expert due to the lowered search cost (Lemma \ref{lem:consumer_return}). However, research in behavioral economics suggests that fairness-driven individuals develop resentment towards parties they perceive as having treated them unfairly (e.g., \cite{charness2002understanding,loch2008social}). In our model, if $k'> k$, it resembles resentment --- a psychological barrier that prevents the consumer from returning to the initial expert if she finds out she is undertreated.

Interestingly, we find the tendency to undertreat consumers depends on whether the experts have access to the consumers' diagnosis history. If the experts have full knowledge of the history (as in the main model), undertreatment will not emerge in equilibrium if the consumers develop resentment. However, if the experts do not have access to consumers' diagnosis history, full undertreatment can still occur in equilibrium. Here undertreatment is mainly serve as a hook to reduce rejection.

\subsubsection{Heterogeneity in Treatment Capability}
\label{subsubsec:heter_ability}

In our main model, the service providers are homogeneous in their ability to provide treatment. Here, we consider the case where the service providers are differentiated in their treatment capabilities. Specifically, there are two types of service providers, the high and the low type. With a probability of $\alpha$, the expert is a high type who is capable of providing both minor and serious treatments. With a probability of $1-\alpha$, the expert is a low type who can only provide a minor treatment. In such a market, the experts are differentiated in both their ethical level and treatment capabilities. The expert knows his type, whereas the consumer only knows the probability of the type. 

Moreover, we assume that the type of ability is independent of the type of honesty, and an honest expert with low capability will reject to treat a consumer with a serious problem, while an opportunistic expert with low capability will recommend a minor treatment to a consumer with a serious problem to gain the profits $p_m-c_m$. Similar to the limited capability in diagnosis, we define inadequate treatment when the expert is capable of serious treatment as {\em{strategic undertreatment}}, and inadequate treatment due to the expert's low type as {\em{non-strategic undertreatment}}. 

Our result shows that when the probability of a serious problem is high (i.e., $\mu$ is low), opportunistic experts with high capability no longer conduct strategic undertreatment. This is because, with the presence of experts with low capabilities, the consumers become more suspicious when the recommended treatment is minor, since it is very likely to be non-strategic undertreatment from an expert with low capability, especially when $\mu$ is low.  As a result, the opportunity expert with high capability will recommend truthfully to a consumer with serious problems since the consumer will accept a serious recommendation for certain while rejecting a minor recommendation due to distrust. Nonetheless, an opportunistic expert with the low ability still undertreats as constrained by his capability. 

\subsubsection{Endogenized Price}
\label{sec:endo_price}
The prices can be endogenously determined by the expert. At the beginning of Period $t$, the honest expert (h) and opportunistic expert (o) decide the prices for a minor and a serious treatment. We use $p_{ig}$ to present the price for condition $i\in \{s, m\}$ and set by the type of the expert $g\in \{h, o\}$. Because of the search cost, the experts are local monopolies \citep{diamond1971model} and there is no direct price competition among experts. In the E-Companion, we show that when experts decide the prices endogenously, undertreatment always occurs with a positive probability (i.e., $T_{s1}^*<1$). 

To gain some more insights about the equilirium outcome when the price is endogenized, we can explore a market fully consists of opportunistic experts (see \cite{fong2014role}). We identify a unique pure strategy equilibrium in which opportunistic experts recommend truthfully on a minor problem but recommend inadequately to a consumer with a serious problem  When the probability of a minor problem is high. The equilibrium price is $p_{mo}^{*}=l_m$ and $p_{so}^{*}=l_m-c_m+c_s$.

When opportunistic experts decide the prices of treatments, they maximize profits and set the price for a minor treatment as high as possible such that $p_m=l_m$. As a result, a consumer may reject a recommendation on the second visit to an expert due to the high price. In this case, when $p_s-c_s>p_m-c_m$, a consumer is suspicious of the expert's motivation when she faces a serious treatment recommendation and chooses to accept the serious treatment with a probability lower than one as a mixed strategy. On the contrary, when $p_s-c_s<p_m-c_m$, the consumer's concern is undertreatment, and the consumer accepts a minor treatment with a probability lower than one as a mixed strategy. Only when $p_s-c_s=p_m-c_m$ the consumers on the second visit believe all recommendations from the experts are truthful and would like to accept the recommendations. When $p_s-c_s=p_m-c_m$, on the first visit to an expert, a consumer accepts a serious treatment with probability one since it is perceived as truthful since the opportunistic expert is not motivated to recommend excessively by the financial incentives. When facing a minor recommendation, a consumer prefers to accept rather than reject it when the probability of a minor problem is high, in which case a minor treatment will be perceived as truthful with a high likelihood. In sum, our results in this extension validate the robustness of our finding and show that undertreatment exists in the unique pure strategy equilibrium if an opportunistic expert can endogenously decide the prices in the market.

\section{Discussions}
\label{sect:discuss}

In credence goods markets, profit-maximizing experts might exploit the informational advantage and provide either overtreatment or undertreatment to the consumers. This paper focuses on the role of the consumer's costly search process and the nature of repeated interactions. Using a dynamic game-theoretic model, our study explores the strategic use of undertreatment by opportunistic service providers and its coexistence with overtreatment in equilibrium. We demonstrate how an opportunistic expert might first recommend a minor and less costly treatment to lure consumers to accept the treatment, and then ``lock-in'' consumers for another more costly treatment because of the search cost of finding a new expert. The tendency to offer an inadequate treatment is influenced by consumer search costs, their beliefs about the problem's severity, and the overall ethics of the market. %In service markets, ``band-aid'' solutions are widely observed. These occur when providers offer quick fixes instead of comprehensive solutions. Our analysis shows that how ``band-aid'' solutions not only fail to address the core problem but also strategically set the stage for future service demands. 

%Combining the analysis in the benchmark model with the results in extensions, we identify different mechanisms that lead to the undertreatment in the credence good marketplace. In the benchmark model whereby consumers return to the previous expert after undertreatment, the direct financial incentives drive the opportunistic experts to recommend inadequately to benefit from two treatments, and the diagnosis uncertainty provides a perfect excuse for liability (Extension 5.1). When experts face capacity constraints, opportunistic experts recommend inadequately to cash out under limited capacity (Extension 5.2). When the consumer does not return to the expert after undertreatment,  opportunistic experts recommend inadequate as a strategic balance of undertreatment (Extension 5.3). Our results in the extensions highlight the importance of understanding the different mechanisms behind the undertreatment phenomenon in the credence good market. 

\subsection{Managerial and Policy Implications}
\label{sect:policyimpl}

The findings can shed light on the following questions:

\textit{How do the practices encouraging ethical behavior affect consumer welfare?} Governments and professional bodies develop and enforce codes of conduct that encourage ethical behavior among service providers, including training and education focused on ethical decision-making and the consequences of fraudulent practices. These practices can potentially increase the ethical level of the market. However, as shown in Proposition \ref{prop:cw_ethic}, consumer welfare can decrease as the proportion of ethical experts increases. This is because when consumers are concerned about expert fraud, they might reject the recommendation and seek a second opinion. As the market becomes more ethical, the consumers tend to be more willing to accept the recommendation. This gives the unethical expert more incentives to defraud and results in lower consumer welfare.

\textit{Does improvement in service providers' capability and capacity reduce undertreatment problems?} Undertreatment in the market is largely due to non-strategic reasons, especially service providers' limited ability and capacity. Interestingly, while improving the service provider's diagnostic capability and capacity reduces the non-strategic undertreatment, we found improving diagnostic capability and capacity can increase the tendency of strategic undertreatment (see Proposition \ref{prop:ability} and \ref{prop:capacity}).

\textit{How do privacy and data sharing regulations affect service providers' behavior and consumers?} Concerns have been raised that providing service providers with access to consumers' diagnostic history could be harmful to consumers, because the shared information can be potentially exploited by opportunistic actors. However, our findings suggest otherwise. We found that when the uncertainty in the nature of the consumer's problem is moderate, consumers can receive more efficient treatment when the opportunistic provider can access the consumers' diagnosis history (Proposition \ref{prop:diag_history}). Policymakers must carefully weigh the potential benefits of information sharing against the associated privacy concerns.

\subsection{Limitations and Future Research}
\label{sect:limitation}

There are several limitations in our paper that leave avenues for future research. First, we did not explicitly model the reputational and liability concerns faced by service providers. In certain contexts, strict liability is challenging to enforce \citep{dulleck2006doctors} and may introduce moral hazard issues \citep{taylor1995economics}. Providers who strategically undertreat consumers might face reputation loss or legal repercussions if such behavior is uncovered, potentially mitigating their incentive to undertreat. Conversely, treatments involving significant risks or invasive procedures, such as certain medical interventions, may lead providers to adopt more conservative approaches due to reputation or liability concerns, potentially resulting in increased undertreatment. Future research could extend the model to explicitly incorporate these reputational and liability concerns. 

Second, we focused on a market where there is no competition among service providers. However, service providers can compete in terms of prices and quality of service. Earlier research found that competition can result in more expert fraud, lower service quality \citep{mimra2016price}, and lower consumer welfare \citep{wolinsky1993competition}. Future work can explore how competition affects consumer welfare in our setting. 

Last, other market participants, such as information intermediaries (e.g., financial advisors, brokers) and self-regulatory bodies (e.g., American Veterinary Medical Association), can also affect the service provision quality (see \cite{inderst2012competition} and \cite{xiao2020half}). Future studies could examine how these parties could moderate market efficiency and consumer welfare.

%%REFERENCES%%
%%%%%%%%%%%%%%%%%%%%%%%%%%%%%%%%%%%%%%%%%%%%%%%%%%%%%%%%%%%%%%%%%%%%%%%%%%%%%%%%%%%%%%%%%%%%%%%%%%%%%%%%%%%%%%%%%%%%%%%%%%%%%%%%%%%%
%% This template complies references using bibtex. You will need to use pomsref.bst file for biblography style.
%REFERENCES USING BIBTEX FILES
%%%%%%%%%%%%%%%%%%%%%%%%%%%%%%%%%%%%%%%%%%%%%%%%%%%%%%%%%%%%%%%%%%%%%%%%%%%%%%%%%%%%%%%%%%%%%%%%%%%%%%%%%%%%%%%%%%%%%%%%%%%%%%%%%%%%

\bibliographystyle{pomsref}

 \let\oldbibliography\thebibliography
 \renewcommand{\thebibliography}[1]{%
    \oldbibliography{#1}%
    \baselineskip14pt %Change this for line spacing within the same reference
    \setlength{\itemsep}{10pt}% %Change this for spacing between two referneces
 }
\bibliography{References}

\appendix

\section*{APPENDIX: Mathematical Proofs}
\setcounter{equation}{0}       % reset equation counter
\renewcommand{\theequation}{A\arabic{equation}}  % change equation numbering

\section*{Proof of Lemma \ref{lem:period2_strat}}

In Period $(t+1)$, if it is a serious recommendation, by accepting it, the consumer pays $p_s$ with her problem resolved; by rejecting it, the consumer gets $-l_m$ if the problem is minor, and gets $-l_s$ if the problem is serious. When $p_s<l_m<l_s$, it is a dominant strategy for a consumer to accept a serious recommendation in Period $(t+1)$. On the other hand, facing a minor recommendation, a consumer can infer it is a truthful recommendation, and the consumer is also willing to accept it. Thus, $A_{m2}=A_{s2}=1$. From the opportunistic expert's perspective, recommending a serious treatment to a consumer is a dominant strategy, since $p_s-c_s>p_m-c_m$ and the consumer will always accept a serious treatment. \hspace{0.3cm} $\Box$

%\par\noindent
%\textbf{Proof of Lemma \ref{lem:consumer_return}}. If the problem remains unresolved after a minor treatment, the consumer knows that her problem is serious. By visiting a new expert, the consumer incurs a cost $k$ and gets a serious treatment with price $p_s$; by returning to the previous expert, the consumer incurs a cost $k^{'}<k$ and gets a serious treatment with price $p_s$. As a result, returning to the previous expert is a dominant strategy. \hspace{0.3cm}$\Box$

\par\noindent
\section*{Proof of Proposition \ref{prop:MainEqbm}}

First, we write down the consumer's expected payoff on the first visit to an expert. Note from Lemma \ref{lem:period2_strat} that a consumer accepts any treatment recommendation on the second visit to an expert, by adopting the backward induction, we specify the consumer's expected payoff from accepting and rejecting a minor treatment recommendation on the first visit to an expert as (\ref{eq:am}) and (\ref{eq:rm}) respectively.
\begin{align}
& \mbox{By accepting ($A_{m1}=1$): } -\tau_m p_m-(1-\tau_m)(p_s+p_m+k^{'}) \label{eq:am} \\
& \mbox{By rejecting ($A_{m1}=0$): } -\tau_m h(p_m+k)-\tau_m (1-h)(p_s+k)-(1-\tau_m)(p_s+k) \label{eq:rm} 
\end{align}

A consumer strictly prefers to accept a minor recommendation if (\ref{eq:am})$>$(\ref{eq:rm}). When (\ref{eq:am})$<$(\ref{eq:rm}), a consumer strictly prefers to reject a minor recommendation. When (\ref{eq:am})$=$(\ref{eq:rm}), the consumer plays a mixed strategy randomizing between accepting and rejecting a minor treatment recommendation.

Similarly,  the consumer's expected payoff from accepting and rejecting a serious recommendation is specified in (\ref{eq:as}) and (\ref{eq:rs}), respectively. 
\begin{align}
& \mbox{By accepting ($A_{s1}=1$): } -p_s \label{eq:as} \\
& \mbox{By rejecting ($A_{s1}=0$): } -\tau_s (p_s+k)-(1-\tau_s)h(p_m+k)-(1-\tau_s)(1-h)(p_s+k)\label{eq:rs}
\end{align}

The consumer strictly prefers to accept a serious recommendation if (\ref{eq:as})$>$(\ref{eq:rs}). A consumer strictly prefers to reject a serious recommendation if (\ref{eq:as})$<$(\ref{eq:rs}). The consumer plays a mixed strategy randomizing between accepting and rejecting a serious treatment recommendation if (\ref{eq:as})$=$(\ref{eq:rs}) and $k=(1-\tau_s)h(p_s-p_m)$.

The opportunistic expert determines his recommendation strategy based on the expected payoff. The incentive compatibility conditions are in IC (\ref{eqn:ss}) and IC (\ref{eqn:sm}). 
%\begin{align}
%& IC(Inadequate): (p_s-c_s+p_m-c_m)A_{m1} \geq (p_s-c_s)A_{s1}  \label{eqn:ss}\\
%& IC(Excessive):  (p_s-c_s)A_{s1} \geq (p_m-c_m)A_{m1}. \label{eqn:sm}
%\end{align}
First, we can show that  $T_{s1}<1$ and $T_{m1}<1$ by contradiction. Suppose the opportunistic expert recommends truthfully to a first-visit consumer with a minor problem such that $T_{m1}=1$, and there is no overtreatment (i.e., $\tau_s=1$). Facing a serious treatment recommendation, it is a dominant strategy for a consumer to accept a serious recommendation $A_{s1}=1$ since (\ref{eq:as})$>$(\ref{eq:rs}). In this case, the opportunistic expert has incentives to deviate from $T_{m1}=1$, since the profit from overtreatment is higher than that from a truthful recommendation such that $(p_s-c_s)A_{s1}=(p_s-c_s)>(p_m-c_m)\geq(p_m-c_m)A_{m1}$. Similarly, suppose the opportunistic expert recommends truthfully on a serious problem such that $T_{s1}=1$, and there is no undertreatment (i.e., $\tau_m=1$). Then, it is a dominant strategy for a consumer to accept a minor recommendation $A_{m1}=1$ since (\ref{eq:am})$>$(\ref{eq:rm}). In this case, the opportunistic expert has incentives to deviate from $T_{s1}=1$, since the profit from undertreatment is strictly higher than that from the truthful recommendation on a serious problem such that $(p_s-c_s+p_m-c_m)A_{m1}=(p_s-c_s+p_m-c_m)>(p_s-c_s)\geq(p_s-c_s)A_{s1}$. 

Since $T_{s1}<1$ and $T_{m1}<1$, there are four possible equilibrium strategy pairs: $(T_{s1}=0,T_{m1}\in (0,1))$, $(T_{s1}=0,T_{m1}=0)$, $(T_{s1}\in (0,1),T_{m1}=0)$; and $(T_{s1}\in (0,1), T_{m1}\in (0,1))$. First, we show that $T_{s1}\in (0,1)$  and $T_{m1}\in (0,1)$ cannot constitute an equilibrium by contradiction. Suppose $T_{s1}\in (0,1)$ and $T_{m1}\in (0,1)$ constitutes an equilibrium. From our analysis earlier, the two conditions in (\ref{eqn:ss}) and (\ref{eqn:sm}) must be binding such that $(p_s-c_s+p_m-c_m)A_{m1}=(p_s-c_s)A_{s1}$ and $(p_s-c_s)A_{s1}=(p_m-c_m)A_{m1}$. However, there is no strategy pair $(A_{s1},A_{m1})$ satisfies the conditions except  $A_{m1}=A_{s1}=0$. However, $(T_{m1}\in (0,1),T_{s1}\in (0,1),A_{m1}=0,A_{s1}=0)$ cannot constitute an equilibrium since when $A_{m1}=A_{s1}=0$, an opportunistic expert gets zero expected payoff and has incentives to deviate to truthful recommending such that $T_{s1}=1$ or $T_{m1}=1$. As a result, $T_{s1}\in (0,1)$  and $T_{m1}\in (0,1)$ cannot be an equilibrium strategy.

Next, we examine the conditions when $T_{s1}=0$  and $T_{m1}=0$ constitute an equilibrium. When $T_{s1}=0$ and $T_{m1}=0$ is the equilibrium strategy, (\ref{eqn:ss}) and (\ref{eqn:sm}) hold with strict inequality, and it indicates $A_{m1}>0$,  $A_{s1}>0$. In addition, when $T_{s1}=0$  and $T_{m1}=0$, $\tau_m=\frac{\mu h}{\mu h+(1-\mu)(1-h)}$ and $\tau_s=\frac{(1-\mu)h}{(1-\mu) h+\mu (1-h)}$ are deterministic values, and the condition under which  $A_{m1}\in(0,1)$ constitutes an equilibrium strategy such that (\ref{eq:am})$=$(\ref{eq:rm}) only holds for a knife-edge $k+\frac{\mu h}{\mu h+(1-\mu)(1-h)}(1-h)(p_s-p_m)=\frac{(1-\mu) (1-h)}{\mu h+(1-\mu)(1-h)}(p_m+k^{'})$; 
similarly, $A_{s1}\in(0,1)$ constitutes an equilibrium strategy such that (\ref{eq:as})$=$(\ref{eq:rs}) only holds when $k=\frac{\mu(1-h)}{\mu(1-h)+(1-\mu)h}h(p_s-p_m)$. Our analysis only focuses on non-trivial cases, and then we examine the equilibrium when $A_{m1}=1$ and $A_{s1}=1$. When $(T_{m1}=0,T_{s1}=0, A_{m1}=1, A_{s1}=1)$ constitutes an equilibrium, it is straightforward to see that (\ref{eqn:ss}) and (\ref{eqn:sm}) hold with strict inequality, and conditions satisfy when (\ref{eq:as})$>$(\ref{eq:rs}) and (\ref{eq:am})$>$(\ref{eq:rm}), which can be reduced to $k>(1-\tau_s)h(p_s-p_m)$ and $k+\tau_m(1-h)(p_s-p_m)>(1-\tau_m)(p_m+k^{'})$. By plugging  $\tau_s=\frac{(1-\mu)h}{(1-\mu) h+\mu (1-h)}$ into $k>(1-\tau_s)h(p_s-p_m)$, the condition can be reduced to $(1-\mu)hk>\mu(1-h)(h(p_s-p_m)-k)$. Note that when $h\leq\frac{k}{p_s-p_m}$, the inequality holds since $\mu(1-h)(h(p_s-p_m)-k)\leq0$; when $h>\frac{k}{p_s-p_m}$, the inequality holds if and only if $\mu<\frac{hk}{hk+(1-h)(h(p_s-p_m)-k)}$. Then we define $\mu^*_2$ as $\mu^*_2=1$ when $h < \frac{k}{p_s - p_m}$ while $\mu^*_2=\frac{hk}{hk + (1-h)h(p_s - p_m) - (1-h)k}$ otherwise, and we have (\ref{eq:as})$>$(\ref{eq:rs}) holds when $\mu<\mu^*_2$. Similarly, by plugging $\tau_m=\frac{\mu h}{\mu h+(1-\mu)(1-h)}$ into $k+\tau_m(1-h)(p_s-p_m)>(1-\tau_m)(p_m+k^{'})$, we find that this inequality can be reduced to $\mu h((1-h)(p_s-p_m)+k)>(1-\mu)(1-h)(p_m+k^{'}-k)$, and this condition holds if and only if $\mu>\frac{(1-h)(p_m+k^{'}-k)}{hk+(1-h)h(p_s-p_m)+(1-h)(p_m+k^{'}-k)}=\mu^*_1>0$. In sum, we have shown that $(T_{m1}=0,T_{s1}=0,A_{m1}=1,A_{s1}=1)$ constitutes an equilibrium if and only if $\mu\in(\mu_1^*,\mu_2^*)$.

%\[
%\mu^*_2 =
%\begin{cases} 
%1 & \text{if } h < \frac{k}{p_s - p_m}, \\
%\frac{hk}{hk + (1-h)h(p_s - p_m) - (1-h)k} & \text{otherwise}.
%\end{cases}
%\]

Then, we identify the equilibrium conditions when $T_{s1}=0$ and $T_{m1}\in (0,1)$. In this equilibrium, $(p_s-c_s+p_m-c_m)A_{m1}>(p_s-c_s)A_{s1}$ and $(p_s-c_s)A_{s1}=(p_m-c_m)A_{m1}$, and we can deduce that $A_{m1}=1$ and $A_{s1}=\frac{p_m-c_m}{p_s-c_s}\in(0,1)$. Note from our earlier analysis, when $A_{m1}=1$, (\ref{eq:am})$>$(\ref{eq:rm}), which is equivalent to $k+\tau_m(1-h)(p_s-p_m)>(1-\tau_m)(p_m+k^{'})$. By plugging $T_{s1}=0$  and $T_{m1}\in (0,1)$ into $\tau_m$, we have $\tau_m=\frac{\mu \bar{T}_m}{\mu \bar{T}_m+(1-\mu)(1-h)}$, and the inequality can be reduced to $\mu \bar{T}_m((1-h)(p_s-p_m)+k)>(1-\mu)(1-h)(p_m+k^{'}-k)$. To simplify, this condition is satisfied when $\mu>\mu^*_1$ since $\frac{(1-h)(p_m+k^{'}-k)}{h(1-h)(p_s-p_m)+kh+(1-h)(p_m+\hat{k}-k)}=\mu^*_1>\frac{(1-h)(p_m+k^{'}-k)}{\bar{T}_m(1-h)(p_s-p_m)+k\bar{T}_m+(1-h)(p_m+k^{'}-k)}$ as $\bar{T}_m>h$. Similarly, when $A_{s1}\in(0,1)$, (\ref{eq:as})$=$(\ref{eq:rs}) is equivalent to $k=(1-\tau_s)h(p_s-p_m)$. By plugging $T_{s1}=0$  and $T_{m1}\in (0,1)$ into $\tau_s$, we have $\tau_s=\frac{(1-\mu)h}{(1-\mu) h+\mu (1-\bar{T}_m)}$, and the equation can be reduced to $k((1-\mu) h+\mu (1-\bar{T}_m))=\mu (1-\bar{T}_m)h(p_s-p_m)$. Solving this equation, we have $\bar{T}_m=1-\frac{(1-\mu)hk}{\mu (h(p_s-p_m)-k)}$ and $T_{m1}=1-\frac{(1-\mu)hk}{\mu(1-h)[h(p_s-p_m)-k]}$. To ensure $T_{m1}\in(0,1)$, we have $\mu>\mu^*_2$. In sum, we have shown that when $\mu>\mu^*_2$, $(T_{m1}=1-\frac{(1-\mu)hk}{\mu(1-h)(h(p_s-p_m)-k)},T_{s1}=0,A_{m1}=1,A_{s1}=\frac{p_m-c_m}{p_s-c_s})$ constitutes an equilibrium.  

Last, we identify the equilibrium condition when $T_{s1}\in (0,1)$  and $T_{m1}=0$. In this equilibrium, $(p_s-c_s+p_m-c_m)A_{m1}=(p_s-c_s)A_{s1}$ and $(p_s-c_s)A_{s1}>(p_m-c_m)A_{m1}$, and we can deduce that $A_{m1}=\frac{p_s-c_s}{p_s-c_s+p_m-c_m}\in(0,1)$ and $A_{s1}=1$. Note from our earlier analysis, when $A_{m1}\in(0,1)$, (\ref{eq:am})$=$(\ref{eq:rm}) is equivalent to $k+\tau_m(1-h)(p_s-p_m)=(1-\tau_m)(p_m+k^{'})$. By plugging $T_{s1}\in (0,1)$  and $T_{m1}=0$ into $\tau_m$, we have $\tau_m=\frac{\mu h}{\mu h+(1-\mu)(1-\bar{T}_s)}$, and the equality can be reduced to $\mu h((1-h)(p_s-p_m)+k)=(1-\mu)(1-\bar{T}_s)(p_m+k^{'}-k)$. To solve this equation, we have $\bar{T_s}=1-\frac{\mu h((1-h)(p_s-p_m)+k)}{(1-\mu)(p_m+k^{'}-k)}$ and $T_{s1}=1-\frac{\mu h((1-h)(p_s-p_m)+k)}{(1-\mu)(1-h)(p_m+k^{'}-k)}$. To ensure $T_{s1}\in(0,1)$, we have $\mu<\frac{(1-h)(p_m+k^{'}-k)}{h(1-h)(p_s-p_m)+kh+(1-h)(p_m-k+k^{'})}=\mu^*_1$. Similarly, when $A_{s1}=1$, (\ref{eq:as})$>$(\ref{eq:rs}) is equivalent to $k>(1-\tau_s)h(p_s-p_m)$. By plugging $T_{s1}\in (0,1)$  and $T_{m1}=0$ into $\tau_s$, we have $\tau_s=\frac{(1-\mu)\bar{T_s}}{(1-\mu) \bar{T}_s+\mu (1-h)}$, and the inequality can be reduced to $(1-\mu)\bar{T_s}k>\mu (1-h)(h(p_s-p_m)-k)$. To solve this inequality, we find the condition can be satisfied when $\mu\leq\mu^*_2$, since $\frac{kh}{(1-h)h(p_s-p_m)+kh-(1-h)k}<\frac{k\bar{T}_s}{(1-h)h(p_s-p_m)+k\bar{T}_s-(1-h)k}$ and $\bar{T}_s>h$. In sum, we have shown that when $\mu<\mu^*_1$, $(T_{m1}=0,T_{s1}=1-\frac{\mu h(1-h)(p_s-p_m)+k\mu h}{(1-\mu)(1-h)(p_m+k^{'}-k)},A_{m1}=\frac{p_s-c_s}{p_s-c_s+p_m-c_m},A_{s1}=1)$ constitutes an equilibrium.  Note that if $\mu^*_2<\mu^*_1$, both POFU and PUFO are possible. Then, an opportunistic expert can maximize his expected payoff in the equilibrium by taking the first-move advantage. By comparing the expected payoff in POFU and that in PUFO, we find that $\Pi_{POFU}=(1-\mu) (p_s-c_s)+\mu (p_s-c_s)>\Pi_{PUFO}=(1-\mu) (p_s-c_s+p_m-c_m)+\mu (p_m-c_m)$ if and only if $\mu>\frac{p_m-c_m}{p_s-c_s}$. As a result, when $\mu^*_2<\mu^*_1$, PUFO is the equilibrium if $\mu>\frac{p_m-c_m}{p_s-c_s}$; while POFU is the equilibrium if $\mu<\frac{p_m-c_m}{p_s-c_s}$. \hspace{0.3cm} $\Box$

\section*{Proof of Corollary \ref{cor:Equ_kost}}

% This proof examines a few comparative statics of the results in Proposition 1. Part (i) has been addressed in the proof of Proposition 1.

In part (i), we examine the impact of $k$ on equilibrium outcome. First, we examine the impact of $k$ on the boundary of FOFU. Note that $\mu^*_2=1$ when $h\leq\frac{k}{p_s-p_m}$ and $\mu^*_2=\frac{kh}{(1-h)h(p_s-p_m)+kh-(1-h)k}$ when $h>\frac{k}{p_s-p_m}$. As $k$ increases, the threshold $\frac{k}{p_s-p_m}$ below which $\mu^*_2=1$ also increases; in addition, when $h>\frac{k}{p_s-p_m}$, by taking the first order derivative of $\frac{kh}{(1-h)h(p_s-p_m)+kh-(1-h)k}$ with respect to $k$, we have $\frac{\partial \mu^*_2}{\partial k}=\frac{h^2(1-h)(p_s-p_m)}{((1-h)h(p_s-p_m)+kh-(1-h)k)^2}>0$, indicating that the upper bound of  FOFU is weakly increasing in $k$. Similarly, by taking the first order derivative of $\mu^*_1=\frac{(1-h)(p_m+k^{'}-k)}{h(1-h)(p_s-p_m)+kh+(1-h)(p_m-k+k^{'})}$ with respect to $k$, we have $\frac{\partial \mu^*_1}{\partial k}=-\frac{h(1-h)((p_s-p_m)(1-h)+(p_m+k^{'}))}{((1-h)h(p_s-p_m)+kh-(1-h)k)^2}<0$, indicating that the lower bound of FOFU is decreasing in $k$. Then, we examine the impact of $k$ on the mixed strategy adopted by the opportunistic expert in POFU and FOPU. In POFU, $(T_{m1}=1-\frac{(1-\mu)hk}{\mu(1-h)(h(p_s-p_m)-k)},T_{s1}=0,A_{m1}=1,A_{s1}=\frac{p_m-c_m}{p_s-c_s})$, the opportunistic expert recommends excessively to a consumer with a minor problem with a probability $T_{m1}=1-\frac{(1-\mu)hk}{\mu(1-h)(h(p_s-p_m)-k)}$. By taking the first-order derivative of $T_{m1}$ with respect to $k$, we have $\frac{\partial T_{m1}}{\partial k}=-\frac{(1-\mu)\mu h^2(1-h)(p_s-p_m)}{[\mu(1-h)(h(p_s-p_m)-k)]^2}<0$. Similarly, in FOPU, $(T_{m1}=0,T_{s1}=1-\frac{\mu h(1-h)(p_s-p_m)+k\mu h}{(1-\mu)(1-h)(p_m+k^{'}-k)},A_{m1}=\frac{p_s-c_s}{p_s-c_s+p_m-c_m},A_{s1}=1)$, the opportunistic expert recommends to a consumer with a serious problem truthfully with a probability $T_{s1}=1-\frac{\mu h(1-h)(p_s-p_m)+k\mu h}{(1-\mu)(1-h)(p_m+k^{'}-k)}$. By taking the first-order derivative of $T_{s1}$ with respect to $k$, we have $\frac{\partial T_{s1}}{\partial k}=-\frac{\mu h(1-\mu)(1-h)(p_m+k^{'}-k)+(1-\mu)(1-h)(\mu h(1-h)(p_s-p_m)+k\mu h)}{[(1-\mu)(1-h)(p_m+k^{'}-k)]^2}<0$. In sum, we have shown that there will be more undertreatment and overtreatment when $k$ increases.

In part (ii), we examine the impact of $p_s$ on equilibrium outcome. First, we examine the impact of $p_s$ on the boundary of FOFU. When $h\leq\frac{k}{p_s-p_m}$, $\mu^*_2=1$. As $p_s$ increases, the threshold below which $\mu^*_2=1$ decreases. When $h>\frac{k}{p_s-p_m}$, by taking the first order derivative of $\frac{kh}{(1-h)h(p_s-p_m)+kh-(1-h)k}$ with respect to $p_s$, we have $\frac{\partial \mu^*_2}{\partial p_s}=\frac{-h^2(1-h)k}{((1-h)h(p_s-p_m)+kh-(1-h)k)^2}<0$, indicating that the upper bound of FOFU weakly decreases in $p_s$. Similarly, by taking the first order derivative of $\mu^*_1=\frac{(1-h)(p_m+k^{'}-k)}{h(1-h)(p_s-p_m)+kh+(1-h)(p_m-k+k^{'})}$ with respect to $k$, we have $\frac{\partial \mu^*_1}{\partial p_s}=\frac{-h(1-h)^2(p_m+k^{'}-k)}{(h(1-h)(p_s-p_m)+kh+(1-h)(p_m-k+k^{'}))^2}<0$, and the lower bound of FOFU decreases as $p_s$ increases. In the mixed strategy equilibria POFU and FOPU, $T_{m1}$ is increasing in $p_s$ as $\frac{\partial T_{m1}}{\partial p_s}=\frac{(1-\mu)\mu h^2(1-h)k }{[\mu(1-h)(h(p_s-p_m)-k)]^2}$\\$>0$, while $T_{s1}$ is decreasing in $p_s$ as $\frac{\partial T_{s1}}{\partial p_s}=-\frac{\mu(1-\mu)h(1-h)^2(p_m+k^{'}-k) }{((1-\mu)(1-h)(p_m+k^{'}-k))^2}<0$. In sum, we have shown that there will be more undertreatment and less overtreatment when $p_s$ increases.
\hspace{0.3cm} $\Box$

\par\noindent
\section*{Proof of Proposition \ref{prop:profit_ethics}}

In this proof, we examine the impact of $h$ on the expert's profits. In FOFU, an opportunistic expert recommends a minor (serious) treatment to a consumer with a serious (minor) problem and the consumer accepts the recommendation with probability one $(T_{m1}=0,T_{s1}=0,A_{m1}=1,A_{s1}=1)$. In this equilibrium, when the consumer's problem is minor (with probability $\mu$), the opportunistic expert recommends a serious treatment and the consumer always accepts it; the opportunistic expert obtains profits $\mu (p_s-c_s)$ from overtreatment. Similarly, when the consumer's problem is serious(with probability $1-\mu$), the expert recommends inadequately, and the consumer accepts the first minor treatment recommendation and returns to the expert when the minor treatment is insufficient to resolve her problem; in this case, the opportunistic expert gets profits from two treatments, and obtain profits $(1-\mu) (p_s-c_s+p_m-c_m)$. In sum, in FOFU $(T_{m1}=0,T_{s1}=0,A_{m1}=1,A_{s1}=1)$, the opportunistic expert gets expected profits $\Pi_{FOFU}=(1-\mu) (p_s-c_s+p_m-c_m)+\mu (p_s-c_s)$.

In POFU, an opportunistic expert recommends inadequately with probability one and recommends excessively with a probability lower than one as a mixed strategy; the consumer always accepts a minor treatment recommendation, and accepts a serious treatment with probability lower than one $(T_{m1}=1-\frac{(1-\mu)hk}{\mu(1-h)(h(p_s-p_m)-k)},T_{s1}=0,A_{m1}=1,A_{s1}=\frac{p_m-c_m}{p_s-c_s})$. When the consumer's problem is serious (with probability $(1-\mu)$), the opportunistic expert recommends inadequately with probability one, and a consumer always accepts a minor treatment; in this case, the opportunistic expert obtains a payoff $(1-\mu) (p_s-c_s+p_m-c_m)$. When the consumer's problem is minor, the opportunistic expert recommends excessively as a mixed strategy, and a consumer accepts a serious treatment with probability $\frac{p_m-c_m}{p_s-c_s}$. Note that an opportunistic expert is indifferent between truthful recommending and excessive recommending in a mixed-strategy equilibrium; the opportunistic expert obtains a payoff $\mu(p_m-c_m)$ if the consumer's condition is minor. In sum, in POFU, the opportunistic expert gets expected payoff $\Pi_{POFU}=(1-\mu) (p_s-c_s+p_m-c_m)+\mu (p_m-c_m)$.

In FOPU, an opportunistic expert recommends excessively with probability one and recommends inadequately with a probability lower than one as a mixed strategy; the consumer always accepts a serious treatment recommendation, and accepts a minor treatment with probability lower than one $(T_{m1}=0,T_{s1}=1-\frac{\mu h(1-h)(p_s-p_m)+k\mu h}{(1-\mu)(1-h)(p_m+k^{'}-k)},A_{m1}=\frac{p_s-c_s}{p_s-c_s+p_m-c_m},A_{s1}=1)$. When the consumer's condition is minor (with probability $\mu$), the opportunistic expert always recommends excessively, and a consumer always accepts a serious treatment, in which case the opportunistic expert obtains a payoff $\mu (p_s-c_s)$. When the consumer's condition is serious, the opportunistic expert recommends inadequately as a mixed strategy.
Note that an opportunistic expert is indifferent between recommending truthfully and inadequately in this mixed-strategy equilibrium; the opportunistic expert obtains a payoff $(1-\mu)(p_s-c_s)$. In sum, in FOPU, the opportunistic expert gets expected payoff $\Pi_{FOPU}=(1-\mu) (p_s-c_s)+\mu (p_s-c_s)=(p_s-c_s)$.

Next, we examine the impact of $h$ on the opportunistic expert's profit. By comparing the opportunistic expert's expected payoff in FOFU, POFU and FOPU, it is straightforward to see that $\Pi_{FOFU}>\Pi_{FOPU}$ and $\Pi_{FOFU}>\Pi_{POFU}$. Note that the upper bound of FOFU is $\mu^*_2$, and $\mu^*_2=1$ if $h\leq\frac{k}{p_s-p_m}$; while $\mu^*_2=\frac{kh}{(1-h)h(p_s-p_m)+kh-(1-h)k}$ if $h>\frac{k}{p_s-p_m}$. When $h>\frac{k}{p_s-p_m}$, by taking the first order derivative of $\mu^*_2$ with respect to $h$, we have $\frac{\partial \mu^*_2}{\partial h}=\frac{k h^2(p_s-p_m)-k^2}{[(1-h)h(p_s-p_m)+kh-(1-h)k]^2}$, and $\mu^*_2$ is minimized when $\frac{\partial \mu^*_2}{\partial h}=0$ and $h^*=\frac{\sqrt{k}}{\sqrt{p_s-p_m}}$.  Plugging $h^*=\frac{\sqrt{k}}{\sqrt{p_s-p_m}}$ into $\mu^*_2$, we get the minimal value of $\mu^*_2$ as $\mu^*_3=\frac{k}{p_s-p_m-2\sqrt{k(p_s-p_m)}+2k}$. On the other hand, the lower bound of FOFU is $\mu^*_1=\frac{(1-h)(p_m+k^{'}-k)}{h(1-h)(p_s-p_m)+kh+(1-h)(p_m-k+k^{'})}$, and $\mu^*_1$ is decreasing in $h$ since $\frac{\partial \mu^*_1}{\partial h}=-\frac{(p_m+k^{'}-k)((p_s-p_m)(1-h)^2+k)}{(h(1-h)(p_s-p_m)+kh+(1-h)(p_m-k+k^{'}))^2}<0$. Note that $\mu^*_1|_{h=0}=1$ and  $\mu^*_1|_{h=1}=0$, $\mu^*_1$ exhausts all the parameter spaces when $h\in [0,1]$. When $\mu<\mu^*_3$, as $h$ increases, the equilibrium moves from FOPU to FOFU, the opportunistic expert's profit increases from  $\Pi_{FOPU}$ to $\Pi_{FOFU}$, and the opportunistic expert's profit weakly increases as $h$ increases. When $\mu>\mu^*_3$, as $h$ increases, the equilibrium moves from FOPU to FOFU, and then moves from FOFU to POFU and lastly moves back to  FOFU, and the expert's profits first increase, then decrease and last increase again as $h$ increases. \hspace{0.3cm} $\Box$

\par\noindent
\section*{Proof of Proposition \ref{prop:profit_probminor}}

In this part, we examine the impact of $\mu$ on the opportunistic expert's profit. Note from our analysis that when $h\leq\frac{k}{p_s-p_m}$, $\mu^*_2=1$ and FUFO constitutes an equilibrium when $\mu\geq \mu^*_1$, while FUPO is the equilibrium when $\mu<\mu^*_1$. When $\mu<\mu^*_1$ and FUPO is the equilibrium, the opportunistic expert gets expected payoff $\Pi_{FUPO}=(p_s-c_s)$, and $\Pi_{FUPO}$ is independent of $\mu$; when $\mu\geq \mu^*_1$, FUFO is the equilibrium, and the opportunistic expert obtains an expected payoff $\Pi_{FUFO}=(1-\mu) (p_s-c_s+p_m-c_m)+\mu (p_s-c_s)$, and $\Pi_{FUFO}$ is decreasing in $\mu$. As a result, when $h<\frac{k}{p_s-p_m}$, the opportunistic expert's profit first weakly increases and then decreases in $\mu$. On the other hand, when $h>\frac{k}{p_s-p_m}$, as $\mu$ increases, the equilibrium moves from FUPO to FUFO, and then moves from FUFO to PUFO when $\mu\geq\mu^*_2$. As alluded to earlier, the opportunistic expert's profits first weakly increase and then decrease as $\mu$ increases when equilibrium shifts from FUPO to FUFO. When $\mu\geq\mu^*_2$, the opportunistic expert gets expected payoff $\Pi_{PUFO}=(1-\mu) (p_s-c_s+p_m-c_m)+\mu (p_m-c_m)$, and $\Pi_{PUFO}$ is decreasing in $\mu$. Also note that when the equilibrium moves from FUFO to PUFO, there is a jump down in the opportunistic expert's payoff since $\Pi_{FUFO}>\Pi_{PUFO}$. As a result, when $h>\frac{k}{p_s-p_m}$, the expert's payoff also first weakly increases and then decreases in $\mu$. In sum, we have shown that the opportunistic expert's profit first weakly increases and then decreases in $\mu$. \hspace{0.3cm} $\Box$

\par\noindent
\section*{Proof of Proposition \ref{prop:cw_ethic}}

In this part, we examine consumer welfare and the impact of $h$ on consumer welfare. Similar to the proof in Proposition $2$, we first define and calculate consumer welfare in each equilibrium. 

We define consumer welfare as the expected payoff of the consumer in each equilibrium. In the pure strategy equilibrium FOFU, an opportunistic expert recommends excessively and inadequately with probability one, while a consumer accepts any recommendation on the first visit to an expert $(T_{m1}=0,T_{s1}=0,A_{m1}=1,A_{s1}=1)$. When the consumer is with a minor problem and encounters an honest expert (with probability $\mu h$), the consumer is recommended to a minor treatment and the consumer accepts it with total payoff $-p_m-k$. When the consumer is with a minor problem and encounters an opportunistic expert (with probability $\mu (1-h)$), the consumer is recommended to a serious treatment and the consumer accepts it with a total payoff $-p_s-k$. On the other hand, when the consumer is with a serious problem and encounters an honest expert (with probability $(1-\mu) h$), the consumer is recommended to a serious treatment and the consumer accepts it with a total payoff $-p_s-k$. When the consumer is with a serious problem and encounters an opportunistic expert (with probability $(1-\mu) (1-h)$), the consumer is recommended to a minor treatment and the consumer accepts it with a total payoff $-p_m-p_s-k^{'}-k$. In sum, consumer welfare in FOFU is $CW_{FOFU}=-\mu h(p_m+k)-\mu (1-h)(p_s+k)-(1-\mu)h(p_s+k)-(1-\mu)(1-h)(p_m+p_s+k+k^{'})$.

 In the mixed strategy equilibrium POFU, an opportunistic expert recommends inadequately with probability one and recommends excessively with a probability lower than one as a mixed strategy; the consumer always accepts a minor treatment recommendation and accepts a serious treatment with a probability lower than one $(T_{m1}=1-\frac{(1-\mu)hk}{\mu(1-h)(h(p_s-p_m)-k)},T_{s1}=0,A_{m1}=1,A_{s1}=\frac{p_m-c_m}{p_s-c_s})$. When the recommendation is a minor treatment, the recommendation can be truthful with probability $\mu (h+(1-h)T_{m1})$ if it is from either an honest expert or an opportunistic expert who recommends truthfully; by accepting the minor recommendation, the consumer gets a payoff $-p_m-k$. Otherwise, a minor treatment recommendation can be an inadequate recommendation from an opportunistic expert(with probability $(1-\mu) (1-h)$); by accepting the minor (inadequate) recommendation, the consumer needs to return to the previous expert when the first treatment is insufficient and gets a payoff $-p_s-p_m-k-k^{'}$. When the recommendation is a serious treatment, the recommendation can be a truthful one from an honest expert (with probability $(1-\mu)h$) or an excessive one from an opportunistic expert (with probability $\mu(1-h)(1-T_{m1})$). By the payoff-invariant nature of the mixed strategy equilibrium, the consumer is indifferent between accepting and rejecting a serious recommendation, and the consumer gets the same expected payoff $-p_s-k$ when the first recommended treatment is serious. In sum, consumer welfare in POFU is $CW_{POFU}=-\mu \bar{T}_m (p_m+k)-\mu (1- \bar{T}_m)(p_s+k)-(1-\mu) h (p_s+k)-(1-\mu)(1-h)(p_m+p_s+k+k^{'})$, in which $\bar{T}_m=h+(1-h)T_{m1}=1-\frac{(1-\mu)hk}{\mu(h(p_s-p_m)-k)}$.

 In the mixed strategy equilibrium FOPU, an opportunistic expert recommends excessively with probability one and recommends inadequately with a probability lower than one as a mixed strategy; the consumer always accepts a serious treatment recommendation, and accepts a minor treatment with a probability lower than one $(T_{m1}=0,T_{s1}=1-\frac{\mu h(1-h)(p_s-p_m)+k\mu h}{(1-\mu)(1-h)(p_m+k^{'}-k)},A_{m1}=\frac{p_s-c_s}{p_s-c_s+p_m-c_m},A_{s1}=1)$. When the recommendation is a serious treatment, the recommendation can be truthful with probability $\mu (h+(1-h)T_{s1})$ if it is from either an honest expert or an opportunistic expert who recommends truthfully; by accepting the serious recommendation, the consumer gets a payoff $-p_s-k$. Otherwise, the serious treatment recommendation can also be an excessive one from an opportunistic expert (with probability $\mu (1-h)$); by accepting it, the consumer obtains a payoff $-p_s-k$. When the recommendation is a minor treatment, the recommendation can be truthful (with probability $\mu h$) if it is from an honest expert, or it can be an inadequate one (with probability $(1-\mu)(1-h)(1-T_{s1})$from an opportunistic expert. By the payoff-invariant nature of the mixed strategy equilibrium, a consumer gets the same expected payoff by accepting and rejecting a recommendation when the recommended treatment is minor, resulting in a payoff $-\mu h(p_m+k)-(1-\mu)(1-h)(1-T_{s1})(p_m+p_s+k+k^{'})$. In sum,  consumer welfare in  FOPU is $CW_{FOPU}=-\mu(1-h)(p_s+k)-(1-\mu)\bar{T_s}(p_s+k)-\mu h(p_m+k)-(1-\mu)(1-\bar{T_s})(p_m+p_s+k+k^{'})$, in which $\bar{T_s}=(1-h)T_{s1}+h=1-\frac{\mu h(1-h)(p_s-p_m)+\mu h k}{(1-\mu)(p_m+k^{'}-k)}$. 
 
 Last, we examine the impact of  $h$ on consumer welfare. Similar to the proof of Proposition $2$, when $\mu\leq \mu^*_3=\frac{k}{p_s-p_m+2k-2\sqrt{k(p_s-p_m)}}$, equilibrium moves from FOPU to FOFU as $h$ increases. By comparing $CW_{FOPU}$ with $CW_{FOFU}$, we find that $CW_{FOPU}>CW_{FOFU}$, since $CW_{FOPU}-CW_{FOFU}=(1-\mu) (\bar{T}_s-h)(p_m+k^{'})>0$. When  $\mu\leq \mu^*_3$ and $h\leq \frac{\mu (p_s-p_m+k)+(1-\mu)(p_m+k^{'}-k)}{2\mu(p_s-p_m)}-\frac{\sqrt{[\mu(p_s-p_m+k)+(1-\mu)(p_m+k^{'}-k)]^2-4\mu(1-\mu)(p_s-p_m)(p_m+k^{'}-k)}}{2\mu(p_s-p_m)}=h^*_1$, $\mu\leq \mu^*_1$ and the equilibrium is FOPU. By taking the first-order derivative of $CW_{FOPU}$ with respect to $h$, we have $\frac{\partial CW_{FOPU}}{\partial h}=\mu (p_s-p_m)+(1-\mu) \frac{\partial \bar{T}_s}{\partial h}(p_m+k^{'})$. By plugging $\bar{T}_s=1-\frac{\mu h(1-h)(p_s-p_m)+\mu h k}{(1-\mu)(p_m+k^{'}-k)}$ into the first order derivative, we have $\frac{\partial \bar{T}_s}{\partial h}=-\frac{\mu((1-2h)(p_s-p_m)+k)}{(1-\mu)(p_m+k^{'}-k)}$ and $\frac{\partial CW_{FOPU}}{\partial h}=\mu\frac{2h(p_s-p_m)(p_m+k^{'})-k(p_s+k^{'})}{p_m+k^{'}-k}$, and $\frac{\partial CW_{FOPU}}{\partial h}>0$ if and only if $h\geq h^*_2=\frac{(p_s+k^{'})k}{2(p_s-p_m)(p_m+k^{'})}$. When $h\geq h^*_1$, the equilibrium is FOFU, and the consumer surplus is $CW_{FOFU}=-\mu h(p_m+k)-\mu (1-h)(p_s+k)-(1-\mu)h(p_s+k)-(1-\mu)(p_m+p_s+k+k^{'})$. By taking the first order derivative of $CW_{FOFU}$ with respect to $h$, we have $\frac{\partial CW_{FOFU}}{\partial h}=\mu(p_s-p_m)+(1-\mu)(p_m+k^{'})>0$. As a result, when $\mu\leq \mu^*_3$, consumer surplus decreases in $h$ when $h\leq min\{h^*_1,h^*_2\}$; otherwise the consumer surplus increases in $h$, except at the point $h=h^*_1$, there will be a decrease of jump in consumer surplus when the equilibrium moves from FOPU to FOFU. When $\mu>\mu^*_3$, as $h$ increases, the equilibrium changes from FOPU to FOFU, FOFU to POFU, and then POFU to FOFU. By comparing the consumer surplus in FOFU with that in POFU, we find that $CW_{FUFO}<CW_{POFU}$, since $CW_{FOFU}-CW_{POFU}=-(1-\mu)(\bar{T}_m-h)(p_s-p_m)<0$. By taking the first-order derivative of $CW_{POFU}$ with respect to $h$, we have $\frac{\partial CW_{POFU}}{\partial h}=(1-\mu)(p_m+k^{'})+\mu \frac{\partial \bar{T}_m}{\partial h}(p_s-p_m)$. By plugging $\bar{T}_m=1-\frac{(1-\mu)h k}{\mu (h(p_s-p_m)-k)}$, we have $\frac{\partial \bar{T}_m}{\partial h}=\frac{(1-\mu) k^2}{\mu[h(p_s-p_m)-k]^2}>0$ and $\frac{\partial CW_{POFU}}{\partial h}>0$. In sum, we find that when $\mu\geq \mu^*_3$, consumer surplus is decreasing in $h$ when $h\leq\min\{h^*_1,h^*_2\}$; when  $h=h^*_1$ consumer surplus decreases as a jump when equilibrium switches from FOPU to FOFU; in other cases, consumer welfare is increasing in $h$. \hspace{0.3cm} $\Box$

\par\noindent
\section*{Proof of Proposition \ref{prop:mu_cw}}

 In this proof, we examine the impact of $\mu$ on consumer welfare. When $h\leq\frac{k}{p_s-p_m}$, the equilibrium moves from FOPU to FOFU as $\mu$ increases. In FOPU,  the consumer surplus is $CW_{FOPU}=-\mu(1-h)(p_s+k)-(1-\mu)\bar{T}_s(p_s+k)-\mu h(p_m+k)-(1-\mu)(1-\bar{T_s})(p_m+p_s+k+k^{'})$. By taking the first-order derivative of $CW_{FOPU}$ with respect to $\mu$, we have $\frac{\partial CW_{FOPU}}{\partial \mu}=-(1-h)(p_s+k)+\bar{T}_s(p_s+k)-(1-\mu)\frac{\partial \bar{T}_s}{\partial \mu}(p_s+k)-h(p_m+k)+(1-\bar{T}_s)(p_s+p_m+k+k^{'})+(1-\mu)\frac{\partial \bar{T}_s}{\partial \mu}(p_s+p_m+k+k^{'})$, simplifying, we have $\frac{\partial CW_{FOPU}}{\partial \mu}=(1-\bar{T_s})(p_m+k^{'})+h(p_s-p_m)+(1-\mu)\frac{\partial \bar{T}_s}{\partial \mu}(p_m+k^{'})$. By plugging $\bar{T}_s=1-\frac{\mu h((1-h)(p_s-p_m)+k)}{(1-\mu)(p_m+k^{'}-k)}$ into the partial derivative, we have $\frac{\partial \bar{T}_s}{\partial \mu}=-\frac{h[(1-h)(p_s-p_m)+k]}{(1-\mu)^2(p_m+k^{'}+k)}$, and  $\frac{\partial CW_{FOPU}}{\partial \mu}=\frac{h[h(p_s-p_m)(p_m+k^{'})-k(p_s+k^{'})]}{p_m+k+k^{'}}$, and $\frac{\partial CW_{FOPU}}{\partial \mu}<0$ if and only if $h<\frac{k(p_s+k^{'})}{(p_s-p_m)(p_m+k^{'})}=h^*_3$. Note that FOPU constitutes an equilibrium when $\mu\leq \mu^*_1$, as a result, we have shown that consumer welfare decreases in $\mu$ when $\mu\leq \mu^*_1$ and $h\leq h^*_3$. When $h<\frac{k}{p_s-p_m}$, as $\mu$ increases, the equilibrium switches from FOPU to FOFU when $\mu>\mu^*_1$. By taking the first-order derivative of $CW_{FOFU}$ with respect to $\mu$, we have $\frac{\partial CW_{FOFU}}{\partial \mu}=(1-h)(p_m+k^{'})+h(p_s-p_m)>0$. As a result, when $h\leq \frac{k}{p_s-p_m}$, equilibrium switches from FOPU to FOFU, and consumer welfare decreases in $\mu$ when $h\leq h^*_3$, while consumer welfare decreases in $\mu$ if $h>h^*_3$ except at the point $\mu=\mu^*_1$, consumer surplus incurs a decrease of jump when equilibrium switches from FOPU to FOFU. When $h\geq \frac{k}{p_s-p_m}$, equilibrium first switches from FOPU to FOFU, and them switches from FOFU to POFU. Note that consumer welfare in POFU is $CW_{POFU}=-\mu \bar{T}_m (p_m+k)-\mu (1-\bar{T}_m)(p_s+k)-(1-\mu) h (p_s+k)-(1-\mu)(1-h)(p_m+p_s+k+k^{'})$.  By taking the first-order derivative to $CW_{POFU}$ with respect to $\mu$, we have $\frac{\partial CW_{POFU}}{\partial \mu}=(1-h)(p_m+k^{'})+(\bar{T}_m+\mu\frac{\partial \bar{T}_m}{\partial \mu})(p_s-p_m)$. By plugging  $\bar{T}_m=1-\frac{(1-\mu)h k}{\mu [h(p_s-p_m)-k]}$ into the partial derivative, we have $\frac{\partial \bar{T}_m}{\partial \mu}=\frac{hk}{\mu^2(h(p_s-p_m)-k)}>0$ and $\frac{\partial CW_{POFU}}{\partial \mu}>0$. In sum, combined with our analysis earlier, we find that when $h> h^*_3$, consumer welfare is increasing in $\mu$, except at the point $\mu=\mu^*_1$; otherwise, if $ h\leq h^*_3$, consumer welfare decreases with $\mu$ when $\mu \leq \mu^*_1$ while increases with $\mu$ when $\mu >\mu^*_1$. \hspace{0.3cm} $\Box$

\par\noindent
\section*{Proof of Proposition \ref{prop:ability}}

(i). In this proof, we discuss the case whereby the expert cannot get perfect information from diagnosis. In other words, uncertainty occurs during the diagnosis process. In this proof, we use $\epsilon$ to denote the diagnosis uncertainty, that is, when the diagnosis outcome shows that the consumer is with a minor (serious) problem, with probability $1-\epsilon$ the actual nature of the consumer's problem is indeed minor (serious), while with probability  $\epsilon$ the consumer's problem is serious (minor). 

By adopting the backward induction analysis strategy, we first examine the strategic interaction between consumers and experts when the consumer is on the second visit to an expert. Similar to the analysis in the proof of Lemma 1, facing a serious treatment recommendation, a consumer on the second visit to an expert chooses to accept the recommendation rather than leave the problem unresolved. As a result, an opportunistic expert will recommend serious treatment to a consumer invariantly if the consumer is on the second visit to an expert. On the other hand, an honest expert will make recommendations based on the diagnosis outcome. The difference between this extension and the base model is that, due to the diagnosis uncertainty, a second-visit consumer might reject a minor treatment recommendation, since it can be an undertreatment due to diagnosis error. 

On the first visit to an expert, facing a serious treatment recommendation, the recommendation is truthful when a). the recommendation is a truthful treatment recommendation from an honest expert whose diagnosis outcome is accurate (with probability $(1-\mu)(1-\epsilon)h$); b). the recommendation is a truthful recommendation from an opportunistic expert whose diagnosis outcome is accurate (with probability $(1-\mu)(1-\epsilon)(1-h)T_{s1}$); c). the recommendation is an excessive recommendation from an opportunistic expert whose diagnosis outcome is inaccurate (with probability $(1-\mu)\epsilon(1-h)(1-T_{m1})$). Otherwise, the recommendation can be excessive when d). the recommendation is from an honest expert whose diagnosis outcome is inaccurate (with probability $\mu h \epsilon$); e). the recommendation is an excessive recommendation from an opportunistic expert whose diagnosis outcome is accurate (with probability $\mu (1-h) (1-\epsilon)(1-T_{m1})$; f). the recommendation is a truthful recommendation from an opportunistic expert whose diagnosis outcome is inaccurate $\mu (1-h) \epsilon T_{s1}$. By using the same notation as in the base model, facing a serious treatment recommendation, the posterior belief that the recommendation is truthful is $\tau_s=\frac{(1-\mu)(1-\epsilon)\bar{T}_s+(1-\mu)(1-\bar{T}_m)\epsilon}{(1-\mu)(1-\epsilon)\bar{T}_s+(1-\mu)(1-\bar{T}_m)\epsilon+\mu \bar{T}_s \epsilon+\mu(1-\bar{T}_m)(1-\epsilon)}$, in which $\bar{T}_s=h+(1-h)T_{s1}$ and $\bar{T}_m=h+(1-h)T_{m1}$.  

Similarly, facing a minor recommendation, it is truthful when a). the recommendation is a truthful minor treatment recommendation from an honest expert whose diagnosis outcome is accurate (with probability $\mu(1-\epsilon)h$); b). the recommendation is a truthful recommendation from an opportunistic expert whose diagnosis outcome is accurate (with probability $\mu(1-\epsilon)(1-h)T_{s1}$); c). it is an inadequate recommendation from an opportunistic expert whose diagnosis outcome is inaccurate (with probability $\mu\epsilon(1-h)(1-T_{s1})$). Otherwise, the recommendation can be inadequate when d). the recommendation is from an honest expert whose diagnosis outcome is inaccurate (with probability $(1-\mu) h \epsilon$); e). it is an inadequate recommendation from an opportunistic expert whose diagnosis outcome is accurate (with probability $(1-\mu)(1-h) (1-\epsilon)(1-T_{s1})$; f). the recommendation is a truthful recommendation from an opportunistic expert whose diagnosis outcome is inaccurate $(1-\mu) (1-h) \epsilon T_{m1}$. By using the same notation as in the base model, facing a minor treatment recommendation, the posterior belief that the minor recommendation is truthful is $\tau_m=\frac{\mu (1-\epsilon)\bar{T}_m+\mu \epsilon(1-\bar{T}_s)}{\mu (1-\epsilon)\bar{T}_m+\mu \epsilon(1-\bar{T}_s)+(1-\mu)\epsilon\bar{T}_m+(1-\mu)(1-\epsilon)(1-\bar{T}_s)}$, in which $\bar{T}_s=h+(1-h)T_{s1}$ and $\bar{T}_m=h+(1-h)T_{m1}$.  

In sum, with the presence of a diagnosis error, the consumer might reject a minor treatment recommendation on the second visit since it can be the outcome of the misdiagnosis by an honest expert. On the other hand, the consumer accepts any serious treatment recommendation on the second visit to an expert since a serious treatment resolves the consumer's problem for certain. In the following analysis, we take the possibility that the consumer can reject a minor recommendation on the second visit to an expert into our analysis.

Next, we analyze the strategic interaction between consumers and experts on the consumer's first visit to an expert. By accepting a serious recommendation, the consumer pays the price $p_s$ with the problem resolved for certain. If the consumer rejects the first serious recommendation and visits a new expert, by accepting the serious treatment, the consumer gets utility $-k-p_s$. The second-visit consumer is recommended to a serious treatment when a). the consumer encounters an opportunistic expert on the second visit (with probability $1-h$);  b). the consumer is with a minor problem but encounters an honest expert whose diagnosis outcome is inaccurate (with probability $(1-\tau_s)h\epsilon $); c). the consumer is with a serious problem and encounters an honest expert whose diagnosis outcome is accurate (with probability $\tau_s h (1-\epsilon)$). When the recommendation on the second visit is a minor treatment, by accepting the minor recommendation, the consumer gets her problem resolved with a payoff $-p_m-k$ if the treatment is sufficient (with probability $(1-\tau_s)h(1-\epsilon)$), while the consumer's problem remains unresolved with a payoff $-l_s-k-p_m$ if the minor recommendation is the outcome from an honest expert who diagnoses inaccurately (with probability $\tau_s h \epsilon$). By rejecting the minor treatment recommendation, the consumer's problem remains unresolved, and the consumer obtains payoff $-l_m-k$ or $-l_s-k$ if her problem is minor or serious, respectively. In sum, we summarize the consumer's expected utility by accepting a serious recommendation (\ref{eq:Unsa}), rejecting a serious treatment and accepting a minor treatment on the second visit (\ref{eq:Unsama}) and rejecting a serious treatment and rejecting a minor treatment on the second visit (\ref{eq:Unsamr}) as follows:

\begin{align}
& -p_s \label{eq:Unsa} \\
& -(1-h)(p_s+k)-\tau_s h(1-\epsilon)(p_s+k)-\tau_s h\epsilon(p_m+k+l_s)-(1-\tau_s)h(1-\epsilon)(p_m+k)-(1-\tau_s)h\epsilon(p_s+k) \label{eq:Unsama} \\
& -(1-h)(p_s+k)-\tau_s h(1-\epsilon)(p_s+k)-\tau_s h\epsilon(k+l_s)-(1-\tau_s)h(1-\epsilon)(l_m+k)-(1-\tau_s)h\epsilon(p_s+k) \label{eq:Unsamr}   
\end{align}
 
By comparing the utility when the consumer accepts a serious treatment (\ref{eq:Unsa}) with that in (\ref{eq:Unsamr}), we find  (\ref{eq:Unsa})$>$(\ref{eq:Unsamr}). This result indicates that rejecting a minor treatment after rejecting a serious treatment on the first visit is a dominated strategy. 

% that it is a dominant strategy for a consumer to accept a serious treatment recommendation rather than rejecting a serious treatment recommendation on the first visit if she also rejects a minor treatment on the second visit to an expert. 

On the other hand, when facing a minor treatment recommendation on the first visit to an expert, the consumer decides whether to accept the minor treatment from the expert. Similar to the analysis earlier, the consumer has three strategies: a). accept the minor treatment recommendation; b). reject the minor treatment and search for a second opinion; on the second visit to an expert, the consumer accepts both serious recommendations and minor recommendations; c). reject the minor treatment and search for a second opinion; on the second visit, the consumer accepts a serious treatment recommendation but rejects a minor recommendation.  The consumer expected utility of a),b) and c) are summarized as (\ref{eq:Unma}), (\ref{eq:Unmrma}) and (\ref{eq:Unmrmr}):

\begin{align}
& -\tau_m p_m -(1-\tau_m)(p_m+p_s+k^{'}) \label{eq:Unma} \\
& -(1-h)(p_s+k)-\tau_m h(1-\epsilon)(p_m+k)-\tau_m h\epsilon(p_s+k)-(1-\tau_m)h(1-\epsilon)(p_s+k)-(1-\tau_m)h\epsilon(p_m+l_s+k) \label{eq:Unmrma} \\
& -(1-h)(p_s+k)-\tau_m h(1-\epsilon)(l_m+k)-\tau_m h\epsilon(p_s+k)-(1-\tau_m)h(1-\epsilon)(p_s+k)-(1-\tau_m)h\epsilon(l_s+k) \label{eq:Unmrmr}   
\end{align}

Opportunistic experts make decisions based on expected payoff. When facing a second-visit consumer, opportunistic experts always recommend a serious treatment since a consumer on the second visit always accepts a serious treatment. When facing a first-visit consumer, the opportunistic experts' recommendations depend on the expected payoff. We summarize the incentive-compatibility constraint of the opportunistic experts to recommend excessively and inadequately as follows:

\vspace{-0.5cm}
\begin{align}
& IC(Excessive): (p_s-c_s+p_m-c_m)A_{m1}-(p_s-c_s)A_{s1} \geq 0 \label{eqn:unss}\\
& IC(Inadequate):  (p_s-c_s)A_{s1}-(p_m-c_m)A_{m1} \geq 0. \label{eqn:unsm}
\end{align}
\vspace{-1.0cm}

Next, we show that $T_{s1}=1$ or $T_{m1}=1$ can not constitute an equilibrium when $\epsilon\leq$ min \\$\{\frac{(1-\mu)k}{(1-\mu)k+\mu h(p_s-p_m-k)},\frac{\mu (1-h)(p_s-p_m+k)}{\mu(1-h)(p_s-p_m+k)+(1-\mu)(p_m+k^{'}-k)}\}$ by contradiction. Suppose $T_{m1}=1$ in equilibrium, then $\tau_s=\frac{(1-\mu)(1-\epsilon)}{(1-\mu)(1-\epsilon)+\mu\epsilon}$. By plugging $\tau_s$ into (\ref{eq:Unsa}) and (\ref{eq:Unsama}), we find (\ref{eq:Unsa})$>$(\ref{eq:Unsama}) can be reduced to $((1-\mu)(1-\epsilon)+\mu\epsilon)(1-h)k+(1-\mu)(1-\epsilon)^2hk+(1-\mu)(1-\epsilon)\epsilon h(l_s+k+p_m-p_s)+\mu\epsilon^2 hk>\mu \epsilon(1-\epsilon)h(p_s-p_m-k)$. When  $\epsilon<\frac{(1-\mu)k}{(1-\mu)k+\mu h(p_s-p_m-k)}$, $LHS>(1-\mu)(1-h)(1-\epsilon)k+(1-\mu)(1-\epsilon)^2hk>\mu\epsilon (1-\epsilon)h(p_s-p_m-k)=RHS$, indicating that accepting a serious treatment is a dominant strategy for the consumer when $T_{m1}=1$ such that $A_{s1}=1$. Then the opportunistic expert has incentives to deviate to $T_{m1}=0$. Similarly, when $T_{s1}=1$, $\tau_m=\frac{\mu(1-\epsilon)}{\mu(1-\epsilon)+(1-\mu)\epsilon}$. By plugging $\tau_m$ into (\ref{eq:Unma}), (\ref{eq:Unmrma}) and (\ref{eq:Unmrmr}), we find that (\ref{eq:Unma})$>$(\ref{eq:Unmrma}) is equivalent to $\mu(1-\epsilon)(1-h)(p_s-p_m+k)+\mu(1-\epsilon)h\epsilon(p_s-p_m+k)+\mu(1-\epsilon)^2 hk+(1-\mu)\epsilon^2h(l_s+k-p_s-k^{'})>(1-\mu)\epsilon((1-h)+h(1-\epsilon))(p_m+k^{'}-k)$. When $\epsilon<\frac{\mu (1-h)(p_s-p_m+k)}{\mu(1-h)(p_s-p_m+k)+(1-\mu)(p_m+k^{'}-k)}$, $LHS>\mu(1-\epsilon)(1-h)(p_s-p_m+k)>(1-\mu)\epsilon(p_m+k-k^{'})>RHS$. By the same token, (\ref{eq:Unma})$>$(\ref{eq:Unmrmr}) is equivalent to $\mu(1-\epsilon)(1-h)(p_s-p_m+k)+\mu(1-\epsilon)h\epsilon(p_s-p_m+k)+\mu(1-\epsilon)^2 h(l_m+k-p_m)+(1-\mu)\epsilon^2h(l_s+k-p_s-p_m-k^{'})>(1-\mu)\epsilon((1-h)+h(1-\epsilon))(p_m+k^{'}-k)$. When $\epsilon<\frac{\mu (1-h)(p_s-p_m+k)}{\mu(1-h)(p_s-p_m+k)+(1-\mu)(p_m+k^{'}-k)}$, $LHS>\mu(1-\epsilon)(1-h)(p_s-p_m+k)>(1-\mu)\epsilon(p_m+k-k^{'})>RHS$. In sum, we have shown that $T_{s1}=1$ or $T_{m1}=1$ can not constitute an equilibrium when $\epsilon\leq \{\frac{(1-\mu)k}{(1-\mu)k+\mu h(p_s-p_m-k)},\frac{\mu (1-h)(p_s-p_m+k)}{\mu(1-h)(p_s-p_m+k)+(1-\mu)(p_m+k^{'}-k)}\}$. Next, we examine the equilibrium condition when $T_{s1}=0$ or $T_{s1}\in(0,1)$, $T_{m1}=0$ or $T_{m1}\in(0,1)$.

First, we show that  $T_{s1}\in(0,1)$ and $T_{m1}\in(0,1)$ never constitutes an equilibrium by contradiction. Suppose $T_{s1}\in(0,1)$ and $T_{m1}\in(0,1)$ constitutes an equilibrium, the incentive-compatibility constraints of the opportunistic expert in (\ref{eqn:unss}) and (\ref{eqn:unsm}) are binding such that $(p_s-c_s+p_m-c_m)A_{m1}=(p_s-c_s)A_{s1}$ and $(p_s-c_s)A_{s1}=(p_m-c_m)A_{m1}$. It is straightforward to see that the two conditions cannot hold simultaneously. 

Then, we examine the equilibrium outcome when $T_{s1}=0$ and $T_{m1}=0$. By checking the incentive compatibility constraints of the opportunistic experts in (\ref{eqn:unss})  and (\ref{eqn:unsm}), $T_{s1}=0$ and $T_{m1}=0$ hold only if $A_{s1}=1$ and $A_{m1}=1$. Then, we check the incentive compatibility constraints for the consumers. By plugging $T_{s1}=0$ and $T_{m1}=0$ into $\tau_m$ and $\tau_s$, we have $\tau_m=\frac{\mu (1-\epsilon)h+\mu \epsilon(1-h)}{\mu (1-\epsilon)h+\mu \epsilon(1-h)+(1-\mu)\epsilon h+(1-\mu)(1-\epsilon)(1-h)}$ and $\tau_s=\frac{(1-\mu)(1-\epsilon)h+(1-\mu)(1-h)\epsilon}{(1-\mu)(1-\epsilon)h+(1-\mu)(1-h)\epsilon+\mu h \epsilon+\mu(1-h)(1-\epsilon)}$. By checking the incentive-compatibility constraints for consumers to accept any treatment recommendation on the first visit such that (\ref{eq:Unsa})$>$(\ref{eq:Unsama}), (\ref{eq:Unma})$>$(\ref{eq:Unmrma}) and (\ref{eq:Unma})$>$(\ref{eq:Unmrmr}), we find that $(T_{m1}=0, T_{s1}=0,A_{s1}=1,A_{m1}=1)$ constitutes an equilibrium when $\mu^{\epsilon}_2\geq\mu\geq max\{\mu^{\epsilon}_1,\mu^{\epsilon}_3\}$, in which $\mu^{\epsilon}_1=\frac{[\epsilon h+(1-\epsilon)(1-h)][(1-h\epsilon)(p_m+k^{'}-k)-h\epsilon(l_s+k-p_s-k^{'})]}{[\epsilon h+(1-\epsilon)(1-h)][(1-h\epsilon)(p_m+k^{'}-k)-h\epsilon(l_s+k-p_s-k^{'})]+[(1-\epsilon) h+\epsilon(1-h)][(p_s-p_m+k)(1-h+h\epsilon)+h(1-\epsilon)k]}$, $\mu^{\epsilon}_2=1$ when $h\leq\frac{k}{(p_s-p_m)(1-\epsilon)}$ $\mu^{\epsilon}_2=\frac{[h(1-\epsilon)+\epsilon(1-h)][k+h\epsilon(l_s+p_m-p_s)]}{[h(1-\epsilon)+\epsilon(1-h)][k+h\epsilon(l_s+p_m-p_s)]+[\epsilon h+(1-\epsilon)(1-h)][h(p_s-p_m)(1-\epsilon)-k]}$  when $h>\frac{k}{(p_s-p_m)(1-\epsilon)}$ and $\mu^{\epsilon}_3=\frac{[\epsilon h+(1-\epsilon)(1-h)][(p_m+k^{'})-h\epsilon(l_s-p_s)-k]}{[\epsilon h+(1-\epsilon)(1-h)][(p_m+k^{'})-h\epsilon(l_s-p_s)-k]+[(1-\epsilon) h+\epsilon(1-h)][(p_s-p_m+k)(1-h+h\epsilon)+h(1-\epsilon)(l_m-p_m+k)]}$.  

Next, we examine the equilibrium outcome when $T_{s1}\in(0,1)$ and $T_{m1}=0$. When $T_{s1}\in(0,1)$ and $T_{m1}=0$ constitute an equilibrium, $(p_s-c_s+p_m-c_m)A_{m1}=(p_s-c_s)A_{s1}$ and $(p_s-c_s)A_{s1}>(p_m-c_m)A_{m1}$, and we can deduce that $A_{s1}=1$ and $A_{m1}=\frac{p_s-c_s}{p_s-c_s+p_m-c_m}$. By checking the incentive-compatibility constraints for the consumers, we have (\ref{eq:Unsa})$>$(\ref{eq:Unsama}), (\ref{eq:Unma})$=$(\ref{eq:Unmrma}) and  (\ref{eq:Unma})$>$(\ref{eq:Unmrmr}) or (\ref{eq:Unma})$>$(\ref{eq:Unmrma}) and  (\ref{eq:Unma})$=$(\ref{eq:Unmrmr}). Plugging $\tau_m=\frac{\mu (1-\epsilon)h+\mu \epsilon(1-\bar{T}_s)}{\mu (1-\epsilon)h+\mu \epsilon(1-\bar{T}_s)+(1-\mu)\epsilon h+(1-\mu)(1-\epsilon)(1-\bar{T}_s)}$ and $\tau_s=\frac{(1-\mu)(1-\epsilon)\bar{T}_s+(1-\mu)(1-h)\epsilon}{(1-\mu)(1-\epsilon)\bar{T}_s+(1-\mu)(1-h)\epsilon+\mu  \epsilon\bar{T}_s+\mu(1-h)(1-\epsilon)}$ into the incentive compatibility constraint of the consumers, we find that $(T_{m1}=0,T_{s1}\in(0,1),A_{m1}\in(0,1),A_{s1}=1)$ constitutes an equilibrium when $\mu\leq max\{\mu^{\gamma}_1,\mu^{\gamma}_3\}$.

Then, we examine the equilibrium when $T_{s1}=0$ and $T_{m1}\in(0,1)$. When  $T_{s1}=0$ and $T_{m1}\in(0,1)$ constitute an equilibrium, $(p_s-c_s+p_m-c_m)A_{m1}>(p_s-c_s)A_{s1}$ and $(p_s-c_s)A_{s1}=(p_m-c_m)A_{m1}$, we can deduce $A_{s1}=\frac{p_m-c_m}{p_s-c_s}$ and $A_{m1}=1$. Then, we have (\ref{eq:Unsa})$=$(\ref{eq:Unsama}), (\ref{eq:Unma})$>$(\ref{eqn:unss}) and (\ref{eq:Unma})$>$(\ref{eq:Unmrmr}). By plugging $\tau_m=\frac{\mu (1-\epsilon)\bar{T}_m+\mu \epsilon(1-h)}{\mu (1-\epsilon)\bar{T}_m+\mu \epsilon(1-h)+(1-\mu)\epsilon \bar{T}_m+(1-\mu)(1-\epsilon)(1-h)}$ and $\tau_s=\frac{(1-\mu)(1-\epsilon)h+(1-\mu)(1-\bar{T}_m)\epsilon}{(1-\mu)(1-\epsilon)h+(1-\mu)(1-\bar{T}_m)\epsilon+\mu  \epsilon h+\mu(1-\bar{T}_m)(1-\epsilon)}$ into the incentive compatibility constraints of the consumers, we have $k+\tau_s h\epsilon (l_s+p_m-p_s)=h(1-\epsilon)(1-\tau_s)(p_s-p_m)$ and this condition can be satisfied if $\mu\geq\mu^{\epsilon}_2$. In sum, we have shown that when $\mu\geq\mu^{\epsilon}_2$, $(T_{m1}\in(0,1),T_{s1}=0,A_{m1}=1,A_{s1}\in(0,1))$ constitutes an equilibrium. 

Lastly, we examine how the boundary $\mu^{\epsilon}_1$, $\mu^{\epsilon}_2$ and $\mu^{\epsilon}_3$ changes as $\epsilon$ increases. To keep the tractability of the model, we only focus on the case when $\epsilon=0$. By taking the first order derivative of $\mu^{\epsilon}_2$ with respect to $\epsilon$ when $\epsilon=0$, we have $\frac{\partial \mu^{\epsilon}_2}{\partial \epsilon}|_{\epsilon=0}=\frac{(1-2h)k(h(p_s-p_m)-k)+h^2(1-h)((l_s-p_s+p_m)(h(p_s-p_m)-k)+(p_s-p_m)k)}{[hk+(1-h)(h(p_s-p_m)-k)]^2}$, and $\frac{\partial \mu^{\epsilon}_2}{\partial \epsilon}|_{\epsilon=0}>0$ when $h<\frac{1}{2}$, while $\frac{\partial \mu^{\epsilon}_2}{\partial \epsilon}|_{\epsilon=0}<0$ when $h$ approaches $1$.Similarly, we take the first-order derivative of $\mu^{\epsilon}_2$ with respect to $\epsilon$, we have $\frac{\partial \mu^{\epsilon}_1}{\partial \epsilon}|_{\epsilon=0}=-\frac{(1-2h)(p_m+k^{'}-k)((p_s-p_m)(1-h)+k)}{[(1-h)((1-h)(p_m+k^{'})-k)+h(k+(1-h)(p_s-p_m))]^2}$ \\ $-\frac{(1-h)h^2(l_s-p_s+p_m)(p_s-p_m)(p_m+k^{'}-hk)}{[(1-h)((1-h)(p_m+k^{'})-k)+h(k+(1-h)(p_s-p_m))]^2}$, and $\frac{\partial \mu^{\epsilon}_1}{\partial \epsilon}|_{\epsilon=0}<0$ when $h<\frac{1}{2}$, while $\frac{\partial \mu^{\epsilon}_2}{\partial \epsilon}|_{\epsilon=0}>0$ when $h$ approaches $1$. Lastly, by taking the first order derivative of $\mu^{\epsilon}_3$ with respect to $\epsilon$ when $\epsilon=0$, we have $\frac{\partial \mu^{\epsilon}_3}{\partial \epsilon}|_{\epsilon=0}=-\frac{(1-2h)(p_m+k^{'}-k)((p_s-p_m)(1-h)+k+h(l_m-p_m))+(1-h)h^2((p_s-p_m)(1-h)+h(l_m-p_m)+k-(l_m-p_s)(p_m+k^{'}-k)}{[(1-h)(p_m+k^{'}-k)+(1-h)((p_s-p_m)(1-h)+h(l_m-p_m)+k)]^2}$, and $\frac{\partial \mu^{\epsilon}_3}{\partial \epsilon}|_{\epsilon=0}<0$ when $h<\frac{1}{2}$, while $\frac{\partial \mu^{\epsilon}_3}{\partial \epsilon}|_{\epsilon=0}>0$ when $h$ approaches $1$. \hspace{0.3cm} $\Box$

\par\noindent
\section*{Proof of Proposition \ref{prop:capacity}}

In this section, we examine the case whereby an expert faces capacity constraints. Specifically, we use $\chi$ to denote the probability that an expert faces limited capacity. Under limited capacity, an expert is only capable of a minor treatment but not a serious treatment. In our analysis, we assume the capacity constraint is an exogenous shock that occurs independently across periods. Next, we examine the expert's decision under capacity constraints. 

When facing a consumer with a serious problem under limited capacity, an honest expert who always recommends truthfully chooses to reject treating the consumer due to limited capacity. On the other hand, an opportunistic expert makes decisions based on his payoff, and it is a weakly dominant strategy for the expert to recommend a minor treatment to a consumer with a serious problem under limited capacity since the opportunistic expert gets $(p_m-c_m)$ if the consumer accepts the minor treatment recommendation, while the opportunistic expert gets utility $0$ otherwise. Note that in this case, an opportunistic expert may recommend a minor treatment to a consumer with a serious problem even if the consumer is on the second visit to an expert; there is a possibility that a second-visit consumer may choose to reject a minor treatment recommendation due to the undertreatment driven by the capacity constraints.

Next, we examine the strategic interactions between consumers and experts. Facing a serious treatment recommendation on the first visit to an expert, a consumer has three options: a). accept the recommendation; b). reject the recommendation and search for another expert; when the second expert recommends a minor treatment, the consumer accepts the recommendation; c). reject the recommendation and search for another expert; when the second expert recommends a minor treatment, the consumer rejects the recommendation. In a), the consumer pays the price $p_s$, and a serious treatment resolves her problem. In b) and c), the consumer faces the following possibility when searching for a second opinion: d). The consumer's problem is serious, and the second expert recommends truthfully, which occurs when the second expert does not have limited capacity (with probability $\tau_s(1-\chi)$); e). The consumer's problem is serious, but she meets an honest expert with a limited capacity (with probability $\tau_s \chi h$); f). The consumer's problem is serious, but she meets an opportunistic expert with limited capacity who recommends inadequately (with probability $\tau_s \chi (1-h)$); g). The consumer's problem is minor, but she gets a serious treatment recommendation from an opportunistic expert without limited capacity (with probability $(1-\tau_s)(1-h)(1-\chi)$);  h). The consumer's problem is minor, and she meets an expert who recommends truthfully, which occurs when the second expert is honest or the second expert is an opportunistic one with limited capacity (with probability $(1-\tau_s)(h+(1-h)\chi)$). In b), a consumer accepts the minor treatment recommendations in f) and h), while in c) a consumer rejects the minor treatment recommendation in f) and h). We summarize the consumer's expected utility of a), b) and c) as follows:

\begin{align}
    & -p_s \label{eq:Limsa} \\
    & -\tau_s(1-\chi)(p_s+k) - \tau_s \chi h(l_s+k) 
    - \tau_s \chi (1-h)(l_s+k+p_m) \nonumber \\
    & \quad - (1-\tau_s)(1-h)(1-\chi)(p_s+k) 
    - (1-\tau_s)(h+(1-h)\chi)(p_m+k) \label{eq:Limsrma} \\
    & -\tau_s(1-\chi)(p_s+k) - \tau_s \chi h(l_s+k) - \tau_s \chi (1-h)(l_s+k) 
    \nonumber \\
    & \quad  - (1-\tau_s)(1-h)(1-\chi)(p_s+k)
    - (1-\tau_s)(h+(1-h)\chi)(l_m+k) \label{eq:Limsrmr}
\end{align}

In the expression above, $\tau_s=\frac{(1-\mu)\bar{T}_s}{(1-\mu)\bar{T}_s+u(1-\bar{T}_m)}$ denotes the posterior belief that the serious treatment recommendation is truthful, in which $\bar{T}_s=h+(1-h)T_{s1}$ and $\bar{T}_m=h+(1-h)T_{m1}$. By comparing the payoff in (\ref{eq:Limsa}) and (\ref{eq:Limsrmr}), it is straightforward to see that (\ref{eq:Limsa})$>$(\ref{eq:Limsrmr}), indicating that rejecting a minor treatment on the second visit after rejecting a serious treatment on the first visit is a dominated strategy. 

Similarly, when facing a minor treatment recommendation on the first visit to an expert, a consumer has three options: a). accept the recommendation; b). reject the recommendation and search for another expert; when the second expert recommends a minor treatment, the consumer accepts the recommendation; c). reject the recommendation and search for another expert; when the second expert recommends a minor treatment, the consumer rejects the recommendation. In a), the consumer first pays the price $p_m$. If the minor treatment is insufficient to resolve the consumer's problem, it is a dominant strategy for the consumer to return to the previous expert since every expert suffers from limited capacity with the same probability $\chi$ while the search cost is lower such that $k^{'}<k$. If the consumer with a serious problem encounters an expert with limited capacity, the consumer's problem gets unresolved with a loss $l_s$. In b) and c), the consumer faces the following possibility when searching for a second opinion: d). The consumer's problem is minor and the second expert recommends truthfully, which occurs when the consumer meets an honest expert or an opportunistic expert with limited capacity (with probability $\tau_m(h+(1-h)\chi)$); e). The consumer's problem is minor, but she gets overtreatment when the second expert is opportunistic without limited capacity (with probability $\tau_m(1-h)(1-\chi)$); f). The consumer's problem is serious, but she meets an opportunistic expert who recommends inadequately under limited capacity (with probability $(1-\tau_m)(1-h)\chi$); g). The consumer's problem is serious, but she meets an honest expert who rejects to treat her due to limited capacity (with probability $(1-\tau_m)h \chi$); h). The consumer's problem is serious, and she meets an expert who recommends truthfully, which occurs when the second expert is not constrained by limited capacity (with probability $(1-\tau_m)(1-\chi)$). In b), a consumer accepts the minor treatment recommendations in d) and f); while in c), a consumer rejects the minor treatment recommendations in d) and f). We summarize the consumer's expected utility of a), b), and c) as follows: 

\begin{align}
& -\tau_m p_m-(1-\tau_m)\chi(p_m+k^{'}+ l_s)-(1-\tau_m)(1-\chi)(p_m+k^{'}+p_s) \label{eq:Limma} \\
& -\tau_m(h+(1-h)\chi)(p_m+k)-\tau_m(1-h)(1-\chi)(p_s+k) \nonumber \\
 & \quad -(1-\tau_m)(1-h)\chi(l_s+p_m+k)-(1-\tau_m)h \chi(l_s+k)-(1-\tau_m)(1-\chi)(p_s+k) \label{eq:Limmrma} \\
& -\tau_m(h+(1-h)\chi)(l_m+k)-\tau_m(1-h)(1-\chi)(p_s+k) \nonumber \\
& \quad -(1-\tau_m)(1-h)\chi(l_s+k)-(1-\tau_m)h \chi(l_s+k)-(1-\tau_m)(1-\chi)(p_s+k)\label{eq:Limmrmr}   
\end{align}

In the expression above, $\tau_m$ denotes the posterior belief that the minor treatment recommendation is truthful, and $\tau_m=\frac{\mu h +\mu (1-h)(\chi+(1-\chi)T_{m1})}{\mu h +\mu (1-h)(\chi+(1-\chi)T_{m1})+(1-\mu)(1-h)(\chi+(1-\chi)(1-T_{s1}))}$. 

Next, we examine the incentive compatibility constraints for overtreatment and undertreatment when the opportunistic expert is not constrained by limited capacity in the current period. Facing a consumer with a serious problem, by recommending truthfully, an opportunistic expert gets $(p_s-c_s)$ if the consumer accepts the serious treatment recommendation (with probability $A_{s1}$); by recommending inadequately, conditional on the acceptance of a minor treatment (with probability $A_{m1}$), the expert gets $p_m-c_m$ if the expert faces limited capacity in the next period while the expert gets $p_m-c_m+p_s-c_s$ otherwise. Then, we summarize the incentive compatibility constraints for opportunistic experts who do not face limited capacity in the current period as follows:

\vspace{-0.5cm}
\begin{align}
& IC(Inadequate): A_{m1}((p_m-c_m)+(p_s-c_s)(1-\chi))-(p_s-c_s)A_{s1} \geq 0 \label{eqn:limss}\\
& IC(Excessive):  (p_s-c_s)A_{s1}-(p_m-c_m)A_{m1} \geq 0. \label{eqn:limsm}
\end{align}
\vspace{-1.0cm}

Next, we show that $T_{m1}<1$ by contradiction. Suppose an opportunistic expert always recommends truthfully to the first-visit consumer with a minor problem such that $T_{m1}=1$, then the consumer perceives any serious recommendation to be truthful $\tau_s=1$ and accepts any serious recommendation on the visit ($A_{s1}=1$). Note from (\ref{eqn:limsm}) immediately, an opportunistic expert will deviate from a truthful recommendation since $(p_s-c_s)A_{s1}-(p_m-c_m)A_{m1} > 0$. However, being different from the base model, $T_{s1}=1$ can be possible in an equilibrium. In addition, note from (\ref{eqn:limss}) when $T_{s1}>0$, $A_{m1}((p_m-c_m)+(p_s-c_s)(1-\chi))\leq(p_s-c_s)A_{s1}$ must hold, then we can deduce that $T_{m1}=0$ since $(p_m-c_m)A_{m1}<A_{m1}((p_m-c_m)+(p_s-c_s)(1-\chi))\leq(p_s-c_s)A_{s1}$. In sum, there are four possible strategy pairs for an opportunistic expert in this extension: $(T_{s1}=1, T_{m1}=0)$, $(T_{s1}\in(0,1), T_{m1}=0)$, $(T_{s1}=0, T_{m1}\in(0,1))$ and $(T_{s1}=0, T_{m1}=0)$. 

First, we identify the conditions under which $(T_{s1}=0, T_{m1}=0)$ constitutes an equilibrium. When $T_{s1}=0$ and $T_{m1}=0$, $A_{m1}((p_m-c_m)+(p_s-c_s)(1-\chi))>(p_s-c_s)A_{s1}$ and $(p_s-c_s)A_{s1}>(p_m-c_m)A_{m1}$ hold with strict inequality, indicating  $A_{m1}=1$ and $A_{s1}=1$. Furthermore, to ensure $(p_m-c_m)+(p_s-c_s)(1-\chi)>(p_s-c_s)$, we have $\chi< \chi^*=\frac{p_m-c_m}{p_s-c_s}$. Then, we check the incentive compatibility constraints when $(T_{s1}=0,T_{m1}=0,A_{m1}=1,A_{s1}=1)$. When  $T_{s1}=0$ and $T_{m1}=0$, the posterior belief of a truthful serious recommendation is $\tau_s=\frac{(1-\mu)h}{(1-\mu)h+\mu(1-h)}$, and the posterior belief of a truthful minor recommendation is  $\tau_m=\frac{\mu h +\mu (1-h)\chi}{\mu h +\mu (1-h)\chi+(1-\mu)(1-h)}$. When $A_{s1}=1$, (\ref{eq:Limsa})$>$(\ref{eq:Limsrma}). By plugging $\tau_s=\frac{(1-\mu)h}{(1-\mu)h+\mu(1-h)}$ into the inequality, we find the condition holds if and only if $\mu\leq\mu^{\chi}_2=\frac{hk+h^2\chi(l_s-p_s)+h\chi(1-h)(l_s-p_s+p_m)}{hk+h^2\chi(l_s-p_s)+h\chi(1-h)(l_s-p_s+p_m)-(1-2h)k+(1-h)(\chi+h(1-\chi))(p_s-p_m)}$. When $A_{m1}=1$, (\ref{eq:Limma})$>$(\ref{eq:Limmrma}) and (\ref{eq:Limma})$>$(\ref{eq:Limmrmr}) hold with strict inequality. By plugging $\tau_m=\frac{\mu h +\mu (1-h)\chi}{\mu h +\mu (1-h)\chi+(1-\mu)(1-h)}$ into the conditions, we have $\mu\geq\mu^{\chi}_1=\frac{(1-h)(k^{'}-k)+(1-h)(h+(1-h)(1-\chi))p_m}{(1-h)(k^{'}-k)+(1-h)(h+(1-h)(1-\chi))p_m+(h+(1-h)\chi)(1-h)(1-\chi)(p_s-p_m)+(h+(1-h)\chi)k}$ and $\mu\geq\mu^{\chi}_3=\frac{(p_m+k^{'}-k)(1-h)}{(p_m+k^{'}-k)(1-h)+(h+(1-h)\chi)k+(h+(1-h)\chi)^2(l_m-p_m)+(h+(1-h)\chi)(1-h)(1-\chi)(p_s-p_m)}$.  In sum, we have shown that the pure strategy equilibrium $(T_{s1}=0, T_{m1}=0,A_{s1}=1, A_{m1}=1)$ constitutes an equilibrium when $\chi< \chi^*$ and $max\{\mu^{\chi}_1,\mu^{\chi}_3\}\leq\mu\leq\mu^{\chi}_2$.

Next, we identify the conditions under which $(T_{s1}=1, T_{m1}=0)$ constitutes an equilibrium. When $T_{s1}=1$ and $T_{m1}=0$, $A_{m1}((p_m-c_m)+(p_s-c_s)(1-\chi))<(p_s-c_s)A_{s1}$ and $(p_s-c_s)A_{s1}>(p_m-c_m)A_{m1}$ hold simultaneously, which can be deduced to $A_{m1}((p_m-c_m)+(p_s-c_s)(1-\chi))<(p_s-c_s)A_{s1}$. Note that when $T_{s1}=1$ and $T_{m1}=0$, both $\tau_m$ and $\tau_s$ are deterministic values, the incentive-compatibility constraints of consumers cannot be binding, indicating the strategies of the consumers must be pure strategies. As a result, there are two possible equilibrium outcomes: $(T_{s1}=1,T_{m1}=0,A_{s1}=1,A_{m1}=1)$ and  $(T_{s1}=1,T_{m1}=0,A_{s1}=1,A_{m1}=0)$. 

When $(T_{s1}=1,T_{m1}=0,A_{s1}=1,A_{m1}=1)$ constitutes an equilibrium,  $\tau_m=\frac{\mu h +\mu (1-h)\chi}{\mu h +\mu (1-h)\chi+(1-\mu)(1-h)\chi}$ and $\tau_s=\frac{1-\mu}{1-\mu h}$. When $A_{s1}=1$, (\ref{eq:Limsa})$>$(\ref{eq:Limsrma}), by plugging $\tau_s$ into this inequality, we have $\mu\leq \mu^{\chi}_4=\frac{k+h\chi(l_s-p_s)+(1-h)\chi(l_s-p_s+p_m)}{hk+h\chi(l_s-p_s)+(1-h)\chi(l_s-p_s+p_m)+(1-h)(\chi+h(1-\chi))(p_s-p_m)}$. When  $A_{m1}=1$, (\ref{eq:Limma})$>$(\ref{eq:Limmrma}) and (\ref{eq:Limma})$>$(\ref{eq:Limmrmr}) hold with strict inequality. By plugging $\tau_m$ into the conditions, we have $\mu\geq\mu^{\chi}_5=\frac{(1-h)\chi(k^{'}-k)+(1-h)((1-h)(1-\chi)+h)p_m}{hk+(1-h)\chi k^{'}+(1-h)((1-h)(1-\chi)+h)p_m+(h+(1-h)\chi)(1-h)(1-\chi)(p_s-p_m)}$ and $\mu\geq\mu^{\chi}_6$ and \\ $\mu^{\chi}_6=\frac{(1-h)\chi(p_m+k^{'}-k)}{(1-h)\chi(p_m+k^{'})+hk+(h+(1-h)\chi)(1-h)(1-\chi)(p_s-p_m)+(h+(1-h)\chi)^2(l_m-p_m)}$. In sum, we have shown that $(T_{s1}=1, T_{m1}=0,A_{s1}=1, A_{m1}=1)$ constitutes an equilibrium when $\chi> \chi^*$ and $max\{\mu^{\chi}_5,\mu^{\chi}_6\}\leq\mu\leq\mu^{\chi}_4$.

When $(T_{s1}=1,T_{m1}=0,A_{s1}=1,A_{m1}=0)$ constitutes an equilibrium,  $\tau_m=\frac{\mu h +\mu (1-h)\chi}{\mu h +\mu (1-h)\chi+(1-\mu)(1-h)\chi}$ and $\tau_s=\frac{1-\mu}{1-\mu h}$. Similar to the analysis previously, $A_{s1}=1$ holds if and only if (\ref{eq:Limsa})$>$(\ref{eq:Limsrma}), which can be reduced to $\mu\leq\mu^{\chi}_4$. On the other hand, the first-visit consumer rejects a minor treatment recommendation when (\ref{eq:Limma})$<$(\ref{eq:Limmrma}) or (\ref{eq:Limma})$<$(\ref{eq:Limmrmr}), which can be reduced to $\mu<\mu^{\chi}_5$ or $\mu<\mu^{\chi}_6$. Also note that the incentive-compatibility constraints for the opportunistic experts such $A_{m1}((p_m-c_m)+(p_s-c_s)(1-\chi))<(p_s-c_s)A_{s1}$ and $(p_s-c_s)A_{s1}>(p_m-c_m)A_{m1}$ always hold when $A_{s1}=1$ and $A_{m1}=0$, regardless the value of $\chi$. In sum, we have shown that $(T_{s1}=1,T_{m1}=0,A_{s1}=1,A_{m1}=0)$ constitutes an equilibrium when $\mu\leq max\{\mu^{\chi}_5,\mu^{\chi}_6\}$.

Then we identify the equilibrium conditions for the mixed strategy equilibiria. When $(T_{s1}=0, T_{m1}\in(0,1))$ constitutes an equilibrium,  $A_{m1}((p_m-c_m)+(p_s-c_s)(1-\chi))>(p_s-c_s)A_{s1}$ and $(p_s-c_s)A_{s1}=(p_m-c_m)A_{m1}$, which can be reduced to $A_{m1}=1$ and $A_{s1}=\frac{p_m-c_m}{p_s-c_s}$. In this equilibrium, $\tau_m=\frac{\mu h +\mu (1-h)(\chi+(1-\chi)T_{m1})}{\mu h +\mu (1-h)(\chi+(1-\chi)T_{m1})+(1-\mu)(1-h)}$ and $\tau_s=\frac{(1-\mu)h}{(1-\mu)h+\mu(1-h)(1-T_{m1})}$. When $A_{s1}=\frac{p_m-c_m}{p_s-c_s}$, (\ref{eq:Limsa})$=$(\ref{eq:Limsrma}). By plugging $\tau_s$ into this equation, we find that (\ref{eq:Limsa})$=$(\ref{eq:Limsrma}) has a feasible solution when $T_{m1}>0$ if and only if $\mu\geq \mu^{\chi}_s$. On the other hand, when $A_{m1}=1$, (\ref{eq:Limma})$>$(\ref{eq:Limmrma}) and (\ref{eq:Limma})$>$(\ref{eq:Limmrmr}) hold with strict inequality. By plugging $\tau_m$ into the conditions, we find that $\mu\geq max\{\mu^{\chi}_3,\mu^{\chi}_1\}$. In sum, we have shown that $(T_{s1}=0, T_{m1}\in(0,1),A_{s1}\in(0,1), A_{m1}=1)$ constitutes an equilibrium when $\mu\geq \mu^{\chi}_2$.

Lastly, When $(T_{s1}\in(0,1), T_{m1}=0)$ constitutes an equilibrium,  $A_{m1}((p_m-c_m)+(p_s-c_s)(1-\chi))=(p_s-c_s)A_{s1}$ and $(p_s-c_s)A_{s1}>(p_m-c_m)A_{m1}$, in which $A_{m1}=\frac{(p_s-c_s)}{((p_m-c_m)+(p_s-c_s)(1-\chi))}$ and $A_{s1}=1$. To ensure that the equilibrium has a feasible solution, we have $(p_m-c_m)+(p_s-c_s)(1-\chi)<(p_s-c_s)$, which can be reduced to $\chi\leq\chi^*$. Next, we examine the incentive compatiblility conditions for the consumers when $A_{m1}\in(0,1)$ and $A_{s1}=1$. Note that when $T_{s1}\in(0,1)$ and $T_{m1}=0$, $\tau_m=\frac{\mu h +\mu (1-h)\chi}{\mu h +\mu (1-h)\chi+(1-\mu)(1-h)(\chi+(1-\chi)(1-T_{s1}))}$ and $\tau_s=\frac{(1-\mu)(h+(1-h)T_{s1})}{(1-\mu)(h+(1-h)T_{s1})+\mu(1-h)}$. When $A_{s1}=1$, (\ref{eq:Limsa})$>$(\ref{eq:Limsrma}), which can be ensured by $\mu\leq\mu^{\chi}_2$. On the other hand, when $A_{m1}\in(0,1)$, either (\ref{eq:Limma})$=$(\ref{eq:Limmrma}) or (\ref{eq:Limma})$=$(\ref{eq:Limmrmr}) is binding. By plugging $\tau_m$ into the conditions, we find $\mu\leq max\{\mu^{\chi}_3,\mu^{\chi}_1\}$. In sum, we have shown that $(T_{s1}\in(0,1), T_{m1}=0,A_{s1}=1, A_{m1}\in(0,1))$ constitutes an equilibrium when $\mu\leq \max\{\mu^{\chi}_3,\mu^{\chi}_1\}$ and $\chi\leq\chi^*$.

In summary, our analysis in this proof shows that when $\chi\leq\chi^*$, POFU $(T_{s1}=0, T_{m1}\in(0,1),A_{s1}\in(0,1), A_{m1}=1)$ constitutes a mixed strategy equilibrium when $\mu\geq \mu^{\chi}_2$, FOFU $(T_{s1}=0, T_{m1}=0,A_{s1}=1, A_{m1}=1)$ constitutes a pure strategy equilibrium when $max\{\mu^{\chi}_1,\mu^{\chi}_3\}\leq\mu\leq\mu^{\chi}_2$ while FOPU $(T_{s1}\in(0,1), T_{m1}=0,A_{s1}=1, A_{m1}\in(0,1))$ constitutes a mixed strategy equilibrium when $\mu\leq \max\{\mu^{\chi}_3,\mu^{\chi}_1\}$. When $\chi>\chi^*$, POFU $(T_{s1}=0, T_{m1}\in(0,1),A_{s1}\in(0,1), A_{m1}=1)$ constitutes a mixed strategy equilibrium when $\mu\geq \mu^{\chi}_4$, FONU $(T_{s1}=1, T_{m1}=0,A_{s1}=1, A_{m1}=1)$ constitutes a pure strategy equilibrium when $max\{\mu^{\chi}_6,\mu^{\chi}_5\}\leq\mu\leq\mu^{\chi}_4$ and PONU $(T_{s1}=1,T_{m1}=0,A_{s1}=1,A_{m1}=0)$ constitutes a pure strategy equilibrium when $\mu\leq max\{\mu^{\chi}_5,\mu^{\chi}_6\}$. \hspace{0.3cm} $\Box$

\par\noindent
\section*{Proof of Proposition \ref{prop:diag_history}}

i). In this section, we examine the case whereby a consumer's search history is not known to the experts. Specifically, upon consumer consulting, an expert cannot tell whether the consumer is visiting the first or second expert, and an opportunistic expert adopts the same strategy across periods such that $T_{m1}=T_{m2}=T_m$ and $T_{s1}=T_{s2}=T_s$. Similar to the base model, a second-visit consumer accepts a serious treatment recommendation when $l_m>p_s$; however, the consumer may choose to reject a minor treatment recommendation on the second visit since the minor treatment recommendation can be inadequate from an opportunistic expert.  

Next, we examine the strategic interactions between consumers and experts. Facing a serious treatment recommendation on the first visit to an expert, a consumer has three options: a). accept the recommendation; b). reject the recommendation and search for another expert; on the second visit to an expert, the consumer accepts both minor and serious treatments. c). reject the recommendation and search for another expert; on the second visit to an expert, the consumer accepts serious treatment and rejects a minor treatment. In a), the consumer pays $p_s$ with her problem fully resolved by a serious treatment. If the consumer rejects the recommendation, the consumer faces four possibilities on the second visit to an expert: d). The consumer's problem is serious, and the second expert recommends truthfully, and it occurs with probability $\tau_s(h+(1-h)T_{s})$; e). The consumer's problem is serious, and the second expert recommends inadequately, and it occurs with probability $\tau_s(1-h)(1-T_{s})$; f). The consumer's problem is minor, and the second expert recommends truthfully, and it occurs with probability $(1-\tau_s)(h+(1-h)T_{m})$; g). The consumer's problem is minor, and the second expert recommends excessively, and it occurs with probability $(1-\tau_s)(1-h)(1-T_{m})$.  In b), a consumer accepts the minor treatment in d) and e); while in c), the consumer rejects the minor treatment in d) and e). We summarize the expected utility of a), b) and c) as follows: 

\begin{align}
& -p_s \label{eq:gammasa} \\
& -\tau_s\bar{T}_s(p_s+k)-\tau_s(1-\bar{T}_s)(l_s+p_m+k)-(1-\tau_s)\bar{T}_m(p_m+k)-(1-\tau_s)(1-\bar{T}_m)(p_s+k) \label{eq:gammasrma} \\
& -\tau_s\bar{T}_s(p_s+k)-\tau_s(1-\bar{T}_s)(l_s+k)-(1-\tau_s)\bar{T}_m(l_m+k)-(1-\tau_s)(1-\bar{T}_m)(p_s+k) \label{eq:gammasrmr}   
\end{align}

In the expression above, $\tau_s=\frac{(1-\mu)\bar{T}_s}{(1-\mu)\bar{T}_s+\mu(1-\bar{T}_m)}$, $\bar{T}_s=(1-h)T_{s}+h$ and $\bar{T}_m=(1-h)T_{m}+h$. By comparing the consumer payoff in (\ref{eq:gammasa}) and (\ref{eq:gammasrmr}), we find that (\ref{eq:gammasa})$>$(\ref{eq:gammasrmr}), indicating that rejecting a minor treatment after rejecting a serious treatment is a dominated strategy. 

Similarly, when the first recommendation is a minor treatment, a consumer has three options: a). accept the minor treatment recommendation; b). reject the recommendation and search for another expert; on the second visit to an expert, the consumer accepts both minor and serious treatments; c). reject the recommendation and search for another expert; on the second visit to an expert, the consumer accepts a serious treatment and rejects a minor treatment.  However, there is a possibility in a) that the consumer's problem does not get resolved after receiving a minor treatment on the first visit, in which case, the consumer returns to the previous expert after undertreatment due to the lower cost $k^{'}<k$. Then, we summarize the expected utility of a), b), and c) when the first recommendation is a minor treatment as follows:

\begin{align}
& -\tau_m p_m-(1-\tau_m)(p_m+p_s+k^{'}) \label{eq:gammama} \\
& -\tau_m \bar{T}_m (p_m+k)-\tau_m(1-\bar{T}_m)(p_s+k)-(1-\tau_m)\bar{T}_s(p_s+k)-(1-\tau_m)(1-\bar{T}_s)(p_m+l_s+k) \label{eq:gammamrma} \\
& -\tau_m \bar{T}_m (l_m+k)-\tau_m(1-\bar{T}_m)(p_s+k)-(1-\tau_m)\bar{T}_s(p_s+k)-(1-\tau_m)(1-\bar{T}_s)(l_s+k)\label{eq:gammamrmr}   
\end{align}

In the expression above, $\tau_m=\frac{\mu\bar{T}_m}{\mu\bar{T}_m+(1-\mu)(1-\bar{T}_s)}$, $\bar{T}_m=h+(1-h)T_{m}$ and $\bar{T}_s=h+(1-h)T_{s}$.

Then, we examine the incentive compatibility constraints for opportunistic experts to make a recommendation. Note that a consumer may reject a minor treatment recommendation on the second visit to an expert by denoting the probability that the consumer is on the first visit, conditional on the consumer having a minor (serious) problem as $\gamma_m$ ($\gamma_s$), we spell out the incentive compatibility constraints for the opportunistic experts to recommend excessively and inadequately as follows:

\vspace{-0.5cm}
\begin{align}
& IC(Excessive): \gamma_m A_{s1}(p_s-c_s)+(1-\gamma_m)(p_s-c_s) \nonumber \\
& \quad \geq\gamma_m A_{m1}(p_m-c_m)+(1-\gamma_m)\gamma_{mm}A_{m2}(p_m-c_m)+(1-\gamma_m)(1-\gamma_{mm})(p_m-c_m) \label{eq:gammams} \\
& IC(Inadequate):  \gamma_s A_{m1}(p_s-c_s+p_m-c_m)+(1-\gamma_s)\gamma_{sm}A_{m2}(p_m-c_m)+(1-\gamma_s)(1-\gamma_{sm})(p_m-c_m)  \nonumber \\
& \quad \geq \gamma_s A_{s1}(p_s-c_s)+(1-\gamma_s)(p_s-c_s). \label{eq:gammasm}
\end{align}
\vspace{-1.0cm}

In the expressions above, $\gamma_m=\frac{1}{1+\bar{T}_m(1-A_{m1})+(1-\bar{T}_m)(1-A_{s1})}$ and $\gamma_s=\frac{1}{1+\bar{T}_s(1-A_{s1})+(1-\bar{T}_s)(1-A_{m1})}$. By normalizing the new consumers in each period to measure one, conditional on the consumer's problem being minor, only the consumer who rejects a truthful minor (with $\bar{T}_m(1-A_{m1})$) or the consumer who rejects an excessive serious treatment (with $(1-\bar{T}_m)(1-A_{s1})$) will proceed to visit a second expert. Similarly,  conditional on the consumer's problem being serious, only the consumer who rejects a truthful serious (with $\bar{T}_s(1-A_{s1})$) or the consumer who rejects an inadequate minor treatment (with $(1-\bar{T}_s)(1-A_{m1})$) will proceed to visit a second expert. In (\ref{eq:gammams}), the left-hand side indicates the expected payoff when an opportunistic expert recommends excessively; when the consumer is visiting the first expert (with $\gamma_m$), the consumer accepts the recommendation with probability $A_{s1}$; when visiting the second expert (with $(1-\gamma_m)$), the consumer accepts a serious treatment with probability one. On the right-hand side of (\ref{eq:gammams}), an opportunistic expert realizes a payoff $p_m-c_m$ by recommending truthfully when a). a consumer on the first visit to an expert (with probability $\gamma_m$) accepts a minor treatment (with probability $A_{m1}$); b) a consumer on the second visit to an expert (with probability $1-\gamma_m$) who rejects a minor treatment before (with probability $\gamma_{mm}=\frac{\bar{T}_m(1-A_{m1})}{\bar{T}_m(1-A_{m1})+(1-\bar{T}_m)(1-A_{s1})}$) will accept a minor treatment with probability $A_{m2}$; c). a consumer on the second visit to an expert (with probability $1-\gamma_m$) who rejects a serious treatment before (with probability $1-\gamma_{mm}=\frac{(1-\bar{T}_m)(1-A_{s1})}{\bar{T}_m(1-A_{m1})+(1-\bar{T}_m)(1-A_{s1})}$) will accept a minor treatment with probability one ((\ref{eq:gammasa})$>$(\ref{eq:gammasrmr})). Similarly, in (\ref{eq:gammasm}), $\gamma_{sm}=\frac{(1-\bar{T}_s)(1-A_{m1})}{\bar{T}_s(1-A_{s1})+(1-\bar{T}_s)(1-A_{m1})}$ denotes the probability that the consumer rejects a minor treatment before, conditional on the consumer with a serious problem is visiting the second expert, and a consumer may accept a minor treatment recommendation probabilistic ($A_{m2}$) if the consumer rejects a minor treatment on the first visit to an expert.

Next, we examine the possible equilibrium outcomes in this extension. Similar to the rationale in the base model, it is straightforward to see that $T_{s}<1$ and $T_{m}<1$ at equilibrium. We show this argument by contradiction. Suppose $T_{s}=1$, then any minor recommendation is truthful ($\tau_m=1$), and it is a dominant strategy for a consumer to accept a minor treatment recommendation such that $A_{m1}=A_{m2}=1$. Note from (\ref{eq:gammasm}) that $\gamma_s(p_s-c_s+p_m-c_m)A_{m1}+(1-\gamma_s)(p_m-c_m)A_{m2}>\gamma_s(p_s-c_s)A_{s1}+(1-\gamma_s)(p_s-c_s)$ holds with strict inequality and then the opportunistic experts have incentives to deviate to $T_{s}=0$. By the same token, $T_{m}=1$ never constitutes an equilibrium. The possible equilibrium strategy pairs are $(T_{s}=0, T_{m}=0)$, $(T_{s}=0, T_{m}\in(0,1))$, $(T_{s}\in(0,1), T_{m}=0)$ and $(T_{s}\in(0,1), T_{m}\in(0,1))$.

First, we show that  $T_{s}\in(0,1)$ and $T_{m}\in(0,1)$ cannot be an equilibrium by contradiction.  Suppose $T_{s}\in(0,1)$ and $T_{m}\in(0,1)$ in equilibrium, then the conditions in (\ref{eq:gammasm}) and (\ref{eq:gammasm}) must be binding such that $(1-\gamma_m+\gamma_m A_{s1})(p_s-c_s)=(\gamma_m A_{m1} +(1-\gamma_m)\gamma_{mm}A_{m2}+(1-\gamma_m)(1-\gamma_{mm}))(p_m-c_m)$ and $ \gamma_s(p_s-c_s+p_m-c_m)A_{m1}+(1-\gamma_s)(\gamma_{sm}A_{m2}+(1-\gamma_{sm}))(p_m-c_m)=\gamma_s(p_s-c_s)A_{s1}+(1-\gamma_s)(p_s-c_s)$. By plugging $\gamma_s=\frac{1}{1+\bar{T}_s(1-A_{s1})+(1-\bar{T}_s)(1-A_{m1})}$, $\gamma_m=\frac{1}{1+\bar{T}_m(1-A_{m1})+(1-\bar{T}_m)(1-A_{s1})}$, $\gamma_{sm}=\frac{(1-\bar{T}_s)(1-A_{m1})}{\bar{T}_s(1-A_{s1})+(1-\bar{T}_s)(1-A_{m1})}$ and $\gamma_{mm}=\frac{\bar{T}_m(1-A_{M1})}{\bar{T}_m(1-A_{m1})+(1-\bar{T}_m)(1-A_{s1})}$ into the two conditions, we have $(\bar{T}_m(1-A_{m1})+(1-\bar{T}_m)(1-A_{s1})+A_{s1})(p_s-c_s)=(A_{m1}+(1-\bar{T}_m)(1-A_s)+\bar{T}_m(1-A_{m1})A_{m2})(p_m-c_m)$ and $A_{m1}(p_s-c_s+p_m-c_m)+((1-\bar{T}_s)(1-A_{m1})A_{m2}+\bar{T}_s(1-A_s))(p_m-c_m)=(A_s+\bar{T}_s (1-A_s)+(1-\bar{T}_s)(1-A_{m1}))(p_s-c_s)$, which can be deduced to $\frac{p_m-c_m}{p_s-c_s}=\frac{\bar{T}_m(1-A_{m1})+(1-\bar{T}_m)(1-A_{s1})+A_{s1}}{A_{m1}+(1-\bar{T}_m)(1-A_{s1})+\bar{T}_m(1-A_{m1})A_{m2}}=\frac{(A_{s1}-A_{m1}+\bar{T}_s (1-A_{s1})+(1-\bar{T}_s)(1-A_{m1}))}{(1-\bar{T}_s)(1-A_{m1})A_{m2}+\bar{T}_s(1-A_{s1})+A_{m1}}<1$, and $1\geq A_{m1}>A_{s1}$. Next, we show that when  $A_{s1}<1$, $A_{m1}=1$ by checking the incentive compatibility constraints of the consumers. When $A_{s1}<1$, $(\ref{eq:gammasa})\leq (\ref{eq:gammasrma})$ can be reduced to $(1-\mu)\bar{T}^2_s k+(1-\mu)\bar{T}_s(1-\bar{T}_s) (l_s+k+p_m-p_s)+\mu (1-\bar{T}_m)^2k\leq \mu(1-\bar{T}_m)\bar{T}_m(p_s-p_m-k)$, and we have $\mu(1-\bar{T}_m)\bar{T}_m(p_s-p_m-k)>(1-\mu)\bar{T}_s(1-\bar{T}_s) (l_s+k+p_m-p_s)>(1-\mu)\bar{T}_s(1-\bar{T}_s)p_m$. Then, it is straightforward to see that (\ref{eq:gammama})$>$(\ref{eq:gammamrma}), since (\ref{eq:gammama})$>$(\ref{eq:gammamrma}) is equivalent to $\mu\bar{T}^2_m k+\mu\bar{T}_m(1-\bar{T}_m)(p_s-p_m+k)+(1-\mu)(1-\bar{T}^2_s)(l_s+k-p_s-k^{'})>(1-\mu)(1-\bar{T}_s)\bar{T}_s(p_m-k+k^{'})$, and it always holds with strict inequality since $LHD=(1-\mu)(1-\bar{T}_s)\bar{T}_s(p_m-k+k^{'})<(1-\mu)(1-\bar{T}_s)\bar{T}_s p_m<\mu(1-\bar{T}_m)\bar{T}_m(p_s-p_m-k)<\mu(1-\bar{T}_m)\bar{T}_m(p_s-p_m+k)<RHS$, indicating that $A_{m1}=1$. However, when $A_{m1}=1$, the incentive compatibility constraints of the opportunisic experts reduce to $\frac{p_m-c_m}{p_s-c_s}=\frac{(A_{s1}-A_{m1}+\bar{T}_s (1-A_{s1})+(1-\bar{T}_s)(1-A_{m1}))}{(1-\bar{T}_s)(1-A_{m1})A_{m2}+\bar{T}_s(1-A_{s1})+A_{m1}}=-\frac{(1-\bar{T}_s)(1-A_{s1})}{1+\bar{T}_s(1-A_{s1})}<0$, which does not have a feasible solution. In sum, we have shown that $T_{s}\in(0,1)$ and $T_{m}\in(0,1)$ cannot be an equilibrium by contradiction.

Next, we identify the conditions under which ($T_{s}=0$, $T_{m}=0)$ constitutes an equilibrium. When $T_{s}=0$ and $T_{m}=0$, the conditions in  (\ref{eq:gammams}) and (\ref{eq:gammasm}) can be reduced to $(h(1-A_{m1})+(1-h)(1-A_{s1})+A_{s1})(p_s-c_s)>(A_{m1}+(1-h)(1-A_{s1})+h(1-A_{m1})A_{m2})(p_m-c_m)$ and $A_{m1}(p_s-c_s+p_m-c_m)+((1-h)(1-A_{m1})A_{m2}+h(1-A_{s1}))(p_m-c_m)>(A_{s1}+h (1-A_{s1})+(1-h)(1-A_{m1}))(p_s-c_s)$. Note that the second inequality can be written as $(A_{m1}+h(1-A_{s1})+(1-h)(1-A_{m1})A_{m2})(p_m-c_m)>(A_{s1}-A_{m1}+h (1-A_{s1})+(1-h)(1-A_{m1}))(p_s-c_s)$, and it can be inferred that $2A_{m1}>A_{s1}\geq0$. Also, note that when $T_{s}=0$ and $T_{m}=0$, the posterior belief $\tau_s=\frac{(1-\mu)h}{(1-\mu)h+\mu(1-h)}$ and $\tau_m=\frac{\mu h}{\mu h+(1-\mu)(1-h)}$, and the incentive-compatibility constraints of the consumer must hold with strict inequality except for the knife-edge values, indicating that when $T_{s}=0$ and $T_{m}=0$, the equilibrium outcome is pure strategy. Note that when $T_{s}=0$ and $T_{m}=0$, we have $2A_{m1}>A_{s1}\geq0$, and $A_{m1}=1$ in a pure strategy equilibrium. Next, we examine the conditions under which $(T_{m}=0,T_{s}=0,A_{m1}=1,A_{s1}=1)$ or $(T_{m}=0,T_{s}=0,A_{m1}=1,A_{s1}=0)$ constitute an equilibrium, respectively. When $(T_{m}=0,T_{s}=0,A_{m1}=1,A_{s1}=1)$  constitutes an equilibrium, the two incentive-compatibility constraints of opportunistic experts satisfy automatically such that $(p_s-c_s)>(p_m-c_m)$ and $(p_s-c_s+p_m-c_m)>(p_s-c_s)$, and we need to check the incentive-compatibility constraints of consumers such that (\ref{eq:gammasa})$>$(\ref{eq:gammasrma}), (\ref{eq:gammama})$>$(\ref{eq:gammamrma}) and (\ref{eq:gammama})$>$(\ref{eq:gammamrmr}).  By plugging $\tau_s=\frac{(1-\mu)h}{(1-\mu)h+\mu(1-h)}$ into (\ref{eq:gammasa})$>$(\ref{eq:gammasrma}), we have $\mu\leq\mu^{\gamma}_2$ in which $\mu^{\gamma}_2=1$ when $h\leq\frac{h}{p_s-p_m}$, while $\mu^{\gamma}_2=\frac{h^2k+h(1-h)(l_s-p_s+p_m+k)}{h^2k+h(1-h)(l_s-p_s+p_m+k)+(1-h)(h(p_s-p_m)-d)}$ otherwise. Similarly, by plugging $\tau_m=\frac{\mu h}{\mu h+(1-\mu)(1-h)}$ into (\ref{eq:gammama})$>$(\ref{eq:gammamrma}) and , we have $\mu\geq\mu^{\gamma}_1$, and $\mu^{\gamma}_1=0$ when $h\leq\frac{l_s-p_s+k-k^{'}}{l_s-p_s+p_m}$ while $\mu^{\gamma}_1=\frac{(1-h)(h(p_m+k^{'}-k)-(1-h)(l_s-p_s+k-k^{'}))}{(1-h)(h(p_m+k^{'}-k)-(1-h)(l_s-p_s+k-k^{'}))+h((1-h)(p_s-p_m)+k)}$ otherwise. In addition, 
 (\ref{eq:gammama})$>$(\ref{eq:gammamrmr}) holds when $\mu\geq\mu^{\gamma}_3$ and $\mu^{\gamma}_3=0$ when $h\leq\frac{l_s-p_s-p_m+k-k^{'}}{l_s-p_s}$ while $\mu^{\gamma}_3=\frac{(1-h)((p_m+k^{'}-k)-(1-h)(l_s-p_s))}{(1-h)((p_m+k^{'}-k)-(1-h)(l_s-p_s))+h^2(l_m-p_m+k)+(1-h)h(p_s-p_m+k)}$ otherwise.  In sum, we find FOFU $(T_{m}=0,T_{s}=0,A_{m1}=1,A_{s1}=1)$ constitutes an equilibrium when $max\{\mu^{\gamma}_1,\mu^{\gamma}_3\}\leq \mu\leq \mu^{\gamma}_2$. When $(T_{m}=0,T_{s}=0,A_{m1}=1,A_{s1}=0)$ constitutes an equilibrium, the incentive compatibility constraints for the opportunistic experts can be reduced to $(1-h)(p_s-c_s)>(2-h)(p_m-c_m)$ and $(p_s-c_s+p_m-c_m)+h(p_m-c_m)>h (p_s-c_s)$, and the two conditions satisfy when $h\leq \frac{(p_s-c_s-2(p_m-c_m))}{p_s-c_s-p_m+c_m}$. Then we examine the incentive-compatibility constraints for the consumers. We have (\ref{eq:gammasa})$>$(\ref{eq:gammasrma}), (\ref{eq:gammama})$>$(\ref{eq:gammamrma}) and (\ref{eq:gammama})$>$(\ref{eq:gammamrmr}), and the conditions can be satisfied when $\mu\leq \mu^{\gamma}_2$ and $h\leq \frac{(p_s-c_s-2(p_m-c_m))}{p_s-c_s-p_m+c_m}$. However, note that the threshold $\mu^{\gamma}_2< 1$ if and only if $h\geq\frac{k}{p_s-p_m}$. In sum, we have shown that $(T_{m}=0,T_{s}=0,A_{m1}=1,A_{s1}=0)$ constitutes an equilibrium when $\mu>\mu^{\gamma}_2$ and $\frac{k}{p_s-p_m} \leq h\leq\frac{(p_s-c_s-2(p_m-c_m))}{p_s-c_s-p_m+c_m}$.

Then, we identify the conditions under which $T_{s}=0$ and $T_{m}\in(0,1)$ constitutes a mix-strategy equilibrium. First, we check the incentive-compatibility constraints for the opportunistic experts. When $T_{s}=0$ and $T_{m}\in(0,1)$, $(1-\gamma_m+\gamma_m A_{s1})(p_s-c_s)=(\gamma_m A_{m1} +(1-\gamma_m)\gamma_{mm}A_{m2}+(1-\gamma_m)(1-\gamma_{mm}))(p_m-c_m)$ and $\gamma_s(p_s-c_s+p_m-c_m)A_{m1}+(1-\gamma_s)(\gamma_{sm}A_{m2}+(1-\gamma_{sm}))(p_m-c_m)>\gamma_s(p_s-c_s)A_{s1}+(1-\gamma_s)(p_s-c_s)$, and we can deduce that $A_{m1}>A_{s1}>0$. Note from the argument earlier such that $A_{m1}=1$ when $A_{s1}<1$, the only possible equilibrium when $T_{s}=0$ and $T_{m}\in(0,1)$ is $(T_{s}=0,T_{m}\in(0,1),A_{m1}=1, A_{s1}\in(0,1))$. Next, we investigate the equilibrium conditions when $(T_{m}\in(0,1),T_{s}=0,A_{m1}=1, A_{s1}\in(0,1))$ constitutes an equilibrium. When $A_{m1}=1$ and $A_{s1}\in(0,1)$, the incentive compatibility constraints of the opportunistic experts reduce to $((1-h)(1-A_{s1})+A_{s1})(p_s-c_s)=(1+(1-h)(1-A_{s1}))(p_m-c_m)$ and $(p_s-c_s+p_m-c_m)+h(1-A_{s1})(p_m-c_m)>(A_{s1}+h(1-A_{s1}))(p_s-c_s)$, and $A_{s1}=\frac{(2-h)(p_m-c_m)-(1-h)(p_s-c_s)}{h(p_s-c_s)+(1-h)(p_m-c_m)}\in(0,1)$ if and only if $h>\frac{(p_s-c_s-2(p_m-c_m))}{p_s-c_s-p_m+c_m}$. When $A_{m1}=1$ and $A_{s1}\in(0,1)$, the incentive compatibility-constraints of the consumers are (\ref{eq:gammasa})$=$(\ref{eq:gammasrma}), (\ref{eq:gammama})$>$(\ref{eq:gammamrma}) and (\ref{eq:gammama})$>$(\ref{eq:gammamrmr}). When (\ref{eq:gammasa})$=$(\ref{eq:gammasrma}), we have $(1-\mu)h^2k+\mu(1-\bar{T}_m)^2k+(1-\mu)h(1-h)(l_s-p_s+k-k^{'})=\mu(1-\bar{T}_m)\bar{T}_m(p_s-p_m-k)$ and to ensure the equation has a feasible solution, we have $\mu>\mu^{\gamma}_2$. In addition, (\ref{eq:gammama})$>$(\ref{eq:gammamrma}) and (\ref{eq:gammama})$>$(\ref{eq:gammamrmr}) can be satisfied when $\mu\geq max\{\mu^{\gamma}_3,\mu^{\gamma}_1\}$. In sum, we have shown that FUPO $(T_{m}\in(0,1),T_{s}=0,A_{m1}=1, A_{s1}\in(0,1))$ constitutes an equilibrium when $\mu\geq \mu^{\gamma}_2$ and $h>\frac{(p_s-c_s-2(p_m-c_m))}{p_s-c_s-p_m+c_m}$.

Lastly, we identify the conditions under which $T_{s}\in(0,1)$ and $T_{m}=0$ constitute a mix-strategy equilibrium. First, we check the incentive compatibility constraints for the opportunistic experts. When $T_{s}\in(0,1)$ and $T_{m}=0$, the incentive compatibility constraints of opportunistic experts reduce to $(1-\gamma_m+\gamma_m A_{s1})(p_s-c_s)>(\gamma_m A_{m1} +(1-\gamma_m)\gamma_{mm}A_{m2}+(1-\gamma_m)(1-\gamma_{mm}))(p_m-c_m)$ and $\gamma_s(p_s-c_s+p_m-c_m)A_{m1}+(1-\gamma_s)(\gamma_{sm}A_{m2}+(1-\gamma_{sm}))(p_m-c_m)=\gamma_s(p_s-c_s)A_{s1}+(1-\gamma_s)(p_s-c_s)$. Next we show that $A_{s1}=1$ by contradiction. Suppose $A_{s1}<1$, then we have $A_{m1}=1$ immediately following the argument earlier ($A_{m1}=1$ when $A_{s1}<1$). However, when  $A_{m1}=1$ the condition $\gamma_s(p_s-c_s+p_m-c_m)A_{m1}+(1-\gamma_s)(\gamma_{sm}A_{m2}+(1-\gamma_{sm}))(p_m-c_m)=\gamma_s(p_s-c_s)A_{s1}+(1-\gamma_s)(p_s-c_s)$ can be reduced to $(1+h(1-A_{s1}))(p_m-c_m)=-(1-A_{s1})(1-h)(p_s-c_s)<0$, and it does not have a feasible solution. When $A_s=1$, the incentive-compatibility constraints of opportunistic experts in $\ref{eq:gammasm}$ can be simplified to $((1-h)(1-A_{m1})A_{m2}+A_{m1})(p_m-c_m)=(1-A_{m1})(2-h)(p_s-c_s)$. Next, we investigate the cases when $A_{m2}=0$ or $A_{m2}=1$, respectively. When $A_{m2}=0$, $A_{m1}=\frac{(2-h)(p_s-c_s)}{(2-h)(p_s-c_s)+(p_m-c_m)}\in(0,1)$, the incentive compatibility conditions of the consumers are  (\ref{eq:gammasa})$>$(\ref{eq:gammasrma}), (\ref{eq:gammamrmr})$\geq$(\ref{eq:gammama}) and (\ref{eq:gammama})$=$(\ref{eq:gammamrmr}). Similar to the analysis earlier,  (\ref{eq:gammasa})$>$(\ref{eq:gammasrma}) can be satisfied when  $\mu\leq \mu^{\gamma}_2$; (\ref{eq:gammamrmr})$\geq$(\ref{eq:gammamrma}) and (\ref{eq:gammamrmr})$=$(\ref{eq:gammama}) can be satisfied when $\mu\leq \mu^{\gamma}_3$.  When $A_{m2}=1$, $A_{m1}=\frac{(2-h)(p_s-c_s)-(1-h)(p_m-c_m)}{(2-h)(p_s-c_s)+h(p_m-c_m)}\in(0,1)$, and the incentive-compatibility conditions of the consumers are  (\ref{eq:gammasa})$>$(\ref{eq:gammasrma}), (\ref{eq:gammama})$\geq$(\ref{eq:gammamrmr}) and (\ref{eq:gammama})$=$(\ref{eq:gammasrma}). Similar to the analysis earlier (\ref{eq:gammasa})$>$(\ref{eq:gammasrma}) can be satisfied when  $\mu\leq \mu^{\gamma}_2$. In addition,  (\ref{eq:gammamrma})$=$(\ref{eq:gammama}) and (\ref{eq:gammama})$\geq$(\ref{eq:gammamrmr}) can be satisfied when $\mu\leq \mu^{\gamma}_1$. In sum, we have shown that FUPU $(T_{m}=0,T_{s}\in(0,1),A_{m1}\in(0,1), A_{s1}=1)$ constitutes an equilibrium when $\mu\leq max\{\mu^{\gamma}_1,\mu^{\gamma}_3\}$

In sum, our analysis shows that FOFU $(T_{m}=0,T_{s}=0,A_{m1}=1,A_{s1}=1)$ constitutes an equilibrium when $max\{\mu^{\gamma}_1,\mu^{\gamma}_3\}\leq \mu\leq \mu^{\gamma}_2$. $(T_{m}=0,T_{s}=0,A_{m1}=1,A_{s1}=0)$ constitutes an equilibrium when $\mu>\mu^{\gamma}_2$ and $\frac{k}{p_s-p_m} \leq h\leq\frac{(p_s-c_s-2(p_m-c_m))}{p_s-c_s-p_m+c_m}$, POFU $(T_{m1}\in(0,1),T_{s}=0,A_{m}=1,A_{s1}\in(0,1))$ constitutes an equilibrium when $\mu\geq \mu^{\gamma}_2$ and $h>\frac{(p_s-c_s-2(p_m-c_m))}{p_s-c_s-p_m+c_m}$ and FOPU $(T_{m}=0,T_{s}\in(0,1),A_{m1}\in(0,1),T_{s1}=1,)$ constitutes an equilibrium when $\mu\leq max\{\mu^{\gamma}_3,\mu^{\gamma}_1\}$.

ii). In this part, we show that consumer welfare is worse off with restricted access to consumer diagnosis history (extension). Following the notation in the base model, we define the consumer welfare in the base model as $CW$, while the consumer welfare in the extension model is defined as $CW_{\gamma}$. Next, we show that $\mu^*_2\leq\mu^{\gamma}_2$,  $\mu^{\gamma}_1\leq\mu^*_1$ and $\mu^{\gamma}_3\leq\mu^*_1$. By plugging $\mu^*_2=\frac{hk}{hk+(1-h)(h(p_s-p_m)-k)}$ and $\mu^{\gamma}_2=\frac{h^2k+h(1-h)(l_s-p_s+p_m+k)}{h^2k+h(1-h)(l_s-p_s+p_m+k)+(1-h)(h(p_s-p_m)-d)}$ into analysis, we find that $\mu^{\gamma}_2-\mu^*_2=\frac{h(1-h)^2(l_s-p_s+p_m)(h(p_s-p_m)-k)}{(h^2k+h(1-h)(l_s-p_s+p_m+k)+(1-h)(h(p_s-p_m)-d))(hk+(1-h)(h(p_s-p_m)-k))}>0$ when $h\geq\frac{h}{p_s-p_m}$. By plugging $\mu^*_1=\frac{(1-h)(p_m-k+k^{'})}{h(k+(1-h)(p_s-p_m))+(1-h)(p_m-k+k^{'})}$, $\mu^{\gamma}_1=\frac{(1-h)(h(p_m+k^{'}-k)-(1-h)(l_s-p_s+k-k^{'}))}{(1-h)(h(p_m+k^{'}-k)-(1-h)(l_s-p_s+k-k^{'}))+h((1-h)(p_s-p_m)+k)}$, $\mu^{\gamma}_3=\frac{(1-h)((p_m+k^{'}-k)-(1-h)(l_s-p_s))}{(1-h)((p_m+k^{'}-k)-(1-h)(l_s-p_s))+h^2(l_m-p_m+k)+(1-h)h(p_s-p_m+k)}$ into analysis, we have $\mu^*_1-\mu^{\gamma}_1=\frac{h(1-h)(k+(1-h)(p_s-p_m))(l_s-p_s+p_m)}{((1-h)(h(p_m+k^{'}-k)-(1-h)(l_s-p_s+k-k^{'}))+h((1-h)(p_s-p_m)+k))(h(k+(1-h)(p_s-p_m))+(1-h)(p_m-k+k^{'}))}>0$ and 
$\mu^*_1-\mu^{\gamma}_3=\frac{h(1-h)^2(l_s-p_s)(k+(1-h)(p_s-p_m))}{((1-h)((p_m+k^{'}-k)-(1-h)(l_s-p_s))+h^2(l_m-p_m+k)+(1-h)h(p_s-p_m+k))(h(k+(1-h)(p_s-p_m))+(1-h)(p_m-k+k^{'}))}>0$.

In the base model, $CW=CW_{POFU}=-\mu \bar{T}_m (p_m+k)-\mu (1- \bar{T}_m)(p_s+k)-(1-\mu) h (p_s+k)-(1-\mu)(1-h)(p_m+p_s+k+k^{'})$, in which $\bar{T}_m=h+(1-h)T_{m1}=1-\frac{(1-\mu)hk}{\mu(h(p_s-p_m)-k)}>h$ when $\mu\geq\mu^*_2$; while $CW=CW_{FOPU}=-\mu(1-h)(p_s+k)-(1-\mu)\bar{T_s}(p_s+k)-\mu h(p_m+k)-(1-\mu)(1-\bar{T_s})(p_m+p_s+k+k^{'})$, in which $\bar{T_s}=(1-h)T_{s1}+h=1-\frac{\mu h(1-h)(p_s-p_m)+\mu h k}{(1-\mu)(p_m+k^{'}-k)}>h$ when $\mu<\mu^*_1$. In this extension, $CW_{\gamma}=-\mu(1-h)(p_s+k)-(1-\mu)h(p_s+k)-\mu h(p_m+k)-(1-\mu)(1-h)(p_m+p_s+k+k^{'})$ when $\mu^{\gamma}_2\geq\mu\geq max\{\mu^{\gamma}_3,\mu^{\gamma}_1\}$ in equilibrium $(T_{m1}=0,T_{s1}=0,A_{m1}=1,A_{s1}=1)$. As a result, it is straightforward to see that $CW_{\gamma}>CW$ when $\mu^*_2\leq\mu\leq\mu^{\gamma}_2$ and $\{\mu^{\gamma}_1,\mu^{\gamma}_3\}\leq\mu\leq\mu^*_1$, as $CW_{\gamma}-CW=-\mu(\bar{T}_m-h)(p_s-p_m)<0$ when $\mu^*_2\leq\mu\leq\mu^{\gamma}_2$ and $CW_{\gamma}-CW=-(1-\mu)(\bar{T}_s-h)(p_m+k^{'})<0$ when $\{\mu^{\gamma}_1,\mu^{\gamma}_3\}\leq\mu\leq\mu^*_1$. \hspace{0.3cm} $\Box$

\section*{Alternative contract agreement}

In this section, we examine the equilibrium outcome when the expert refunds a consumer the fee for the initial treatment, that is, a consumer only needs to pay $p_s-p_m$ for the second treatment if the first minor treatment is insufficient to resolve her problem. Note from Lemma 2: when $k>k^{'}$, a consumer with a serious problem still returns to the previous expert when the first treatment is insufficient to resolve her problem, and the only difference is that the consumer pays less for the second treatment, while the expert gets lower payoff from undertreatment. Next, we examine equilibrium outcomes under the alternative contract agreement. 

First, we spell out the expected payoff of a consumer on the first visit to an expert. Facing a serious treatment recommendation, the consumer can choose to a). accept the recommendation or b) reject the recommendation and search. In b), after the consumer rejects the first recommendation and searches for a second expert, note that an opportunistic expert recommends a serious treatment invariantly to a consumer (Lemma 1), then the second-visit consumer faces c). a truthful serious recommendation from an honest expert or an opportunistic expert (with probability $\tau_s$); d) an excessive recommendation from an opportunistic expert (with probability $(1-\tau_s)(1-h)$); e) a truthful minor recommendation from an honest expert (with probability $(1-\tau_s)h$). We summarize the consumer's expected payoff of accepting and rejecting the first serious treatment as follows:

\begin{align}
& -p_s \label{eq:altersa} \\
& -\tau_s (p_s+k)-(1-\tau_s)h(p_m+k)-(1-\tau_s)(1-h)(p_s+k) \label{eq:altersr} 
\end{align}

In the expressions above, $\tau_s=\frac{(1-\mu)\bar{T}_s}{(1-\mu)\bar{T}_s+\mu(1-\bar{T}_m)}$, $\bar{T}_s=h+(1-h)T_{s1}$ and $\bar{T}_m=h+(1-h)T_{m1}$.

Similarly, facing a minor treatment recommendation, the consumer can choose to a). accept the recommendation or b) reject the recommendation and search. In b), after the consumer rejects the first recommendation and searches for a second expert, the consumer faces c). a truthful minor recommendation from an honest expert (with probability $\tau_m h$); d) an excessive recommendation from an opportunistic expert (with probability $\tau_m(1-h)$); e) a truthful serious recommendation from an honest expert or an opportunistic expert (with probability $(1-\tau_m)$).  Note that a consumer only needs to pay $p_s-p_m$ for the second treatment if the first treatment is insufficient to resolve her problem. We summarize the consumer's expected payoff of accepting and rejecting the first minor treatment as follows: 

\begin{align}
& -\tau_m p_m-(1-\tau_m)(p_s+k^{'}) \label{eq:alterma} \\
& -\tau_m h (p_m+k)-\tau_m(1-h)(p_s+k)-(1-\tau_m)(p_s+k) \label{eq:altermr} 
\end{align}

In the expressions above, $\tau_m=\frac{\mu\bar{T}_m}{\mu\bar{T}_m+(1-\mu)(1-\bar{T}_s)}$, $\bar{T}_s=h+(1-h)T_{s1}$ and $\bar{T}_m=h+(1-h)T_{m1}$. Different from the base, we find that (\ref{eq:alterma})$>$(\ref{eq:altermr}), indicating that it is a dominant strategy for a consumer to accept a minor recommendation on the first visit to an expert such that $A_{m1}=1$. 

Next, we specify the incentive-compatibility constraints for an opportunistic expert to recommend inadequately and excessively as in (\ref{eqn:altersm}) and (\ref{eqn:alterms}), respectively. 

\vspace{-0.5cm}
\begin{align}
& IC(Inadequate): ((p_m-c_m)+(p_s-p_m-c_s))A_{m1}\geq(p_s-c_s)A_{s1} \label{eqn:altersm}\\
& IC(Excessive):  (p_s-c_s)A_{s1}\geq(p_m-c_m)A_{m1}   \label{eqn:alterms}
\end{align}
\vspace{-1.0cm}  

Next, we show that $A_{s1}>0$ by contradiction. Suppose  $A_{s1}=0$ in equilibrium, then $(p_s-c_s)A_{s1}<(p_m-c_m)A_{m1}$ holds with strict inequality, then the opportunistic expert recommends truthfully to a consumer with a minor problem, and thus $T_{m1}=1$ and $\tau_s=1$. However, when $\tau_s=1$, (\ref{eq:altersa})$>$(\ref{eq:altersr}) holds with strict inequality, indicating that a consumer accepts a serious treatment and $A_{s1}=1$, leading to contradiction.  As a result, the only possible equilibrium strategy pairs are $(A_{s1}=1,A_{m1}=1)$ and $(A_{s1}\in(0,1),A_{m1}=1)$.

When $(A_{s1}=1,A_{m1}=1)$ constitutes an equilibrium, the incentive compatibility constraints for the consumers must hold with strict inequality such that (\ref{eq:alterma})$>$(\ref{eq:altermr}) and (\ref{eq:altersa})$>$(\ref{eq:altersr}). Note from the incentive compatibility constraints of the opportunistic expert, when  $A_{s1}=1$ and $A_{m1}=1$,  $(p_s-c_s)A_{s1}>((p_m-c_m)+(p_s-p_m-c_s))A_{m1}$ and $(p_s-c_s)A_{s1}>(p_m-c_m)A_{m1}$ hold with strict inequality, and we have $T_{s1}=1$ and $T_{m1}=0$. By plugging $\tau_s=\frac{(1-\mu)}{(1-\mu)+\mu(1-h)}$ and $\tau_m=1$ into the incentive-compatibility constraints of the consumers, we have $\mu\leq\frac{k}{(1-h)h(p_s-p_m)+hk}$. In sum, we have shown that $(T_{m1}=0,T_{s1}=1,A_{m1}=1,A_{s1}=1)$ constitutes an equilibrium when $\mu\leq\frac{k}{(1-h)h(p_s-p_m)+hk}$.

When $(A_{s1}\in(0,1),A_{m1}=1)$ constitutes an equilibrium, the incentive compatibility constraints for the consumers must satisfy (\ref{eq:alterma})$=$(\ref{eq:altermr}) and (\ref{eq:altersa})$>$(\ref{eq:altersr}). When $p_s-p_m<c_s$, $((p_m-c_m)+(p_s-p_m-c_s))A_{m1}<(p_m-c_m)A_{m1}=(p_s-c_s)A_{s1}$, and $T_{s1}=1$ when $T_{m1}\in(0,1)$. By plugging  $\tau_s=\frac{(1-\mu)}{(1-\mu)+\mu(1-\bar{T}_m)}$ and $\tau_m=1$ into the incentive-compatibility constraints of the consumers, we have $T_{m1}=1-\frac{(1-\mu)k}{\mu(1-h)(h(p_s-p_m)-k)}$, in which $\mu\geq\frac{k}{h(1-h)(p_s-p_m)+hk}$. When $p_s-p_m>c_s$, $((p_m-c_m)+(p_s-p_m-c_s))A_{m1}>(p_m-c_m)A_{m1}=(p_s-c_s)A_{s1}$, and $T_{s1}=0$ when $T_{m1}\in(0,1)$. By plugging  $\tau_s=\frac{(1-\mu)(1-h)}{(1-\mu)(1-h)+\mu(1-\bar{T}_m))}$ and $\tau_m=\frac{\mu \bar{T}_m}{\mu\bar{T}_m+(1-\mu)(1-h)}$ into the incentive compatibility constraints of the consumers, we have $T_{m1}=1-\frac{(1-\mu)k}{\mu(1-h)(h(p_s-p_m)-k)} $ in which $\mu>\frac{hk}{hk+(1-h)h(p_s-p_m)-(1-h)k}$. Note that $\frac{hk}{hk+(1-h)h(p_s-p_m)-(1-h)k}<\frac{k}{(1-h)h(p_s-p_m)+hk}$, when $\mu\in(\frac{hk}{hk+(1-h)h(p_s-p_m)-(1-h)k},\frac{k}{(1-h)h(p_s-p_m)+hk})$, the equilibrium is $(T_{m1}=0,T_{s1}=1,A_{m1}=1,A_{s1}=\frac{p_m-c_m}{p_s-c_s})$ since an opportunistic expert can get a higher expected payoff. 

In sum, our analysis shows that when $(p_s-p_m<c_s$, $(T_{m1}=0,T_{s1}=1,A_{m1}=1,A_{s1}=1)$ constitutes an equilibrium when $\mu\leq\frac{k}{(1-h)h(p_s-p_m)+hk}$, while $(T_{m1}=1-\frac{(1-\mu)k}{\mu(1-h)(h(p_s-p_m)-k)},T_{s1}=1,A_{m1}=1,A_{s1}=1)$ constitutes an equilibrium when $\mu>\frac{k}{(1-h)h(p_s-p_m)+hk}$ and there is no undertreatment. When $p_s-p_m>c_s$, $(T_{m1}=0,T_{s1}=1,A_{m1}=1,A_{s1}=\frac{p_m-c_m}{p_s-c_s})$ constitutes an equilibrium when $\mu\leq\frac{k}{(1-h)h(p_s-p_m)+hk}$, while $(T_{m1}=1-\frac{(1-\mu)k}{\mu(1-h)(h(p_s-p_m)-k)},T_{s1}=0,A_{m1}=1,A_{s1}=\frac{p_m-c_m}{p_s-c_s})$ constitutes an equilibrium when $\mu>\frac{k}{(1-h)h(p_s-p_m)+hk}$.  \hspace{0.3cm} $\Box$

\section*{Undertreatment is not discovered in the short run}

In this section, we examine the situation whereby undertreatment cannot be discovered in the short run. We extend our base model by assuming a probability $\delta\in(0,1)$ that a consumer cannot discover her problem is serious after she receives undertreatment in Period $t$. Similar to the intuition in Lemma $1$ and Lemma $2$, we find that an opportunistic expert recommends a serious treatment to a consumer on the second visit to an expert invariantly, and a second-visit consumer accepts any recommendation from an expert. Next, we examine the possible equilibrium outcomes in this extension. 

First, we spell out the expected payoff of a consumer on the first visit to an expert. Facing a serious treatment recommendation, the consumer can choose to a). accept the recommendation or b) reject the recommendation and search. In b), after the consumer rejects the first recommendation and searches for a second expert, the consumer faces c). a truthful serious recommendation from an honest expert or an opportunistic expert (with probability $\tau_s$); d) an excessive recommendation from an opportunistic expert (with probability $(1-\tau_s)(1-h)$); e) a truthful minor recommendation from an honest expert (with probability $(1-\tau_s)h$). We summarize the consumer's expected payoff of accepting and rejecting the first serious treatment as follows: 

\begin{align}
& -p_s \label{eq:deltasa} \\
& -\tau_s (p_s+k)-(1-\tau_s)h(p_m+k)-(1-\tau_s)(1-h)(p_s+k) \label{eq:deltasr} 
\end{align}

In the expressions above, $\tau_s=\frac{(1-\mu)\bar{T}_s}{(1-\mu)\bar{T}_s+\mu(1-\bar{T}_m)}$, $\bar{T}_s=h+(1-h)T_{s1}$ and $\bar{T}_m=h+(1-h)T_{m1}$.

Similarly, facing a minor treatment recommendation, the consumer can choose to a). accept the recommendation or b) reject the recommendation and search. In a).if the minor recommendation is truthful (with probability $\tau_m$), the consumer's problem get resolved with $-p_m$. If the minor treatment is insufficient to resolve the consumer's problem and the consumer identifies the undertreatment in the second period (with probability $(1-\tau_m)(1-\delta)$), the consumer returns to the previous expert and gets her problem resolved with $-p_m-p_s-k^{'}$; otherwise, the consumer cannot identify her problem is serious (with probability $(1-\tau_m)\delta$) and gets a utility $-p_m-l_s$. In b), after the consumer rejects the first recommendation and searches for a second expert, the consumer can get a truthful minor recommendation from an honest expert (with probability $\tau_m h$) with a payoff $-p_m-k$; an excessive recommendation from an opportunistic expert (with probability $\tau_m(1-h)$) with a payoff $-p_s-k$; or a truthful serious recommendation from an honest expert or an opportunistic expert (with probability $(1-\tau_m)$) with a payoff $-p_s-k$. We summarize the consumer's expected payoff of accepting and rejecting the first minor treatment as follows: 

\begin{align}
& -\tau_m p_m-(1-\tau_m)\delta(p_s+p_m+k^{'})- (1-\tau_m)(1-\delta)(l_s+p_m)\label{eq:deltama} \\
& -\tau_m h (p_m+k)-\tau_m(1-h)(p_s+k)-(1-\tau_m)(p_s+k) \label{eq:deltamr} 
\end{align}

In the expressions above, $\tau_m=\frac{\mu\bar{T}_m}{\mu\bar{T}_m+(1-\mu)(1-\bar{T}_s)}$, $\bar{T}_s=h+(1-h)T_{s1}$ and $\bar{T}_m=h+(1-h)T_{m1}$. Next, we specify the incentive-compatibility constraints for an opportunistic expert to recommend inadequately and excessively as in (\ref{eqn:deltasm}) and (\ref{eqn:deltams}), respectively. 

\vspace{-0.5cm}
\begin{align}
& IC(Inadequate): ((p_m-c_m)+(1-\delta)(p_s-c_s))A_{m1}\geq(p_s-c_s)A_{s1} \label{eqn:deltasm}\\
& IC(Excessive):  (p_s-c_s)A_{s1}\geq(p_m-c_m)A_{m1}   \label{eqn:deltams}
\end{align}
\vspace{-1.0cm}  

Next, we show that $T_{m1}<1$ by contradiction. Suppose  $T_{m1}=1$ in equilibrium, then $\tau_s=1$ and a consumer would like to accept any serious treatment in equilibrium and $A_{s1}=1$; in which case $(\ref{eqn:deltams})$ holds with strict inequality, indicating that an opportunistic expert has incentives to deviate to $T_{m1}=0$. In addition, we can also show that $T_{s1}=0$ when $T_{m1}\in(0,1)$. This is because when $T_{m1}\in(0,1)$, the condition in (\ref{eqn:deltams}) is binding and $(p_s-c_s)A_{s1}=(p_m-c_m)A_{m1}$, and the condition in (\ref{eqn:deltams}) must hold with strict inequality. As a result, there are only four possible equilibrium strategy pairs such that $(T_{m1}=0,T_{s1}=0)$, $(T_{m1}=0,T_{s1}=1)$, $(T_{m1}=0,T_{s1}\in(0,1))$ and $(T_{m1}\in(0,1),T_{m1}=0)$.

When $(T_{m1}=0,T_{s1}=0)$ constitutes an equilibrium, the conditions in (\ref{eqn:deltams}) and (\ref{eqn:deltams}) must hold with strict inequality, and from which we can deduce $A_{m1}=1$ and $A_{s1}=1$, and $\delta<\frac{p_m-c_m}{p_s-c_s}$. Next, we examine the incentive compatibility constraints for the consumers (\ref{eq:deltasa})$>$(\ref{eq:deltasr}) and (\ref{eq:deltama})$>$(\ref{eq:deltamr}). By plugging $\tau_m=\frac{\mu h}{\mu h+(1-\mu)(1-h)}$ and $\tau_s=\frac{(1-\mu) h}{(1-\mu) h+\mu(1-h)}$ into the conditions, we have $\mu^{\delta}_1\leq\mu\leq\mu^{\delta}_2$, in which $\mu^{\delta}_2=1$ when $h\leq \frac{k}{p_s-p_m}$ while $\mu^{\delta}_2=\frac{hk}{hk+h(1-h)(p_s-p_m)-(1-h)k}$ when $h>\frac{k}{p_s-p_m}$, while $\mu^{\delta}_1=\frac{(1-h)(1-\delta)(p_m-k^{'}+k)+(1-h)\delta(l_s-p_s+p_m-k)}{hk+(1-h)(p_s-p_m)+(1-h)(1-\delta)(p_m-k^{'}+k)+(1-h)\delta(l_s-p_s+p_m-k)}$. In sum, we have shown that FOFU $(T_{m1}=0,T_{s1}=0,A_{m1}=1,A_{s1}=1)$ constitutes an equilibrium when $\mu^{\delta}_1\leq\mu\leq\mu^{\delta}_2$ and $\delta<\frac{p_m-c_m}{p_s-c_s}$.

When $(T_{m1}=0,T_{s1}=1)$ constitutes an equilibrium, we have $((p_m-c_m)+(1-\delta)(p_s-c_s))A_{m1}<(p_s-c_s)A_{s1}$ and $(p_s-c_s)A_{s1}>(p_m-c_m)A_{m1}$. If an opportunistic expert recommends truthfully to a consumer with a serious problem, then a consumer can deduce that a minor treatment recommendation is truthful and  $\tau_m=1$. Facing a minor treatment recommendation, a consumer would like to accept it and $A_{m1}=1$. To satisfy the incentive compatibility constraints of the opportunistic experts, we have $A_{m1}=A_{s1}=1$ and $\delta>\frac{p_m-c_m}{p_s-c_s}$. When $(T_{m1}=0,T_{s1}=1,A_{m1}=1,A_{s1}=1)$ constitutes an equilibrium, the incentive compatibility constraints for the consumers satisfy when (\ref{eq:deltasa})$>$(\ref{eq:deltasr}) and (\ref{eq:deltama})$>$(\ref{eq:deltamr}). By plugging $\tau_m=1$ and  $\tau_s=\frac{(1-\mu)}{(1-\mu)+\mu(1-h)}$ into the conditions, we have $\mu\leq \mu^{\delta}_3=\frac{k}{hk+(1-h)h(p_s-p_m)}$ and $\mu^{\delta}_3>\mu^{\delta}_2$. In sum, we have shown that FONU $(T_{m1}=0,T_{s1}=1,A_{m1}=1,A_{s1}=1)$ constitutes an equilibrium when $\mu\leq\mu^{\delta}_3$ and $\delta>\frac{p_m-c_m}{p_s-c_s}$.

When $(T_{m1}=0,T_{s1}\in(0,1))$ constitutes an equilibrium, we have $((p_m-c_m)+(1-\delta)(p_s-c_s))A_{m1}=(p_s-c_s)A_{s1}$ and $(p_s-c_s)A_{s1}>(p_m-c_m)A_{m1}$, in which $A_{s1}=1$, $A_{m1}\in(0,1)$ and $\delta<\frac{p_m-c_m}{p_s-c_s}$. When $(T_{m1}=0,T_{s1}\in(0,1),A_{m1}\in(0,1),A_{s1}=1)$ constitutes an equilibrium, the incentive compatibility constraints for the consumers satisfy when (\ref{eq:deltasa})$>$(\ref{eq:deltasr}) and (\ref{eq:deltama})$=$(\ref{eq:deltamr}).  By plugging $\tau_m=\frac{\mu h}{\mu h+(1-\mu)(1-\bar{T}_s)}$ and  $\tau_s=\frac{(1-\mu)\bar{T}_s}{(1-\mu)\bar{T}_s+\mu(1-h)}$ into the conditions, we have $\mu<\mu^{\delta}_1$. In sum, we have shown that FOPU $(T_{m1}=0,T_{s1}\in(0,1),A_{m1}\in(0,1),A_{s1}=1)$ constitutes an equilibrium when $\mu<\mu^{\delta}_1$ and $\delta<\frac{p_m-c_m}{p_s-c_s}$.

When $(T_{m1}\in(0,1),T_{s1}=0)$ constitutes an equilibrium, we have $((p_m-c_m)+(1-\delta)(p_s-c_s))A_{m1}>(p_s-c_s)A_{s1}$ and $(p_s-c_s)A_{s1}=(p_m-c_m)A_{m1}$, in which $A_{s1}\in(0,1)$, $A_{m1}=1$. When $(T_{m1}\in(0,1),T_{s1}=0, A_{m1}=1,A_{s1}\in(0,1))$ constitutes an equilibrium, the incentive compatibility constraints for the consumers satisfy when (\ref{eq:deltasa})$=$(\ref{eq:deltasr}) and (\ref{eq:deltama})$>$(\ref{eq:deltamr}). By plugging $\tau_m=\frac{\mu \bar{T}_m}{\mu \bar{T}_m+(1-\mu)(1-h)}$ and  $\tau_s=\frac{(1-\mu)(1-h)}{(1-\mu)h+\mu(1-\bar{T}_m)}$ into the conditions, we have $\mu>\mu^{\delta}_2$. In sum, we have shown that POFU $(T_{m1}\in(0,1),T_{s1}=0,A_{m1}=1,A_{s1}\in(0,1))$ constitutes an equilibrium when $\mu>\mu^{\delta}_2$.

In summary, when $\delta<\frac{p_m-c_m}{p_s-c_s}$, FOFU $(T_{m1}=0,T_{s1}=0,A_{m1}=1,A_{s1}=1)$ constitutes an equilibrium when $\mu^{\delta}_1\leq\mu\leq\mu^{\delta}_2$, FOPU $(T_{m1}=0,T_{s1}\in(0,1),A_{m1}\in(0,1),A_{s1}=1)$ constitutes an equilibrium when $\mu<\mu^{\delta}_1$, while POFU $(T_{m1}\in(0,1),T_{s1}=0,A_{m1}=1,A_{s1}\in(0,1))$ constitutes an equilibrium when $\mu>\mu^{\delta}_2$. When $\delta>\frac{p_m-c_m}{p_s-c_s}$,  FONU $(T_{m1}=0,T_{s1}=1,A_{m1}=1,A_{s1}=1)$ constitutes an equilibrium when $\mu\leq\mu^{\delta}_3$ and POFU $(T_{m1}\in(0,1),T_{s1}=0,A_{m1}=1,A_{s1}\in(0,1))$ constitutes an equilibrium when $\mu>\mu^{\delta}_3$.  \hspace{0.3cm} $\Box$

\section*{Consumer Resentment After an Undertreatment}

In this section, we first examine the case whereby the experts can observe the consumer's search history. Note that the consumers will not return to the previous expert after undertreatment, the incentive-compatibility constraints of the opportunistic experts when the consumer is searching for the first expert can be specified as follows:

\vspace{-0.5cm}
\begin{align}
& IC(Inadequate): (p_m-c_m)A_{m1}\geq(p_s-c_s)A_{s1} \label{eqn:resentsm1}\\
& IC(Excessive):  (p_s-c_s)A_{s1}\geq(p_m-c_m)A_{m1}   \label{eqn:resentms1}
\end{align}
\vspace{-1.0cm}  

When facing a serious treatment recommendation, a consumer accepts the recommendation and pays for a price $p_s$ with her problem resolved. If the consumer rejects the recommendation, the consumer may meet a). a truthful serious recommendation (with probability $\tau_s$), b) an excessively serious recommendation (with probability $(1-\tau_s)(1-h)$), or c). a truthful minor recommendation (with probability $(1-\tau_s)h$). We spell out the expected payoff of the consumer when accepting and rejecting a serious treatment recommendation as follows: 

\begin{align}
& -p_s \label{eq:resentsa1} \\
& -\tau_s (p_s+k)-(1-\tau_s)h(p_m+k)-(1-\tau_s)(1-h)(p_s+k) \label{eq:resentsr1} 
\end{align}

In the expressions above, $\tau_s=\frac{(1-\mu)\bar{T}_s}{(1-\mu)\bar{T}_s+\mu(1-\bar{T}_m)}$, $\bar{T}_s=h+(1-h)T_{s1}$ and $\bar{T}_m=h+(1-h)T_{m1}$.

Similarly, the consumer can accept and reject a minor recommendation on the first visit to an expert. When the first treatment is undertreatment, the consumer searches for a new expert due to resentment. The expected payoff of the consumer when accepting and rejecting a minor treatment can be specified as follows: 

\begin{align}
& -\tau_m p_m-(1-\tau_m)(p_s+p_m+k) \label{eq:resentma1} \\
& -\tau_m h (p_m+k)-\tau_m(1-h)(p_s+k)-(1-\tau_m)(p_s+k) \label{eq:resentmr1} 
\end{align}

In the expressions above, $\tau_m=\frac{\mu\bar{T}_m}{(1-\mu)(1-\bar{T}_s)+\mu\bar{T}_m}$.

Note from the incentive compatibility constraints of an opportunistic expert in (\ref{eqn:resentsm1}) and (\ref{eqn:resentms1}), there are two possible equilibrium strategy pairs such that $(T_{m1}=0,T_{s1}=1)$ and  $(T_{m1}\in(0,1),T_{s1}\in(0,1))$. However, when the conditions in (\ref{eqn:resentsm1}) and (\ref{eqn:resentms1}) are binding and $(T_{m1}\in(0,1),T_{s1}\in(0,1))$, it is a weakly dominant strategy for an opportunistic expert to truthfully recommend when the consumer is with a serious problem $T_{s1}=1$, since by doing so, the consumer will accept a minor treatment with a probability one when $\tau_m=1$, increasing the overall expected equilibrium payoff for the opportunistic expert. Next, we examine the equilibrium conditions for $(T_{m1}=0,T_{s1}=1)$ and $(T_{m1}\in(0,1),T_{s1}=1)$

When  $(T_{m1}=0,T_{s1}=1)$ constitutes an equilibrium, $A_{m1}=A_{s1}=1$. By plugging $\tau_s=\frac{(1-\mu)}{(1-\mu)+\mu(1-h)}$ and $\tau_m=1$ into (\ref{eq:resentsa1})$>$(\ref{eq:resentsr1}) and (\ref{eq:resentma1})$>$(\ref{eq:resentmr1}), we have $\mu<\frac{k}{k+(1-h)h(p_s-p_m-k)-(1-h)^2k}$. Similarly,  $(T_{m1}\in(0,1),T_{s1}=1)$ constitutes an equilibrium when $A_{m1}=1$ and $A_{s1}\in(0,1)$. By plugging $\tau_s=\frac{(1-\mu)\bar{T}_s}{(1-\mu)\bar{T}_s+\mu(1-h)}$ and $\tau_m=1$ into (\ref{eq:resentsa1})$=$(\ref{eq:resentsr1}) and (\ref{eq:resentma1})$>$(\ref{eq:resentmr1}), we have $(1-\mu)k+\mu(1-\bar{T}_m)(1-h)k=\mu(1-\bar{T}_m)h(p_s-p_m-k)$. Solving this condition, we have $\bar{T}_m=1-\frac{(1-\mu)k}{\mu(h(p_s-p_m)-k)}$. To ensure  $\bar{T}_m>h$, we have $\mu>\frac{k}{k+(1-h)h(p_s-p_m-k)-(1-h)^2k}$. In sum, we have shown that when the consumer exerts resentment and does not return to the previous expert after undertreatment, $(T_{m1}=0,T_{s1}=1,A_{m1}=1,A_{s1}=1)$ constitutes an equilibrium when  $\mu<\frac{k}{k+(1-h)h(p_s-p_m-k)-(1-h)^2k}$; $(T_{m1}\in(0,1),T_{s1}=1,A_{m1}=1,A_{s1}\in(0,1))$ constitutes an equilibrium when  $\mu\geq\frac{k}{k+(1-h)h(p_s-p_m-k)-(1-h)^2k}$.

Next, we identify the conditions under which the undertreatment occurs in equilibrium when the experts are restricted to the consumer's search history. By adopting a similar notation set as in Section 5.3, we use $\gamma_m$ ($\gamma_s$) to denote the probability that a consumer is on the first visit to an expert, conditional on the consumer is with a minor (serious) problem. Then, the incentive compatibility constraints of the opportunistic expert can be specified as follows: 

 \vspace{-0.5cm}
\begin{align}
& IC(Excessive): (1-\gamma_m+\gamma_m A_{s1})(p_s-c_s) \geq (\gamma_m A_{m1} +(1-\gamma_m)\gamma_{mm}A_{m2}+(1-\gamma_m)(1-\gamma_{mm}))(p_m-c_m) \label{eq:resentms2}\\
& IC(Inadequate):  \gamma_s(p_m-c_m)A_{m1}+(1-\gamma_s)(\gamma_{sm}A_{m2}+(1-\gamma_{sm}))(p_m-c_m) \geq(1-\gamma_s+\gamma_s A_{s1})(p_s-c_s). \label{eq:resentsm2}
\end{align}
\vspace{-1.0cm}

In the expressions above, $\gamma_m=\frac{1}{1+\bar{T}_m(1-A_{m1})+(1-\bar{T}_m)(1-A_{s1})}$, $\gamma_s=\frac{1}{1+\bar{T}_s(1-A_{s1})+(1-\bar{T}_s)(1-A_{m1})}$, $\gamma_{mm}=\frac{\bar{T}_m(1-A_{m1})}{\bar{T}_m(1-A_{m1})+(1-\bar{T}_m)(1-A_{s1})}$) and $\gamma_{sm}=\frac{(1-\bar{T}_s)(1-A_{m1})}{\bar{T}_s(1-A_{s1})+(1-\bar{T}_s)(1-A_{m1})}$, in which $\bar{T}_m=h+(1-h)T_{m1}$ and $\bar{T}_s=h+(1-h)T_{s1}$. Next, we examine consumer's incentive compatibility constraints when visiting the first expert. 

When the consumer is visiting the first expert, similar to the analysis in the proof of Proposition 8, facing a serious treatment recommendation, a consumer has three options: a). accept the recommendation; b) reject the recommendation and search for a second expert; on the second visit to an expert, the consumer accepts both minor and serious treatments; c). reject the recommendation and search for a second expert; on the second visit to an expert, the consumer accepts a serious treatment recommendation but rejects a minor treatment recommendation. 

\begin{align}
& -p_s \label{eq:resentsa2} \\
& -\tau_s\bar{T}_s(p_s+k)-\tau_s(1-\bar{T}_s)(l_s+p_m+k)-(1-\tau_s)\bar{T}_m(p_m+k)-(1-\tau_s)(1-\bar{T}_m)(p_s+k) \label{eq:resentsrma2} \\
& -\tau_s\bar{T}_s(p_s+k)-\tau_s(1-\bar{T}_s)(l_s+k)-(1-\tau_s)\bar{T}_m(l_m+k)-(1-\tau_s)(1-\bar{T}_m)(p_s+k) \label{eq:resentsrmr2}   
\end{align}

By the same token, when the consumer is visiting the first expert, facing a minor treatment recommendation, a consumer has three options: a). accept the recommendation; b) reject the recommendation and search for a second expert; on the second visit to an expert, the consumer accepts both minor and serious treatments; c). reject the recommendation and search for a second expert; on the second visit to an expert, the consumer accepts a serious treatment recommendation but rejects a minor treatment recommendation. The difference between this model and that in Section 5.3 is that the consumer will search for another expert after undertreatment, although searching for another expert involves a higher cost $k>k^{'}$.

\begin{align}
& -\tau_m p_m-(1-\tau_m)(p_m+p_s+k) \label{eq:resentma2} \\
& -\tau_m \bar{T}_m (p_m+k)-\tau_m(1-\bar{T}_m)(p_s+k)-(1-\tau_m)\bar{T}_s(p_s+k)-(1-\tau_m)(1-\bar{T}_s)(p_m+l_s+k) \label{eq:resentmrma2} \\
& -\tau_m \bar{T}_m (l_m+k)-\tau_m(1-\bar{T}_m)(p_s+k)-(1-\tau_m)\bar{T}_s(p_s+k)-(1-\tau_m)(1-\bar{T}_s)(l_s+k)\label{eq:resentmrmr2}   
\end{align}

In the expressions above, $\tau_s=\frac{(1-\mu)\bar{T}_s}{(1-\mu)\bar{T}_s+\mu(1-\bar{T}_m)}$, $\tau_m=\frac{\mu\bar{T}_m}{(1-\mu)(1-\bar{T}_s)+\mu\bar{T}_m}$, $\bar{T}_s=h+(1-h)T_{s1}$ and $\bar{T}_m=h+(1-h)T_{m1}$. Next, we examine the conditions under which undertreatment occurs in equilibrium. 

When undertreatment occurs in an equilibrium, (\ref{eq:resentsm2}) holds and $\gamma_s(p_m-c_m)A_{m1}+(1-\gamma_s)(\gamma_{sm}A_{m2}+(1-\gamma_{sm}))(p_m-c_m)\geq(1-\gamma_s+\gamma_s A_{s1})(p_s-c_s)$ from which we can deduce that $1\geq A_{m1}>A_{s1}$. Note that when $A_{s1}<1$, (\ref{eq:resentsa2})$\leq$(\ref{eq:resentsrma2}). By plugging $\tau_s=\frac{(1-\mu)\bar{T}_s}{(1-\mu)\bar{T}_s+\mu(1-\bar{T}_m)}$ into this inequality, we have $(1-\mu)\bar{T}^2_sk+(1-\mu)\bar{T}_s(1-\bar{T}_s)(p_m+k+l_s-p_s)+\mu(1-\bar{T}_m)^2k\leq \mu(1-\bar{T}_m)\bar{T}_m(p_s-p_m-k)$, indicating that $\mu(1-\bar{T}_m)\bar{T}_m(p_s-p_m-k)>(1-\mu)\bar{T}_s(1-\bar{T}_s)(p_m+k+l_s-p_s)$. However, when $\mu(1-\bar{T}_m)\bar{T}_m(p_s-p_m-k)>(1-\mu)\bar{T}_s(1-\bar{T}_s)(p_m+k+l_s-p_s)$, (\ref{eq:resentma2})$<$(\ref{eq:resentmrma2}) and (\ref{eq:resentsa2})$<$(\ref{eq:resentmrmr2}) hold with strict inequality, indicating that $A_{m1}=1$ when $A_{s1}<1$.

When $A_{m1}=1$ and $A_{s1}<1$ constitutes an equilibrium, the incentive compatibility constraints can be reduced to:

 \vspace{-0.5cm}
\begin{align}
& IC(Excessive): (A_{s1}+(1-A_{s1})(1-\bar{T}_m))(p_s-c_s)\geq (1+(1-A_{s1})(1-\bar{T}_m))(p_m-c_m) \label{eq:resentms3}\\
& IC(Inadequate):  (1+(1-A_{s1})\bar{T}_s))(p_m-c_m)\geq(A_{s1}+(1-A_{s1})\bar{T}_s)(p_s-c_s). \label{eq:resentsm3}
\end{align}
\vspace{-1.0cm}

Similar to the argument in the previous proof, an opportunistic expert never recommends truthfully when the consumer is with a minor problem, and $T_{m1}<1$ in equilibrium; when $T_{s1}<1$, overtreatment and undertreat exist concurrently and the conditions in (\ref{eq:resentms3}) and (\ref{eq:resentsm3}) hold simultaneously, from which we can deduce that $\bar{T}_s<1-\bar{T}_m$. There are two possible equilibrium pairs when the conditions in (\ref{eq:resentms3}) and (\ref{eq:resentsm3}) hold, $(T_{m1}=0,T_{s1}\in(0,1))$ and $(T_{m1}\in(0,1),T_{s1}=0$, Next, we examine the conditions for each equilibrium.

When $(T_{m1}=0,T_{s1}\in(0,1),A_{m1}=1,A_{s1}\in(0,1))$ constitutes an equilibrium, the condition in (\ref{eq:resentsm3}) is binding while (\ref{eq:resentsa2})$=$(\ref{eq:resentsrma2}). By plugging $\tau_s=\frac{(1-\mu)\bar{T}_s}{(1-\mu)\bar{T}_s+\mu(1-h)}$ into the conditions, we have $(1-\mu)\bar{T}^2_sk+(1-\mu)\bar{T}_s(1-\bar{T}_s)(p_m+k+l_s-p_s)+\mu(1-h)^2k=\mu(1-h)h(p_s-p_m-k)$. To ensure $\bar{T}_s<1-\bar{T}_m$, we have $\bar{T}_s\in(h,1-h)$. Then, let $f(x)=\mu x^2(p_s-p_m)-\mu x(p_s-p_m+k)+(1-\mu)h^2 k+(1-\mu)h(1-h)(p_m+l_s+k-p_s)+\mu k$, (\ref{eq:resentsa2})$=$(\ref{eq:resentsrma2}) has a feasible solution such that $x\in(h,1-h)$ if and only if $f(h)>0$, $f(1-h)<0$ and $h<1-h$. Solving the conditions above, we have $\mu^{r}_1\leq\mu\leq\mu^{r}_2$ and $1-\frac{k}{p_s-p_m}<h<\frac{1}{2}$, in which $\mu^{r}_1=\frac{h^2 k+h(1-h)(p_m+l_s+k-p_s)}{h^2 k+h(1-h)(p_m+l_s+k-p_s)+(1-h)(p_s-p_m+k)-k-(1-h)^2(p_s-p_m))}$ and $\mu^{r}_2=\frac{h^2 k+h(1-h)(p_m+l_s+k-p_s)}{h^2 k+h(1-h)(p_m+l_s+k-p_s)+h(p_s-p_m+k-k-h^2(p_s-p_m))}$.

When $(T_{m1}\in(0,1),T_{s1}=0,A_{m1}=1,A_{s1}\in(0,1))$ constitutes an equilibrium, the condition in (\ref{eq:resentms3}) is binding while (\ref{eq:resentsa2})$=$(\ref{eq:resentsrma2}). By plugging $\tau_s=\frac{(1-\mu)h}{(1-\mu)h+\mu(1-\bar{T}_m)}$ into the condition, we have $(1-\mu)h^2k+(1-\mu)h(1-h)(p_m+k+l_s-p_s)+\mu(1-\bar{T}_m)^2k=\mu(1-\bar{T}_m)\bar{T}_m(p_s-p_m-k)$. To ensure $\bar{T}_s<1-\bar{T}_m$, we have $\bar{T}_m\in(h,1-h)$. Let $f(x)=(1-\mu)x^2(p_m+l_s-p_s)-(1-\mu)x(p_m+l_s-p_s+k)+\mu(1-h)h(p_s-p_m-k)-\mu(1-h)^2k$, (\ref{eq:resentsa2})$=$(\ref{eq:resentsrma2}) has a feasible solution if and only if $f(h)>0$, $f(1-h)<0$ and $h<1-h$. Solving the conditions above, we have $\mu^{r}_3\leq\mu\leq\mu^{r}_4$ and $\frac{k}{p_s-p_m}<h<\frac{1}{2}$, in which $\mu^{r}_3=\frac{h((p_m+l_s-p_s+k)-h(p_m+l_s-p_s))}{h((p_m+l_s-p_s+k)-h(p_m+l_s-p_s))+(1-h)(h(p_s-p_m)-k)}$ and $\mu^{r}_4=\frac{(1-h)((p_m+l_s-p_s+k)-(1-h)(p_m+l_s-p_s))}{(1-h)((p_m+l_s-p_s+k)-(1-h)(p_m+l_s-p_s))+(1-h)(h(p_s-p_m)-k)}$.

In sum, we have shown that with consumer resentment, a consumer would not return to the previous expert after an undertreatment. Undertreatment will not occur if the consumer search history is known to the experts. When the consumer search history is unknown to the experts, undertreatment occurs when both probability of a minor problem $\mu$ and the market ethics level $h$ are moderate.  \hspace{0.3cm} $\Box$

\section*{Heterogeneity in Treatment Capability}

In this section, we examine the case where the service providers are differentiated by their ability to provide treatment. The markets consist of two types of experts: with a probability $\alpha$; the expert is a high type who is capable of providing both minor and serious treatments; with a probability $1-\alpha$, the expert is a low type who can only provide a minor treatment, and the type of expert ability is independent of the type of ethics level. When a low-type honest expert meets a consumer with a serious problem, the expert will reject to treat the consumer and charge nothing; when a low-type opportunistic expert meets a consumer with a serious problem, the expert will recommend a minor treatment to the consumer invariantly, since by doing so the opportunistic expert realizes a profit $p_s-c_s$; otherwise, the opportunistic expert gets nothing if reject to treat the consumer. Next, we examine the possible equilibrium outcomes in this extension. 

First, we spell out the expected payoff of a consumer on the first visit to an expert. Facing a serious treatment recommendation, the consumer can choose to a). accept the recommendation; b). reject the recommendation and search for another expert; on the second visit to an expert, the consumer accepts both minor and serious treatments. c). reject the recommendation and search for another expert; on the second visit to an expert, the consumer accepts serious treatment and rejects a minor treatment. In b) and c), after the consumer rejects the first recommendation and searches for a second expert, the consumer faces d). a truthful serious recommendation from a high-type honest expert or a high-type opportunistic expert  (with probability $\tau_s \alpha$);  e) an excessive recommendation from a high-type opportunistic expert (with probability $(1-\tau_s)(1-h)\alpha$); f) a truthful minor recommendation either from an honest expert or a low-type opportunistic expert (with probability $(1-\tau_s)(h+(1-h)(1-\alpha))$); g).a rejection from a low-type honest expert (with probability $\tau_s h (1-\alpha)$); h).an undertreatment from a low-type opportunistic expert (with probability $\tau_s (1-h) (1-\alpha)$). The difference is that in b), a second-visit consumer accepts a minor treatment recommendation in f) and h), while in c), a second-visit consumer rejects a minor treatment recommendation in f) and h). We summarize the consumer's expected payoff in a), b) and c) as follows: 

\begin{align}
& -p_s \label{eq:alphasa} \\
& -\tau_s \alpha(p_s+k)-(1-\tau_s)(1-h)\alpha(p_s+k)-(1-\tau_s)(h+(1-h)(1-\alpha))(p_m+k) \nonumber \\
& \quad -\tau_s h (1-\alpha)(l_s+k)-\tau_s(1-h) (1-\alpha)(p_m+l_s+k) \label{eq:alphasrma} \\
& -\tau_s \alpha(p_s+k)-(1-\tau_s)(1-h)\alpha(p_s+k)-(1-\tau_s)(h+(1-h)(1-\alpha))(l_m+k)\nonumber \\ 
& \quad -\tau_s h (1-\alpha)(l_s+k)-\tau_s (1-h) (1-\alpha)(l_s+k) \label{eq:alphasrmr}
\end{align}

In the expressions above, $\tau_s=\frac{(1-\mu)\bar{T}_s}{(1-\mu)\bar{T}_s+\mu(1-\bar{T}_m)}$, $\bar{T}_s=h+(1-h)T_{s1}$ and $\bar{T}_m=h+(1-h)T_{m1}$. By comparing the payoff in (\ref{eq:alphasa}) and (\ref{eq:alphasrmr}), it is straightforward to see that (\ref{eq:alphasa})$>$(\ref{eq:alphasrmr}), indicating that rejecting a minor treatment after rejecting a serious treatment is a dominated strategy. 

Similarly, when facing a minor treatment recommendation, note that a consumer can also update the expert's type of ability from the recommendation result; then, it is not necessary for a consumer to return to the previous expert when the first treatment is undertreatment. Then, in this extension, the consumer has four choices when visiting an expert: a). accept the recommendation; if the recommendation is insufficient, the consumer returns to the previous expert with a cost $k^{'}$; b). accept the recommendation; if the recommendation is insufficient, the consumer searches for a new expert with a cost $k$; c). reject the recommendation and search for another expert; on the second visit to an expert, the consumer accepts both minor and serious treatments. d). reject the recommendation and search for another expert; on the second visit to an expert, the consumer accepts serious treatment and rejects a minor treatment. In c) and d), after the consumer rejects the first recommendation and searches for a second expert, the consumer faces e). a truthful minor recommendation from an honest expert or a low-type opportunistic expert  (with probability $\tau_m (h+(1-h)(1-\alpha))$);  f). an inadequate recommendation from a low-type opportunistic expert (with probability $(1-\tau_m)(1-h)(1-\alpha)$); g). a rejection to treatment from a low-type honest expert (with probability $(1-\tau_m) h \alpha$); h). an excessive recommendation from a high-type opportunistic expert (with probability $\tau_m (1-h) \alpha$); i). a truthful serious recommendation from a high-type expert (with probability $(1-\tau_m) \alpha$). The difference is that in c), a second-visit consumer accepts a minor treatment recommendation in e) and f), while in d), a second-visit consumer rejects a minor treatment recommendation in e) and f). We summarize the consumer's expected payoff in a), b), c) and d) as follows:

\begin{align}
& -\tau_m p_m-(1-\tau_m)(p_m+k^{'}+\tau_h p_s+(1-\tau_h)l_s) \label{eq:alphamar} \\
& -\tau_m p_m-(1-\tau_m)(p_m+k+\alpha p_s+(1-\alpha)l_s) \label{eq:alphamas} \\
& -\tau_m (h+(1-h)(1-\alpha))(p_m+k)-(1-\tau_m)(1-h)(1-\alpha)(p_m+l_s+k)\nonumber \\
& \quad  - (1-\tau_m) h (1-\alpha)(l_s+k)  -\tau_m (1-h) \alpha(p_s+k)-(1-\tau_m) \alpha(p_s+k) \label{eq:alphamrma} \\
& -\tau_m (h+(1-h)(1-\alpha))(l_m+k)-(1-\tau_m)(1-h)(1-\alpha)(l_s+k)\nonumber \\
& \quad - (1-\tau_m) h (1-\alpha)(l_s+k)-\tau_m (1-h) \alpha(p_s+k)-(1-\tau_m) \alpha(p_s+k) \label{eq:alphamrmr} 
\end{align}

In the expressions above, $\tau_m=\frac{\mu\bar{T}_m}{(1-\mu)(1-\bar{T}_s)+\mu\bar{T}_m}$, in which$ \bar{T}_s=h+(1-h)T_{s1}$ and $\bar{T}_m=h+(1-h)T_{m1}$. In addition, we use $\tau_h$ to denote the probability that the first expert is a high-type expert, conditional on the first recommendation being an undertreatment. $\tau_h=\frac{(1-\mu)(1-h)\alpha(1-T_{s1})}{(1-\mu)(1-h)\alpha(1-T_{s1})+(1-\mu)(1-h)(1-\alpha)}=\frac{\alpha(1-T_{s1})}{(1-\alpha)+\alpha(1-T_{s1})}$. In this case, after undertreatment, by returning to first, the consumer gets her problem resolved with price $p_s$ if the opportunistic expert is a high type expert (with probability $\tau_h$), while the consumer's problem gets unresolved with a loss $l_s$ if the opportunistic expert is a low type expert (with probability $1-\tau_h$).

Different from the base model, the incentive compatibility constraints for an opportunistic expert to recommend inadequately depends on whether the consumer returns to the previous expert after undertreatment (the tradeoff between (\ref{eq:alphamar}) and (\ref{eq:alphamas})). If (\ref{eq:alphamar}) $>$ (\ref{eq:alphamas}), which is equivalent to $\frac{\alpha(1-\alpha)T_{s1}}{(1-\alpha)+\alpha(1-T_{s1})}(l_s-p_s)<k-k^{'}$, a consumer returns to the previous expert after undertreatment, and the incentive compatibility constraint for an opportunistic expert are:

\vspace{-0.5cm}
\begin{align}
& IC(Inadequate): ((p_m-c_m)+(p_s-c_s))A_{m1}\geq(p_s-c_s)A_{s1} \label{eqn:alphasm1}\\
& IC(Excessive):  (p_s-c_s)A_{s1}\geq(p_m-c_m)A_{m1}   \label{eqn:alphams1}
\end{align}
\vspace{-1.0cm}  

 If (\ref{eq:alphamar}) $<$ (\ref{eq:alphamas}), which is equivalent to $\frac{\alpha(1-\alpha)T_{s1}}{(1-\alpha)+\alpha(1-T_{s1})}(l_s-p_s)>k-k^{'}$, a consumer searches a new expert after undertreatment, and the incentive compatibility constraint for an opportunistic expert are:

\vspace{-0.5cm}
\begin{align}
& IC(Inadequate): (p_m-c_m)A_{m1}\geq(p_s-c_s)A_{s1} \label{eqn:alphasm2}\\
& IC(Excessive):  (p_s-c_s)A_{s1}\geq(p_m-c_m)A_{m1}   \label{eqn:alphams2}
\end{align}
\vspace{-1.0cm}

Next, we examine the conditions under which the consumers never return to the previous expert after undertreatment, and the opportunistic experts with high capability do not engage in strategic undertreatment ($T_{s1}=1$). Following the incentive compatibility constraints in (\ref{eqn:alphasm2}) and (\ref{eqn:alphams2}), the only equilibrium pair satisfies the condition is $(T_{s1}=1,T_{m1}=0)$. When $(T_{s1}=1,T_{m1}=0)$ constitutes an equilibrium, we have (\ref{eq:alphamar})$>$(\ref{eq:alphamas}), which can be reduced to
$\frac{\alpha(1-\alpha)}{(1-\alpha)}(l_s-p_s)>k-k^{'}$ and $\alpha>\frac{k-k^{'}}{l_s-p_s}$. When $T_{s1}=1$ and $T_{m1}=0$, $(p_s-c_s)A_{s1}>(p_m-c_m)A_{m1}$ holds with strict inequality when $A_{s1}=1$, $A_{m1}=1$ or $A_{m1}=0$. First, we examine the conditions under which $(T_{m1}=0,T_{s1}=1,A_{m1}=1,A_{s1}=1)$ constitutes an equilibrium. By plugging $\tau_m=\frac{\mu (h+(1-h)(1-\alpha))}{\mu (h+(1-h)(1-\alpha))+(1-\mu)(1-h)(1-\alpha)}$ and $\tau_s=\frac{(1-\mu)}{(1-\mu)+\mu(1-h)}$ into the incentive-compatibility conditions such that (\ref{eq:alphasa})$>$(\ref{eq:alphasrma}), (\ref{eq:alphamas})$>$(\ref{eq:alphamrma}) and  (\ref{eq:alphamas})$>$(\ref{eq:alphamrmr}), we find that  $(T_{m1}=0,T_{s1}=1,A_{m1}=1,A_{s1}=1)$ constitutes an euqilibirum when $max\{\mu^{\alpha}_1,\mu^{\alpha}_3\}\leq \mu\leq \mu^{\alpha}_2$, in which $\mu^{\alpha}_1=\frac{(1-h)(1-\alpha)(\alpha+h(1-\alpha))p_m}{(1-h)(1-\alpha)(\alpha+h(1-\alpha))p_m+(h+(1-h)(1-\alpha))((h+(1-h)(1-\alpha))k+(1-h)\alpha(p_s+k-p_m))}$, \\ $\mu^{\alpha}_2=\frac{\alpha k+h\alpha (l_s+k-p_s)}{\alpha k+h\alpha (l_s+k-p_s)+(1-h)(h+(1-h)\alpha)(p_s-p_m-k)-(1-h)h\alpha(l_s+k-p_s)-(1-h)^2(1-\alpha)(l_s+k+p_m-p_s)}$ and $\mu^{\alpha}_3=\frac{(1-h)(1-\alpha)p_m}{(1-h)(1-\alpha)p_m+(h+(1-h)(1-\alpha))((h+(1-h)(1-\alpha))(l_m+k-p_m)+(1-h)\alpha(p_s+k-p_m))}$. Similarly, when $(T_{m1}=0,T_{s1}=1,A_{m1}=0,A_{s1}=1)$ constitutes an equilibirum, we have (\ref{eq:alphasa})$>$(\ref{eq:alphasrma}), (\ref{eq:alphamas})$<$(\ref{eq:alphamrma}) or  (\ref{eq:alphamas})$>$(\ref{eq:alphamrmr}), which can be reduced to $\mu\leq max\{\mu^{\alpha}_1,\mu^{\alpha}_3\}$. 

In sum, we have shown that when $\alpha>\frac{k-k^{'}}{l_s-p_s}$,  $(T_{m1}=0,T_{s1}=1,A_{m1}=1,A_{s1}=1)$ constitutes an euqilibirum when $max\{\mu^{\alpha}_1,\mu^{\alpha}_3\}\leq \mu\leq \mu^{\alpha}_2$, while $(T_{m1}=0,T_{s1}=1,A_{m1}=0,A_{s1}=1)$ constitutes an equilibirum, we have $\mu\leq max\{\mu^{\alpha}_1,\mu^{\alpha}_3\}$. \hspace{0.3cm} $\Box$

\par\noindent
\section*{Endogenized Price}

In this section, we examine the equilibrium outcome when the experts can decide the prices of treatment endogenously. At the beginning of Period t, an expert decides the prices of treatment, and the prices cannot be changed across periods afterward. In this extension, we use $p_{mh}$ and $p_{sh}$ to denote the prices of a minor treatment and a serious treatment set by an honest expert, while $p_{mo}$ and $p_{so}$ are to denote the prices of a minor treatment and a serious treatment set by an opportunistic expert. An honest expert sets the prices based on the principle that maximizes his expected payoff while recommending truthfully; an opportunistic expert sets the prices only to maximize his expected payoff. In addition, we also assume that a consumer cannot observe the prices before visiting an expert. That is, the consumer knows the price information only {\em ex-post}. Next, we show that an opportunistic expert never recommends truthfully to a consumer with a serious problem and $T_{s1}<1$ by contradiction. 

(i) Suppose an opportunistic expert recommends truthfully when the first-visit consumer is with a serious problem and $T_{s1}=1$ at the equilibrium. Then, any minor treatment in the market is truthful. When the consumer is visiting the first expert, it is a dominant strategy for the consumer to accept any minor treatment recommendation and $A_{m1}=1$. In this case, both opportunistic experts and honest experts have incentives to set the price of a minor treatment as $p_{mo}=p_{mh}=l_m$ to maximize the expected payoff from a minor problem. However, note that the search cost is lower when returning to the previous expert than to visit a new one($k^{'}<k$); a consumer always returns to the previous expert after an undertreatment, in which case an opportunistic expert gets a higher profit from recommending inadequately to a consumer with a serious problem since $A_{m1}(p_{so}-c_s+p_{m0}-c_m)=p_{so}-c_s+p_{m0}-c_m>A_{s1}(p_{so}-c_s)$. As a result, the opportunistic expert will deviate from recommending truthfully ($T_{s1}=1$), leading to a contradiction.                                  

(ii) Next, we examine the pure-strategy equilibrium when the market fully consists of opportunistic experts. In this model, we define the opportunistic expert's strategy as $T_{m1}$, $T_{s1}$, $T_{m2}$, $T_{s2}$ when making a recommendation. Also, the opportunistic experts can determine the prices of minor and serious treatment as $p_m\in[c_m,l_m]$ and $p_s\in[c_s,l_s]$. Following the notations in the base model, the consumer's strategy of accepting a minor/serious treatment on the first or second visit can be specified as $A_{m1}$,$A_{s1}$,$A_{m2}$,$A_{s2}$. Next, we analyze the game by using the backward induction analysis. 

First, we spell out the expected payoff of a consumer on the first visit to an expert. Facing a minor treatment recommendation, the consumer can choose to a). accept the recommendation (with a payoff (\ref{eq:endma})); b) reject the recommendation and search; and c). quite the market with her problem unresolved  (with payoff (\ref{eq:endrqm})). In b), after the consumer rejects the first recommendation and searches for a second expert, the consumer can choose to d). accept all recommendations (with a payoff (\ref{eq:endmrmasa})). e.) accept a minor treatment but reject a serious treatment (with a payoff (\ref{eq:endmrmasr})); f). accept a serious treatment but reject a minor treatment (with a payoff (\ref{eq:endmrmrsa})); g). reject all recommendations (with a payoff (\ref{eq:endmrmrsr})). The expected payoff of the consumers can be summarized as follows:

\begin{align}
& -\tau_m p_m-(1-\tau_m)(p_m+p_s+k^{'}) \label{eq:endma} \\
& -\tau_m (l_m)-(1-\tau_m)(l_s) \label{eq:endrqm} \\
& -\tau_m T_{m2}(p_m)-\tau_m (1-T_{m2})(p_s)-(1-\tau_m)T_{s2}(p_s)-(1-\tau_m)(1-T_{s2})(p_m+l_s)-k \label{eq:endmrmasa}  \\
& -\tau_m T_{m2}(p_m)-\tau_m (1-T_{m2})(l_m)-(1-\tau_m)T_{s2}(l_s)-(1-\tau_m)(1-T_{s2})(p_m+l_s)-k \label{eq:endmrmasr}  \\
& -\tau_m T_{m2}(l_m)-\tau_m (1-T_{m2})(p_s)-(1-\tau_m)T_{s2}(p_s)-(1-\tau_m)(1-T_{s2})(l_s)-k \label{eq:endmrmrsa}\\
& -\tau_m (l_m)-(1-\tau_m)(l_s)-k \label{eq:endmrmrsr}
\end{align}

In the expressions above, $\tau_m=\frac{\mu T_{m1}}{\mu T_{m1}+(1-\mu)(1-T_{s1})}$ denotes the conditional probability that the minor treatment recommendation is truthful. Note from the expressions above, it is straightforward to see that (\ref{eq:endrqm})$>$(\ref{eq:endmrmrsr}), indicating that rejecting all recommendations on the second visit is a dominated strategy. 

Similarly, when the consumer is facing a serious treatment recommendation, the consumer can choose to a). accept the recommendation (with a payoff (\ref{eq:endsa})), b).reject the recommendation and search, and c).quite the market with her problem unresolved (with a payoff (\ref{eq:endrqs})). If the consumer rejects the first recommendation and searches for a second expert, the consumer can choose to d). accept all recommendations (with a payoff (\ref{eq:endsrmasa})). e.) accept a minor treatment but reject a serious treatment (with a payoff (\ref{eq:endsrmasr})); f). accept a serious treatment but reject a minor treatment (with a payoff (\ref{eq:endsrmrsa})); g). reject all recommendations on the second visit (with a payoff (\ref{eq:endsrmrsr})). The expected payoff of the consumers can be summarized as follows:

\begin{align}
& -p_s \label{eq:endsa} \\
& -\tau_s (l_s)-(1-\tau_m)(l_m) \label{eq:endrqs} \\
& -\tau_s T_{s2}(p_s)-\tau_s (1-T_{s2})(l_s+p_m)-(1-\tau_s)T_{m2}(p_m)-(1-\tau_2)(1-T_{m2})(p_s)-k \label{eq:endsrmasa}  \\
& -\tau_s T_{s2}(l_s)-\tau_s (1-T_{s2})(l_s+p_m)-(1-\tau_s)T_{m2}(p_m)-(1-\tau_2)(1-T_{m2})(l_m)-k \label{eq:endsrmasr}  \\
& -\tau_s T_{s2}(p_s)-\tau_s (1-T_{s2})(l_s)-(1-\tau_s)T_{m2}(l_m)-(1-\tau_2)(1-T_{m2})(p_s)-k\label{eq:endsrmrsa}\\
& -\tau_s (l_s)-(1-\tau_m)(l_m)-k \label{eq:endsrmrsr}
\end{align}

In the expressions above, $\tau_s=\frac{(1-\mu) T_{s1}}{(1-\mu) T_{s1}+\mu(1-T_{m1})}$ denotes the conditional probability that the serious treatment recommendation is truthful. Note from the expressions above, it is straightforward to see that (\ref{eq:endrqs})$>$(\ref{eq:endsrmrsr}), indicating that rejecting all recommendations on the second visit to an expert is a dominated strategy. 

Then, we examine the incentive compatibility constraints of the opportunistic experts. When the opportunistic experts are facing a consumer on the second visit, the incentive-compatibility constraints of the opportunistic experts are:

\vspace{-0.5cm}
\begin{align}
& IC(Inadequate): (p_m-c_m)A_{m2}\geq(p_s-c_s)A_{s2} \label{eqn:endss2}\\
& IC(Excessive):  (p_m-c_m)A_{m2}\leq(p_s-c_s)A_{s2}   \label{eqn:endsm2}
\end{align}
\vspace{-1.0cm}

When the consumer is visiting the first expert, the incentive-compatibility constraints of the opportunistic experts are:

\vspace{-0.5cm}
\begin{align}
& IC(Inadequate): (p_m-c_m+p_s-c_s)A_{m1}\geq(p_s-c_s)A_{s1} \label{eqn:endss1}\\
& IC(Excessive):  (p_s-c_s)A_{s1}\geq(p_m-c_m)A_{m1}   \label{eqn:endsm1}
\end{align}
\vspace{-1.0cm}

Next, we show that the unique pure strategy equilibrium occurs when $p_s-c_s=p_m-c_m$. By contradiction, suppose $p_s-c_s>p_m-c_m$, facing a consumer with a serious problem on the second visit, the opportunistic expert has incentives to recommend a serious treatment to a consumer since a serious treatment generates a higher profit margin. Then, the consumer on the second visit perceives any minor treatment recommendation as being truthful and would like to accept it such that $A_{m2}=1$.
Note that in a pure strategy equilibrium, an opportunistic expert has incentives to recommend excessively if and only if $ (p_m-c_m)A_{m2}<(p_s-c_s)A_{s2}$, indicating that $A_{s2}=1$. When $A_{m2}=1$ and $A_{s2}=1$ constitute an equilibrium, the incentive-compatibility constraints of the consumers must satisfy (\ref{eq:endmrmasa})$>$(\ref{eq:endmrmasr}), (\ref{eq:endmrmasa})$>$(\ref{eq:endmrmrsa}), (\ref{eq:endsrmasa})$>$(\ref{eq:endsrmasr}) and (\ref{eq:endsrmasa})$>$(\ref{eq:endsrmrsa}). By plugging $\tau_s=\frac{(1-\mu) T_{s1}}{(1-\mu) T_{s1}+\mu(1-T_{m1})}$  into the inequalities above, the conditions can be reduced to $\mu T_{m1}(p_s-l_m)<(1-\mu)(1-T_{s1})(l_s-p_s)$ and $\mu (1-T_{m1})(p_s-l_m)<(1-\mu)T_{s1}(l_s-p_s)$. However, when we focus on pure strategy equilibria, in which $T_{m1}=0$ or $T_{m1}=1$ and $T_{s1}=0$ or $T_{s1}=1$, and the two conditions above cannot hold with strict inequality concurrently, leading to contradiction. Similarly, when $p_s-c_s<p_m-c_m$,  facing a consumer on the second visit, the opportunistic expert has incentives to recommend a minor treatment to a consumer regardless of the consumer's problem since it generates a higher profit margin. In this case, the consumer perceives any serious treatment recommendation as being truthful and would like to accept such that  $A_{s2}=1$.In a pure-strategy equilibrium, the opportunistic expert has the incentives to recommend inadequately if and only if $ (p_m-c_m)A_{m2}>(p_s-c_s)A_{s2}$, which holds when $A_{m2}=1$. When $A_{m2}=1$ and $A_{s2}=1$ constitute an equilibrium, (\ref{eq:endmrmasa})$>$(\ref{eq:endmrmasr}), (\ref{eq:endmrmasa})$>$(\ref{eq:endmrmrsa}), (\ref{eq:endsrmasa})$>$(\ref{eq:endsrmasr}) and (\ref{eq:endsrmasa})$>$(\ref{eq:endsrmrsa}). By plugging $\tau_m=\frac{\mu T_{m1}}{(1-\mu)(1-T_{s1})+\mu T_{m1}}$  into the inequalities above, the conditions can be reduced to $\mu T_{m1}(l_m-p_m)>(1-\mu)(1-T_{s1})p_m$ and $\mu (1-T_{m1})(l_m-p_m)<(1-\mu)T_{s1}p_m$, and the two conditions cannot hold simultaneously for pure strategy pairs such that $T_{m1}=0$ or $T_{m1}=1$ and $T_{s1}=0$ or $T_{s1}=1$.

Last, we investigate the equilibrium conditions when $p_s-c_s=p_m-c_m$ constitutes an equilibrium. When $p_s-c_s=p_m-c_m$, an opportunistic expert gets the same payoff from a minor treatment and a serious treatment, and it is a weakly dominant strategy for an opportunistic expert to recommend truthfully when the consumer is with a minor problem since by doing so, the consumer perceives any serious treatment recommendation to be truthful and accepts any serious treatment recommendation such that $A_{s1}=A_{s2}=1$, which maximizes the expert's expected payoff. Similarly, when the consumer with a minor problem is visiting a second expert, the expert also recommends truthfully, and the consumer accepts a minor treatment recommendation on the second visit $A_{m2}=1$ since the opportunistic has no incentives to recommend inadequately as $p_s-c_s=p_m-c_m$. Next, we identify the conditions when $T_{s1}=0$ constitutes the unique pure strategy equilibrium, since $T_{s1}=1$ never constitutes an  equilibrium as we have shown in part (i). When $T_{s1}=0$, (\ref{eqn:endss1}) holds with strictly inequality in a pure strategy equilibrium only if $A_{m1}=1$. When $A_{m1}=1$ constitutes an equilibrium, we have  (\ref{eq:endma})$>$(\ref{eq:endrqm}) and (\ref{eq:endsa})$>$(\ref{eq:endsrmasa}), and the two conditions can be reduced to $(1-\mu)(l_s-p_s-p_m-k^{'})>\mu(l_m-p_m)$ and $k>(1-\mu)(p_m+k^{'})$. To maxmize his expected payoff, an opportunistic expert sets the price of a minor treatment as $p_m=l_m$, and the two conditions above can be achieved when $\mu>1-\frac{k}{l_m+k^{'}}$. In sum, we have shown that $(T_{m1}=1,T_{s1}=0,T_{m2}=1,T_{s2}=1,A_{m1}=1,A_{s1}=1,A_{m2}=1,T_{s2}=1)$ constitutes the unique pure-strategy equilibrium when $\mu>1-\frac{k}{l_m+k^{'}}$, in which $p_m=l_m$, $p_s=l_m-c_m+c_s$. \hspace{0.3cm} $\Box$

\end{document}